\documentclass{article}

\usepackage{graphicx}
\usepackage{newtxtext}
\usepackage{newtxmath}
\usepackage[square,sort,comma,numbers]{natbib}
\usepackage{hyperref}
\hypersetup{colorlinks,allcolors=black}
\usepackage{bm} % third party packages, Jinghua
\usepackage{caption}
\usepackage{subcaption}
\usepackage{xcolor} % third party packages, Jinghua
\usepackage{authblk}
\hypersetup{
    colorlinks = true,
    urlcolor   = blue,
    citecolor  = black,
}

\newcommand{\RomanNumeralCaps}[1]
\linenumbers

\title{An Enhanced Spectral Boundary Integral Method for Modeling \textcolor{black}{Highly} Nonlinear Water Waves in Variable Depth}

\author[1,2]{Jinghua Wang}

\affil[1]{Department of Civil and Environmental Engineering, The Hong Kong Polytechnic University, Hong Kong}
\affil[2]{Shenzhen Research Institute, The Hong Kong Polytechnic University, Shenzhen, China}
\date{}  
\begin{document}

\maketitle
\begin{abstract}
%All papers should feature a single-paragraph abstract of no more than 250 words 
This paper presents a new numerical model based on the \textcolor{black}{highly} nonlinear potential flow theory for simulating the propagation of water waves in variable depth. A new set of equations for estimating the surface vertical velocity is derived based on the boundary integral equation considering the water depth variability. A successive approximation scheme is also proposed in this study for calculating the surface vertical velocity. With the usage of Fast Fourier Transform, the model can be efficiently used for simulating \textcolor{black}{highly} nonlinear water waves on large spatiotemporal scale in a phase-resolving approach. The new model is comprehensively verified and validated through simulating a variety of nonlinear wave phenomenon including free propagating solitary wave, wave transformations over submerged bar, Bragg reflection over undulating bars, nonlinear evolution of Peregrine breather, \textcolor{black}{obliquely propagating uniform waves} and extreme waves in crossing random seas. Good agreements are achieved between the numerical simulations and laboratory measurements, indicating that the new model is sufficiently accurate. A discussion is presented on the accuracy and efficiency of the present model, which is compared with the Higher-Order \textcolor{black}{Spectral} method. The results show that the present model can be significantly more efficient at the same level of accuracy. It is suggested that the new model developed in the paper can be reliably used to simulate the nonlinear evolution of ocean waves in phase-resolving approach to shed light on the dynamics of nonlinear wave phenomenon taking place on a large spatiotemporal scale, which may be computationally expensive by using other existing methods.     
\end{abstract}

%\begin{keywords}
%Pseudo-spectrum method, Fast Fourier Transform, Boundary integral equation, Wave-seabed interactions.
%\end{keywords}

%{\bf MSC Codes }  {\it(Optional)} Please enter your MSC Codes here

\section{Introduction}
\label{sec:introduction}
Investigation of nonlinear water waves using numerical models has been extensively studied for decades, through either phase-averaged  or phase-resolving approach. A variety of wave models were developed with their own approximations, and a wide range of applications have been carried out based on such models. To date, the phase-averaged approach has been advanced to the third-generation wind wave models, eg., WAM, WAVEWATCHIII, SWAN, etc., which are widely employed to forecast/hindcast the evolution of spectra of wind waves and swells \citep{komen1996dynamics,holthuijsen10,cavaleri2007wave,cavaleri2018wave}. On the other hand, the phase-resolving models are developed to simulate the waves propagation and their interactions with bathymetry and structures with the free surface of each individual wave being resolved. Whereas due to the complexities of the physics on different spatiotemporal scales being involved, there is no single model that can deal with all physical processes at once for 'universal' purposes. For simplicity, a summary of the development of both phase-averaged and phase-resolving approaches is provided in \cite{higuera2023}, only some key phase-resolving models are summarized herein.

With increasing computation power of modern computers in recent decades, simulating waves on a large scale of hundreds of wavelengths in phase-resolving approach becomes popular. Given the phase information, the wave surface and velocity field can be re-constructed, which facilitates the investigations of nonlinear wave phenomenon such as rogue/freak wave mechanisms, water scattering on variable seabed bathymetry or shearing current, wave radiations due to multi-body structures (eg., ocean renewable farms composed of multiple wave energy converters or wind turbines), and so on. Thereafter, only those phase-resolving models developed for such purposes will be briefly reviewed in the followings, which in general can be categorized into depth-integrated models and potential flow models. Nevertheless, the CFD (short for Computational Fluids Mechanics) models that solve the Navier Stokes equations or the recently emerging Lattice-Boltzmann models are not ruled out regarding such purposes, given that the sufficient computational power is provided.     

The depth-integrated models, such as Mild Slope (MS) equation \citep[and so on]{Kirby1984,Kirby1986,Massel993,Sharmaetal2014,Toubouletal2016}, Nonlinear Shallow Water (NLSW) equation \citep[and so on]{xing2017,kurganov2018,castro2019}, Boussinesq-type models \citep[and so on]{Peregrine1967,Liu1995,Wei1995,Chen1995,Hsiao2002,Lynett2004,Lynett2004b,Chen2006,Shi2012,Liu2018} and weighted-residual \textcolor{black}{(WR)} models \citep[and so on]{Green1974,Green1976, Green1976a,Ertekin1986,Shields1988,Lannes2009,yang2020,yang2022depth}, were developed with the vertical dependence being removed through integrating the continuity and momentum equations over the water depth. Among these models, the NLSW model is widely used in simulating long waves such as tides, tsunamis and storm surges on a regional or global scale. Various commercial and open-source packages are readily available, eg., Delft3D, MIKE21/3, FVCOM, MOST, COMCOT, SLOSH, ADCIRC, etc. On the other hand, the MS, Boussinesq and \textcolor{black}{WR} models are preferably employed to study waves on coastal or basin scales with an emphasis on wave scattering problems. Meanwhile, to model complex physics such as wave breaking, bottom friction, subgrid-scale lateral mixing effects, etc., \textit{ad hoc} features are often incorporated. To extend the their range of applications, the Boussinesq models are significantly improved over their accuracy in simulating strong nonlinear and highly dispersive waves in deep waters (water depth greater than half of the characteristic wavelength), such as COULWAVE \citep{Lynett2004,Lynett2004b} and FUNWAVE-TVD \citep{Chen1995,kirby1998funwave,Chen2006,Shi2012}, etc. Meanwhile, the \textcolor{black}{WR} models origin from nonlinear plate and shell theory, which shows equivalence to the fully nonlinear and weakly dispersive Boussinesq models for dealing with water wave problems \citep{Lannes2009}. Besides, they use polynomials to approximate the horizontal velocities, and the residuals are estimated via Galerkin method. For simplicity, a comparison of selected Boussinesq models and \textcolor{black}{WR} models is summarized in \cite{higuera2023}, and they will not be discussed further here as the application by using depth-integrated models in deep waters is very rare. 

Another category of the phase-resolving models is based on the fully nonlinear potential flow (FNPF) theory, which assumes that the fluid is inviscid and irrotational. Depending on the methodologies employed for solving the boundary value problem (the so-called diagnostic equation), the Boundary Element (BE) models \citep[and so on]{longuet1976deformation,tanaka1983stability,grilli1989efficient,dold1992efficient,ferrant1997,celebi1998,henderson1999unsteady,xue2001,grilli2001}, Finite Element (FE) models \citep[and so on]{wu1994,wu1995,ma1997,ma1998,clauss1999,sriram2006,ma2006,yan2006,ma2009,yan2010}, Spectral Element (SE) models \citep[and so on]{patera1984spectral,karniadakis2013spectral,engsig2016stabilised}, Finite Difference (FD) models \citep[and so on]{li1997three,bingham2007,engsig_kurap2009,engsig_kurap2012}, \textcolor{black}{and models based on polynomial expansions \citep[and so on]{kennedy1997fully,shao2014harmonic,yates2015accuracy,raoult2016validation}, } were suggested. The BE models can either solve the full equations in time domain, or they can also be solved in frequency domain, leading to a number of second-order diffraction-refraction models such as WAMIT, ANSYS AQWA, and so on. 
\textcolor{black}{The main drawback of BEM is the necessity to assemble and solve full linear systems, which later leads to several efficient methods through using higher-order element \citep{grilli1997numerical,grilli2001}, matrix decomposition based on Fourier or Wavelet transforms \citep{alpert1993wavelet,potts2003fast} or hierarchical algorithms \citep{goreinov1997matrix,hackbusch1999sparse} including the
Fast Multipole Method (FMM) \citep{korsmeyer1993multipole,fochesato2006fast}. The FMM is advantageous as it can be implemented on large parallel computer architectures with distributed memory \citep{wang1995efficient,harris2022nonlinear}, e.g., the ExaFMM library \citep{yokota2011biomolecular}.}

In contrast to the BE models that require \textcolor{black}{discretization over the boundaries, the other FNPF models adopt volume-based discretization or polynomial representations. For instance,} the FE and SE models utilize the polynomials or trignometric ploynomials as the shape functions to approximate the solutions. \textcolor{black}{With the solution of the Laplace equation being represented in series of harmonic or Chebyshev polynomials, the Harmonic Polynomial Cell (HPC) model \citep{shao2014harmonic} and Whispers3D \citep{raoult2016validation} were proposed.} The FD models employ the \(\sigma\)-coordinate transformation and finite difference scheme for evaluating the spatial gradient, \textcolor{black}{e.g., the open-source tools REEF3D::FNPF \citep{wang2022flexible}, OceanWave3D \citep{engsig_kurap2012}. The latter }is widely used standalone or can be coupled with other CFD tools \citep{jacobsen2017waves2foam}. \textcolor{black}{The} FNPF models can be used in arbitrary water depth to simulate free propagating waves or interactions between waves and (floating) structures. Hence, they can mimic the laboratory settings and are sometimes referred to as Numerical Wave Tank (NWT) \textcolor{black}{\citep{grilli1997numerical}}.  

More recently, it is of interest to model free propagating waves on large spatiotemporal scales for investigating the nonlinear mechanisms of rogue waves \citep{kharif2008rogue,pelinovsky2008extreme}. For the sake of efficiency and accuracy, the Higher-Order Spectral (HOS) method \citep{west1987new,dommermuth1987high} was employed, which applies the perturbation and Taylor series to the velocity potential with the assumption that the wave steepness is small. The diagnostic procedure demands repetitive conversions of the quantities between Fourier and physical spaces. With the usage of Fast Fourier Transform (FFT) and aliasing error treatment, the solution can be derived efficiently, and such a method is often referred to as pseudo-spectrum method. Another alternative method relies on the Dirichlet-Neumann operator or Spectral Continuation method \textcolor{black}{\citep{craig1993numerical,nicholls1998traveling,bateman2001efficient,  xu2009numerical}, which is essentially equivalent with the HOS method regarding the operator expansions \citep{schaffer2008comparison}. The capability to account for variable water depth was later introduced \citep{liu1998generalized,smith1998operator,guyenne2005numerical,gouin2016development,gouin2017propagation}, as well as moving bottom topography \citep{guyenne2008high,hao2022numerical}.} Owing to the FFT, the HOS method can be efficiently used to simulate random directional seas on a scale of \(\sim 10^2 \) peak wavelengths in phase-resolving approach \citep{xiao2013rogue,ducrozet20073}. This method is now openly available and is widely disseminated through two open-source packages known as HOS-NWT \citep{ducrozet2012modified} and HOS-Ocean \citep{ducrozet2016hos}. 

Alternatively, another pseudo-spectrum method called the Enhanced Spectral Boundary Integral (ESBI) method was developed \citep{clamond2001fast,fructus2005,clamond2005efficient,fructus2007,wang2015,wang2018}. This method reformulates the boundary integral equation and estimates the surface vertical velocity in convolutions. The diagnostic procedure is closed by including a small remaining integral term only when wave steepness is sufficiently large, otherwise this term can be ignored. The convolutions can be evaluated through applying FFTs leading to an efficient numerical scheme. It can be shown that the ESBI method can achieve accuracy in same magnitude with the HOS method with perturbation order \(M=7\) for simulating a free propagating Stokes waves of steepness \(\varepsilon=0.35\) (\(\varepsilon=k_pA\) where \(k_p\) is the peak wavenumber and \(A\) is the wave amplitude), but uses much less number of FFT operations. More details will be provided in the section \ref{sec:dis}. 

\textcolor{black}{To summarize, the depth-integrated models are widely used in coastal region over shallow to intermediate water depth,
but require special attention when the water depth becomes too large. Their application on modelling nonlinear (quasi-) resonance interactions and formation of rogue waves in deep water is rare. On the other hand, FNPF models are ideal for study deep water waves and their interactions with floating structures, or wave scattering in coastal region subject to presence of steep bottom obstacles and influences of coastlines. A practical application of the FNPF model in modelling coastal water waves was demonstrated in \cite{wang2022flexible} by using REEF3D::FNPF with a computational domain of 40 $\times$ 20 peak wavelengths (14 km $\times$ 7 km) for a 850 peak period (3.5 hour) sea state. However, they are still computationally expensive for such a large spatiotemporal scale when compared with, e.g., HOS model, which has been used on a scale of 128 $\times$ 128 peak wavelengths \citep{xiao2013rogue}. Hence, to speed up the simulations, parallel computation techniques are often desired. Lastly, the pseudo-spectrum models can be very efficient regarding the purpose of modelling wave propagation from deep water to shallow water. Yet, it is admitted that they are not suitable for complex coastline configurations so far. Moreover, there is currently a lack of efficient tool for simulating highly nonlinear waves propagation from deep to shallow water with emphasis on (quasi-) resonance and wave scattering simultaneously. It becomes the driver of this study, and a new development will be proposed based on a pseudo-spectrum model.  }

The ESBI method has been employed to investigate the statistical and deterministic properties of rogue waves in crossing seas and directional seas subject to shearing current in infinitely deep water \citep{wang2021extreme,wang2021modeling,wang2022two}. It is demonstrated that the ESBI method is computationally efficient for simulating fully nonlinear wave while resolving the wave phases on large spatiotemporal scale. It is reiterated that good efficiency of a numerical model is particularly important for simulating random seas, since a large number of realizations are often required to minimize the uncertainties due to the random phase assumption. Nevertheless, the ESBI method has not been extended to study nonlinear waves in coastal regions so far. Hereby, this study will propose a set of new formulations and a successive approximation scheme for calculating the surface vertical velocity \textcolor{black}{based on the original work of \cite{fructus2007}. In their study, the vertical velocity is estimated with the 3rd order convolutions including two additional integration terms that are costly to compute. The present study will suggest a higher-order formulations to approximate the convolutions up to the 7th order, without the need to consider the integration terms. By doing so, the efficiency of the present model can be substantially improved}. The new method is firstly verified for simulations of a solitary wave propagating over constant and variable depth, and the total energy conservation and shoaling induced amplitude amplification is examined. It is then used to simulate a variety of nonlinear wave phenomenons in finite and deep waters, including wave transformation over submerged bar, Bragg reflection over submerged sinusoidal bars, nonlinear evolution of a Peregrine breather, \textcolor{black}{obliquely propagating uniform waves} and formation of extreme waves in random crossing seas. The simulation results are compared with laboratory measurements, and good agreements are achieved. At last, a discussion on the computational efficiency of the ESBI method is provided, and a comparison with the HOS method is presented. 

\textcolor{black}{It should be addressed that the application of the improved ESBI model is restrictive on simulating wave propagating over steep bottom obstacles. It is because that the seabed surface is single-valued, and the product of characteristic wavenumber and seabed height is assumed to be small to utilize Taylor expansions. Regarding this limitation, a detailed discussion is provided in section \ref{sec:limit}. Moreover, similar with other models of this type, the present ESBI model needs special treatment to adequately represent breaking waves and associated energy dissipation. Many studies have been carried out to extend the FNPF models via physically meaningful approaches relying on analogies with well-known dissipative phenomena with calibrated constant depending on the type of breaker \citep[and so on]{svendsen1978wave,schaffer1993boussinesq,kennedy2000boussinesq,guignard2001modeling,grilli2020fully,mohanlal2023unified}. By doing so, the dissipation can also suppress the overturning of the free surface allowing for stabilized simulations. }      

This paper is structured in such a way that Section \ref{sec:formula} presents the new formulations and the numerical successive approximation scheme. The model verifications and validations are provided in Section \ref{sec:verify-validate}. Then Section \ref{sec:dis} will discuss the efficiency and accuracy of the present model and the comparison with HOS model. At last, the concluding remarks are highlighted in Section \ref{sec:con}. 

\section{Formulations and numerical schemes}\label{sec:formula}
\subsection{Governing equations}
\begin{figure}
  \centerline{\includegraphics[width=120mm]{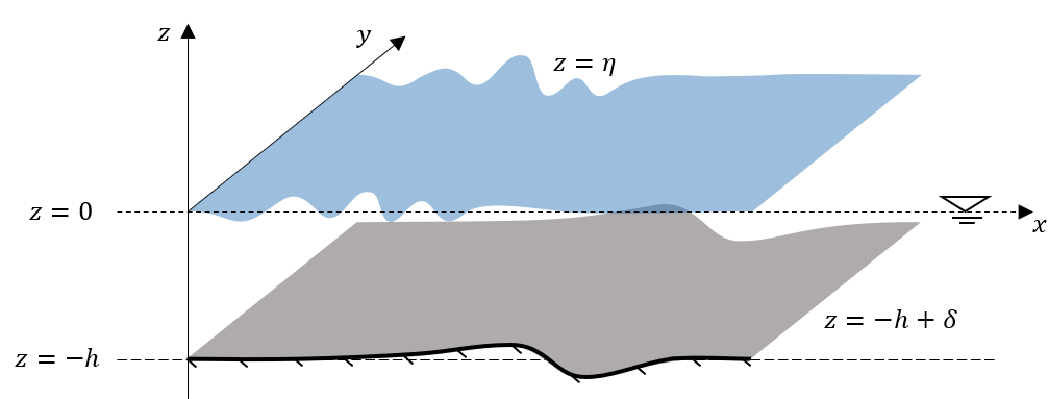}}
  \caption{Description of the problem and Cartesian coordinate system.}
\label{fig:cartesian}
\end{figure}

The Cartesian coordinate system is used here with \(\bm{x} = (x,y)\) \textcolor{black}{being horizontal} and \(z\) being \textcolor{black}{vertical} coordinates, respectively. The still water level is at \(z=0\), and the seabed elevation is denoted by \(z=-h+\delta\), where \(h\) and \(\delta\) are the constant and variable parts, respectively (see figure \ref{fig:cartesian}). Unless otherwise specified, the variables are being non-dimensionalized, i.e., distance (\(\bm{x}\), \(z\), \(h\) and \(\delta\)) and time (\(t\)) are multiplied by the peak wavenumber \(k_p\) and circular frequency \(\omega_p = \sqrt{gk_p}\), respectively, where \(g\) is the gravitational acceleration.

The potential flow theory assumes that the fluid is inviscid and irrotational. This leads to the continuity equation being written as the Laplace equation, i.e.,
\begin{equation}\label{eq:laplace}
    \Delta \phi=0,
\end{equation}

\noindent where $\Delta=\partial^2/\partial x^2+\partial^2/\partial y^2+\partial^2/\partial z^2$ is the Laplacian operator, $\phi$ is the velocity potential (rescaled by \(\sqrt{k_p^3/g}\)). 

On the other hand, the governing equation can also be posed as the boundary integral equation of Green's theorem, i.e.,
\begin{equation}\label{eq:boundaryIntegral}
    \lambda \phi(\bm{x}',z')=\iint_{\Gamma} \left[\phi(\bm{x},z) \frac{\partial \psi}{\partial n} - \psi \frac{\partial \phi}{\partial n}(\bm{x},z) \right]\mathrm{d}s,
\end{equation}

\noindent where $(\bm{x}',z')$ is the position (evaluation point) that can be located on or enclosed by the surface boundary $\Gamma$, $\psi$ is an arbitrary function satisfying Eq.(\ref{eq:laplace}) and is usually expressed as the \textcolor{black}{free space Green's function based on} the distance between $(\bm{x}',z')$ and $(\bm{x},z)$;  $n$ is the unit vector perpendicular to the surface boundary and points outwards. $\lambda$ is \textcolor{black}{the interior solid angle formed by the boundary at $\bm{x}'$, which equals to $2\pi$ for a smooth surface}.
%a constant equal to $4\pi$ if $(\bm{x}',z')$ is an interior point or $2\pi$ if $(\bm{x}',z')$ is located on the boundary. 

The primary advantage of using the velocity potential is that it is a scalar quantity. Therefore, the number of unknowns is reduced compared to the Euler or Navier-Stokes equations, as the velocity vectors can be obtained directly by calculating the gradient of the velocity potential. Eq.(\ref{eq:boundaryIntegral}) can be used to find the solution of velocity on the boundary, which will be demonstrated in Section \ref{sec:BIE}.

The free surface boundary conditions for the potential flow wave theory consists of those on the water free surface \(z=\eta(\bm{x},t)\): 
 \begin{equation}
 \frac{\partial \eta}{\partial{t} } + \nabla \phi \cdot \nabla \eta - \frac{\partial \phi}{\partial z}  = 0,
 \label{eq:kbc}
 \end{equation}
\begin{equation}
 \frac{\partial \phi}{\partial{t} } + \frac{1}{2}\left[ \left|\nabla \phi \right|^2 + \left(\frac{\partial \phi}{\partial z} \right) ^2 \right] + \eta = 0,
 \label{eq:dbc}
 \end{equation}
 and that on seabed surface \(z = -h+ \delta\):
\begin{equation}\label{eq:bottomBC}
    \frac{\partial \phi}{\partial z}-\nabla \phi \cdot \nabla \delta= 0,
 \end{equation}
 
 \noindent where \(\nabla=(\partial_x,\partial_y)\) is the horizontal gradient operator, and the seabed is assumed to be stationary and impermeable.

\subsection {Prognostic equations based on free surface boundary conditions}

Eqs.(\ref{eq:kbc}) and (\ref{eq:dbc}) are identical to the canonical pair derivable from the Hamiltonian system \citep{zakharov1968}. They can be rewritten as a skew-symmetrical form \citep{fructus2005}:
\begin{equation}
\frac{\partial \Psi}{\partial t} + \mathcal{A}\Psi = \mathcal{N},
\label{eq:prognostic}
\end{equation}

\noindent where
\begin{eqnarray}
\Psi = \begin{pmatrix}
k \mathcal{F}\left\{ \eta \right\} \\
k\omega \mathcal{F}\left\{ \widetilde{\phi} \right\} \\
\end{pmatrix},\quad
\mathcal{A} = \begin{bmatrix}
0 & - \omega \\
\omega & 0 \\
\end{bmatrix}, \quad
\mathcal{N} = \begin{pmatrix}
k\left( \mathcal{F}\left\{ V \right\} - k\tanh{(kh)} \mathcal{F}\left\{ \widetilde{\phi} \right\} \right) \\
\frac{k\omega}{2} \mathcal{F}\left\{ \frac{\left( V + \nabla\eta \cdot \nabla\widetilde{\phi} \right)^{2}}{1 + \left| \nabla\eta \right|^{2}} - \left| \nabla\widetilde{\phi} \right|^{2} \right\} \\
\end{pmatrix},
\end{eqnarray}

\noindent and \(\widetilde{\phi}=\phi(\bm{x},z=\eta,t)\) denotes velocity potential on free surface, \(V = \sqrt{1 + |\nabla\eta|^{2}}\partial\phi/\partial{n}\) is the vertical velocity of the surface elevation, while \(\mathcal{F}\{*\}\) is the Fourier transform defined as:
\begin{equation}
\mathcal{F}\{*\} = \iint_{-\infty}^{\infty} * e^{-\mathrm{i} \bm{k} \cdot \bm{x}} \mathrm{d} \bm{x},
\end{equation}
with \(\mathcal{F}^{-1}\{*\}\) being its inverse transform and \(\mathrm{i}=\sqrt{-1}\). The Fourier transform is implemented numerically by using the Fast Fourier Transform (FFT). 

Eq.(\ref{eq:prognostic}) can be further reformulated as:
\begin{eqnarray}
\Psi\left( t \right) = e^{- \mathcal{A}\left( t - t_{0} \right)} \left[   \Psi\left( t_{0} \right) + \int_{t_{0}}^{t}{e^{\mathcal{A}\left( t - t_{0} \right)}\mathcal{N} \mathrm{d} t} \right], 
\end{eqnarray}
and can be used as the prognostic equation for updating unknowns \(\eta\) and \(\widetilde{\phi}\) in time with the integration term evaluated by using a \textcolor{black}{six-stage fifth-order Runge-Kutta method with embedded fourth-order solution \citep{clamond2007note,wang2015}. The time step size is controlled adaptively according to the difference of the solution between fifth- and fourth-order solutions. In general, to keep the difference below 1\%, the time step size is automatically adjusted to about 1/20 peak wave period. This is sufficient for the validation cases in section \ref{sec:submerged} $\sim$ \ref{sec:peregrine}. For large spatiotemporal simulations of strong nonlinear waves, a tolerance of 0.1\% is selected corresponding to a time step size of about 1/50 peak wave period, which applies to the crossing sea simulations in section \ref{sec:crossing}.} 

It is noted that to update the solutions \((\eta,\widetilde{\phi})\), the gradient of \(\phi\), more specifically, velocity \(V\) needs to be diagnosed from solving the boundary value problem of Eq. (\ref{eq:laplace}) or Eq. (\ref{eq:boundaryIntegral}). 
In contrast to the conventional Boussinesq-type models, \textcolor{black}{which usually rely on depth-integrated horizontal momentum equation and the horizontal velocities at a reference level are of interests}, the present method seeks for the solutions of the vertical velocity at the free surface, which is explained in the next section.  

\subsection {Diagnostic equations based on the boundary integral equation}\label{sec:BIE}
To obtain values of \(V\), the boundary integral equations for evaluation points on free surface and seabed are employed. Based on that, a new set of equations expressed in convolutions will be given in the following subsections. 

\subsubsection {Evaluation points on free surface}
For the evaluation points on free surface, the boundary integral equation can be written as \citep{fructus2007}:
\begin{equation}
  \iint_{S + B}^{}{\left( \frac{1}{r_{S}} + \frac{1}{r_{B}} \right)\frac{\partial\phi'}{\partial{n'}}} \mathrm{d} s' = 2\pi\widetilde{\phi} + \iint_{S + B}^{}{\phi'\frac{\partial}{\partial{n}^{'}}\left( \frac{1}{r_{S}} + \frac{1}{r_{B}} \right)} \mathrm{d} s' ,
\label{eq:bi}
\end{equation}

\noindent where \(r_S^2 = R^2+(z'-z)^2\), \(r_B^2 = R^2+(z'+z+2h)^2\), \(R=|\bm{R}|=|\bm{x'}-\bm{x}|\) is the horizontal distance between source and evaluation points, \(S\) and \(B\) are the free surface and seabed surface, respectively, and the prime denotes variables corresponding to the source points.  

Hereby, it is assumed that the waves do not overturn, and the seabed surface is single-valued and stationary. \textcolor{black}{Introducing $V'={\partial\phi'}/{\partial{n'}}\sqrt{1 + |\nabla'\eta'|^{2}\ }$, $\mathrm{d} s'=\sqrt{1 + |\nabla'\eta'|^{2}\ } \mathrm{d}\bm{x}'$ at free surface $S$, and $\mathrm{d} s'=\sqrt{1 + |\nabla'\delta'|^{2}\ } \mathrm{d}\bm{x}'$ at bottom surface $B$,} then Eq.(\ref{eq:bi}) can be reformulated as:
\begin{eqnarray}
\iint_{S_{0}}^{}{\left( \frac{1}{r_{S}} + \frac{1}{r_{B}} \right)V' \mathrm{d} \bm{x}'} &=& 2\pi\widetilde{\phi}  + \iint_{S_{0}}{\widetilde{\phi}'\frac{\partial}{\partial{n'}}\left( \frac{1}{r_{S}} + \frac{1}{r_{B}} \right)\sqrt{1 + |\nabla'\eta'|^{2}\ } \mathrm{d} \bm{x}'} \nonumber\\
&&- \iint_{B_{0}}^{}{{\widetilde{\varphi}}'\frac{\partial}{\partial{n'}}\left( \frac{1}{r_{S}} + \frac{1}{r_{B}} \right)\sqrt{1 + |\nabla'\delta'|^{2}\ } \mathrm{d} \bm{x}'},
\label{eq:bi-s}
\end{eqnarray}

\noindent where  \(\widetilde{\varphi} = \phi(\bm{x},z = - h + \delta,t)\)
is the velocity potential on the seabed surface, \(S_{0}\) and \(B_{0}\) are the projected area of \(S\) and \(B\) to the horizontal plane, respectively.

%Hence, for the integral part on the surface, \(z' = \eta'\), \(z = \eta\), so that \(r_{S}^{2} = R^{2} + \left( \eta' - \eta \right)^{2}\), \(r_{B}^{2} = R^{2} + \left( \eta' + \eta + 2h \right)^{2}\), and

Furthermore, it is note that the leading order of the integrands in Eq. (\ref{eq:bi-s}) can be approximated by \textcolor{black}{applying Taylor expansions about $(\eta,\eta') = 0$ and $(\delta,\delta') = 0$}:
\textcolor{black}{
\begin{eqnarray}\label{eq:bi-s_start}
\frac{1}{r_{S}} + \frac{1}{r_{B}} & = & \sum_{j=0}^{3} \frac{(-1)^j}{(2j)!} \left[ \eta_{-}^{2j} (\nabla' \cdot \nabla')^j \frac{1}{R} + \eta_{+}^{2j} (\nabla' \cdot \nabla')^j \frac{1}{R_{2h}} \right] \nonumber\\
&& + \sum_{j=1}^{3} \frac{(-1)^{j-1}}{(2j-1)!} \eta_{+}^{j} (\nabla' \cdot \nabla')^{j-1} \frac{\partial}{\partial(2h)}\frac{1}{R_{2h}} + O(\varepsilon^7),
%\frac{1}{r_{S}} + \frac{1}{r_{B}} & = & \frac{1}{R} + \frac{1}{R_{2h}} + \left( \eta' + \eta \right)\frac{\partial}{\partial(2h)}\frac{1}{R_{2h}} - \frac{1}{2}\left( \eta' - \eta \right)^{2}\nabla' \cdot \nabla'\frac{1}{R} \nonumber\\
%&&- \frac{1}{2}\left( \eta' + \eta \right)^{2}\nabla' \cdot \nabla'\frac{1}{R_{2h}} - \frac{1}{6}\left( \eta' + \eta \right)^{3}\nabla' \cdot \nabla'\frac{\partial}{\partial(2h)}\frac{1}{R_{2h}} \nonumber\\
%&&+ \frac{1}{24}\left( \eta' - \eta \right)^{4}\left( \nabla' \cdot \nabla' \right)^{2}\frac{1}{R} + \frac{1}{24}\left( \eta' + \eta \right)^{4}\left( \nabla' \cdot \nabla' \right)^{2}\frac{1}{R_{2h}} \nonumber\\
%&&+ \frac{1}{120}\left( \eta' + \eta \right)^{5}\left( \nabla' \cdot \nabla' \right)^{2}\frac{\partial}{\partial(2h)}\frac{1}{R_{2h}} - \frac{1}{720}\left( \eta' - \eta \right)^{6}\left( \nabla' \cdot \nabla' \right)^{3}\frac{1}{R} \nonumber\\
%&&- \frac{1}{720}\left( \eta' + \eta \right)^{6}\left( \nabla' \cdot \nabla' \right)^{3}\frac{1}{R_{2h}} + O(\varepsilon^7),
\end{eqnarray}
}
%\noindent meanwhile,
\textcolor{black}{
\begin{eqnarray}
\frac{\partial}{\partial n'}\bigg(  \frac{1}{r_{S}} &+& \frac{1}{r_{B}} \bigg)\sqrt{1 + |\nabla'\eta'|^{2}\ }  \nonumber\\ 
&=& \frac{\partial}{\partial(2h)}\frac{1}{R_{2h}} +\sum_{j=1}^{3} \frac{(-1)^{j}}{(2j)!} \nabla' \cdot \left( \eta_{+}^{2j} (\nabla' \cdot \nabla')^{j-1} \nabla'\frac{\partial}{\partial(2h)}\frac{1}{R_{2h}} \right) \nonumber\\
&&  +\sum_{j=1}^3 \frac{(-1)^j}{(2j-1)!} \bigg[ \nabla' \cdot \bigg( \eta_{-}^{j} (\nabla' \cdot \nabla')^{j-1} \nabla'\frac{1}{R} +  \eta_{+}^{j} (\nabla' \cdot \nabla')^{j-1} \nabla'\frac{1}{R_{2h}} \bigg) \bigg]  \nonumber\\
&& + O(\varepsilon^7),
%& =& \frac{\partial}{\partial(2h)}\frac{1}{R_{2h}} - \nabla' \cdot \left( \left( \eta' - \eta \right)\nabla'\frac{1}{R} \right) - \nabla' \cdot \left( \left( \eta' + \eta \right)\nabla'\frac{1}{R_{2h}} \right)\nonumber\\
%&&   - \nabla' \cdot \left( \frac{1}{2}\left( \eta' + \eta \right)^{2}\nabla'\frac{\partial}{\partial(2h)}\frac{1}{R_{2h}} \right) + \frac{1}{6}\nabla' \cdot \left( \left( \eta' - \eta \right)^{3}\nabla'\nabla' \cdot \nabla'\frac{1}{R} \right) \nonumber\\
%&& + \frac{1}{6}\nabla' \cdot \left( \left( \eta' + \eta \right)^{3}\nabla'\nabla' \cdot \nabla'\frac{1}{R_{2h}} \right) + \frac{1}{24}\nabla' \cdot \left( \left( \eta' + \eta \right)^{4}\nabla'\nabla' \cdot \nabla'\frac{\partial}{\partial(2h)}\frac{1}{R_{2h}} \right) \nonumber\\
%&& - \frac{1}{120}\nabla' \cdot \left( \left( \eta' - \eta \right)^{5}\nabla'\left( \nabla' \cdot \nabla' \right)^{2}\frac{1}{R} \right) - \frac{1}{120}\nabla' \cdot \left( \left( \eta' + \eta \right)^{5}\nabla'\left( \nabla' \cdot \nabla' \right)^{2}\frac{1}{R_{2h}} \right) \nonumber\\
%&&- \frac{1}{720}\nabla' \cdot \left( \left( \eta' + \eta \right)^{6}\nabla'\left( \nabla' \cdot \nabla' \right)^{2}\frac{\partial}{\partial(2h)}\frac{1}{R_{2h}} \right) + O(\varepsilon^7),
\end{eqnarray}
}
%For the integral part on the bottom, \(z' = \delta' - h\),\(z = \eta\), so that \(r_{S}^{2} = R^{2} + \left( \delta' - \eta - h \right)^{2}\),\(r_{B}^{2} = R^{2} + \left( \delta' + \eta + h \right)^{2}\), and

\noindent and
\begin{eqnarray}\label{eq:bi-s_end}
\frac{\partial}{\partial{n}'}\bigg( \frac{1}{r_{S}} &+& \frac{1}{r_{B}} \bigg)\sqrt{1 + |\nabla'\delta'|^{2}\ } \nonumber\\
 &=&- 2\nabla' \cdot \left( \delta'\nabla'\frac{1}{R_{h}} \right) - 2\eta\nabla' \cdot \left( \delta'\nabla'\frac{\partial}{\partial h}\frac{1}{R_{h}} \right) \nonumber\\
&& + \frac{1}{3}\nabla' \cdot \left( \delta'^{3}\nabla'\nabla' \cdot \nabla'\frac{1}{R_{h}} \right) + \eta^{2}\nabla' \cdot \left( \delta'\nabla'\nabla' \cdot \nabla'\frac{1}{R_{h}} \right) \nonumber\\
&& + \frac{1}{3}\eta\nabla' \cdot \left( \delta'^{3}\nabla'\nabla' \cdot \nabla'\frac{\partial}{\partial h}\frac{1}{R_{h}} \right) + \frac{1}{3}\eta^{3}\nabla' \cdot \left( \delta'\nabla'\nabla' \cdot \nabla'\frac{\partial}{\partial h}\frac{1}{R_{h}} \right) \nonumber\\
&& - \frac{1}{60}\nabla' \cdot \left( \delta'^{5}\nabla'\left( \nabla' \cdot \nabla' \right)^{2}\frac{1}{R_{h}} \right) - \frac{1}{6}\eta^{2}\nabla' \cdot \left( \delta'^{3}\nabla'\left( \nabla' \cdot \nabla' \right)^{2}\frac{1}{R_{h}} \right) \nonumber\\
&&- \frac{1}{12}\eta^{4}\nabla' \cdot \left( \delta'\nabla'\left( \nabla' \cdot \nabla' \right)^{2}\frac{1}{R_{h}} \right) - \frac{1}{60}\eta\nabla' \cdot \left( \delta'^{5}\nabla'\left( \nabla' \cdot \nabla' \right)^{2}\frac{\partial}{\partial h}\frac{1}{R_{h}} \right) \nonumber\\
&&- \frac{1}{18}\eta^{3}\nabla' \cdot \left( \delta'^{3}\nabla'\left( \nabla' \cdot \nabla' \right)^{2}\frac{\partial}{\partial h}\frac{1}{R_{h}} \right) - \frac{1}{60}\eta^{5}\nabla' \cdot \left( \delta'\nabla'\left( \nabla' \cdot \nabla' \right)^{2}\frac{\partial}{\partial h}\frac{1}{R_{h}} \right) \nonumber\\
&&+ O(\varepsilon^m\zeta^{7-m}),
\end{eqnarray}

\noindent where \textcolor{black}{$\eta_{\pm} = \eta' \pm \eta$}, \(R_{nh} = \sqrt{R^2+(nh)^2}\), \(n \in [1,2]\) and \( m \in [1,6]\) are integers,  \(\varepsilon = k_p |\eta|\) and \(\zeta = k_p |\delta|\) represent the magnitudes of wave steepness and seabed slope, respectively (\(|\eta|\) and \(|\delta|\) are the characteristic wave amplitude and seabed height, respectively). 

Applying Fourier transform to Eqs.(\ref{eq:bi-s} $\sim$ \ref{eq:bi-s_end}), the following equation can be derived (refer to Appendix \ref{appA}):
\begin{equation}
V = \sum_{j=1}^{7} V^{(j)} + I^{(S)},
\label{eq:con-V}
\end{equation}
\noindent where it is noted that
\begin{equation}
\mathcal{F}\left\{ \frac{1}{R_{nh}} \right\} = \frac{2\pi}{k}e^{- \mathrm{i}\bm{k} \cdot \bm{x}' - nkh}, \quad n = \text{1 or 2},
\label{eq:hilbert}
\end{equation}

\noindent \(V^{(j)}\) stands for the quantity in terms of \(j^{\text{th}}\) convolution, and \(I^{(S)}\) is the remaining integration term. \textcolor{black}{It is worthy noting that the boundary integrals are implemented throughout the space at infinity. However, the Fourier transform can be applied to the studied area of finite size with given periodic lateral boundary conditions. This is attributed to repetition properties of the Fourier analysis.
%This is because that, due to the repetition of the studied area in space, the resulted Fourier spectrum is equivalent to that at infinity. 
} 

\textcolor{black}{In deep water, the integration term $I^{(S)}$ in Eq. (\ref{eq:con-V}) can be approximated by the 8th order convolutions (refer to Eq. (A8) in \cite{wang2015}}). 

\subsubsection {Evaluation points on seabed surface}
For evaluation points on the seabed surface, the corresponding boundary integral equation can be written as \citep{fructus2007}:
\begin{eqnarray}
2\pi\widetilde{\varphi} & = &\iint_{S_{0}}^{}{\left( \frac{1}{r_{S}} + \frac{1}{r_{C}} \right)V' \mathrm{d} \bm{x}'} - \iint_{S_{0}}^{}{{\widetilde{\phi}}'\frac{\partial}{\partial n'}\left( \frac{1}{r_{S}} + \frac{1}{r_{C}} \right)\sqrt{1 + \left| \nabla'\eta' \right|^{2}\ } \mathrm{d} \bm{x}'} \nonumber\\
&& +\iint_{B_{0}}^{}{{\widetilde{\varphi}}'\frac{\partial}{\partial n'}\left( \frac{1}{r_{S}} + \frac{1}{r_{C}} \right)\sqrt{1 + |\nabla'\delta'|^{2}\ } \mathrm{d} \bm{x}'},
\label{eq:bi-b}
\end{eqnarray}

\noindent where \(r_{C}^{2} = R^{2} + \left( z' + z \right)^{2}\). Similarly, the integrands in Eq.(\ref{eq:bi-b}) can be expanded as:
%For the integral term over the free surface, \(z' = \eta'\) and \(z = \delta - h\), thus \(r_{S}^{2} = R^{2} + \left( \eta' - \delta + h \right)^{2}\), \(r_{C}^{2} = R^{2} + \left( \eta' + \delta - h \right)^{2}\) and
\begin{eqnarray}\label{eq:bi-b_start}
\frac{1}{r_{S}} + \frac{1}{r_{C}} & = &\frac{2}{R_{h}} - 2\delta\frac{\partial}{\partial h}\frac{1}{R_{h}} - \left( \eta'^{2} + \delta^{2} \right)\nabla' \cdot \nabla'\frac{1}{R_{h}} \nonumber\\
&& + \frac{1}{3}\left( \delta^{3} + 3\eta'^{2}\delta \right)\nabla' \cdot \nabla'\frac{\partial}{\partial h}\frac{1}{R_{h}}  +\frac{1}{24}\left( 2\eta'^{4} + 12\eta'^{2}\delta^{2} + 2\delta^{4} \right)\left( \nabla' \cdot \nabla' \right)^{2}\frac{1}{R_{h}} \nonumber\\
&& +\frac{1}{120}\left( - 10\eta'^{4}\delta - 20\eta'^{2}\delta^{3} - 2\delta^{5} \right)\left( \nabla' \cdot \nabla' \right)^{2}\frac{\partial}{\partial h}\frac{1}{R_{h}} + O(\varepsilon^m\zeta^{6-m}) ,\nonumber\\
\end{eqnarray}
%\noindent and
\begin{eqnarray}
\frac{\partial}{\partial{n}'}\bigg( \frac{1}{r_{S}} &+& \frac{1}{r_{C}} \bigg)\sqrt{1 + \left| \nabla'\eta' \right|^{2}\ } \nonumber\\
&=& - 2\nabla' \cdot \left( \eta'\nabla'\frac{1}{R_{h}} \right) + 2\delta\nabla' \cdot \left( \eta'\nabla'\frac{\partial}{\partial h}\frac{1}{R_{h}} \right) + \frac{1}{3}\nabla' \cdot \left( \eta'^{3}\nabla'\nabla' \cdot \nabla'\frac{1}{R_{h}} \right)\nonumber\\
&&  +\delta^{2}\nabla' \cdot \left( \eta'\nabla'\nabla' \cdot \nabla'\frac{1}{R_{h}} \right) - \frac{1}{3}\delta\nabla' \cdot \left( \eta'^{3}\nabla'\nabla' \cdot \nabla'\frac{\partial}{\partial h}\frac{1}{R_{h}} \right)\nonumber\\
&&  -\frac{1}{3}\delta^{3}\nabla' \cdot \left( \eta'\nabla'\nabla' \cdot \nabla'\frac{\partial}{\partial h}\frac{1}{R_{h}} \right)  - \frac{1}{60}\nabla' \cdot \left[ \eta'^{5}\nabla'\left( \nabla' \cdot \nabla' \right)^{2}\frac{1}{R_{h}} \right] \nonumber\\
&& -\frac{1}{6}\delta^{2}\nabla' \cdot \left[ \eta'^{3}\nabla'\left( \nabla' \cdot \nabla' \right)^{2}\frac{1}{R_{h}} \right] - \frac{1}{12}\delta^{4}\nabla' \cdot \left[ \eta'\nabla'\left( \nabla' \cdot \nabla' \right)^{2}\frac{1}{R_{h}} \right] \nonumber\\
&&  + O(\varepsilon^m\zeta^{6-m}),
\end{eqnarray}
%For the integral term over the bottom, \(z' = \delta' - h\) and \(z = \delta - h\), thus \(r_{S}^{2} = R^{2} + \left( \delta^{'} - \delta \right)^{2}\), \(r_{C}^{2} = R^{2} + \left( \delta^{'} + \delta - 2h \right)^{2}\) and
\noindent and
\textcolor{black}{
\begin{eqnarray}\label{eq:bi-b_end}
\frac{\partial}{\partial{n}'}\bigg( \frac{1}{r_{S}} &+& \frac{1}{r_{C}} \bigg)\sqrt{1 + |\nabla'\delta'|^{2}\ } \nonumber\\
&=&- \frac{\partial}{\partial(2h)}\frac{1}{R_{2h}} + \sum_{j=1}^{2} \frac{(-1)^{j-1}}{(2j)!}\nabla'\cdot\left[ \delta_{+}^{2j} \nabla'(\nabla' \cdot \nabla')^{j-1} \frac{\partial}{\partial(2h)}\frac{1}{R_{2h}} \right] \nonumber\\
&&+ \sum_{j=1}^{3} \frac{(-1)^j}{(2j-1)!} \nabla'\cdot\left[ \delta_{-}^{2j-1} \nabla'(\nabla' \cdot \nabla')^{j-1} \frac{1}{R} + \delta_{+}^{2j-1} \nabla'(\nabla' \cdot \nabla')^{j-1} \frac{1}{R_{2h}} \right] \nonumber\\
%\frac{\partial}{\partial{n}'}\bigg( \frac{1}{r_{S}} &+& \frac{1}{r_{C}} \bigg)\sqrt{1 + |\nabla'\delta'|^{2}\ } \nonumber\\
%&=&- \frac{\partial}{\partial(2h)}\frac{1}{R_{2h}} - \nabla' \cdot \left[ \left( \delta' - \delta \right)\nabla'\frac{1}{R} \right] - \nabla' \cdot \left[ \left( \delta' + \delta \right)\nabla'\frac{1}{R_{2h}} \right] \nonumber\\
%&& +\frac{1}{2}\nabla' \cdot \left[ \left( \delta' + \delta \right)^{2}\nabla'\frac{\partial}{\partial(2h)}\frac{1}{R_{2h}} \right] + \frac{1}{6}\nabla' \cdot \left[ \left( \delta' - \delta \right)^{3}\nabla'\nabla' \cdot \nabla'\frac{1}{R} \right]\nonumber\\
%&&   +\frac{1}{6}\nabla' \cdot \left[ \left( \delta' + \delta \right)^{3}\nabla'\nabla' \cdot \nabla'\frac{1}{R_{2h}} \right] + \frac{1}{24}\nabla' \cdot \left[ \left( \delta' + \delta \right)^{4}\nabla'\nabla' \cdot \nabla'\frac{\partial}{\partial( - 2h)}\frac{1}{R_{2h}} \right] \nonumber\\
%&&  -\frac{1}{120}\nabla' \cdot \left[ \left( \delta' - \delta \right)^{5}\nabla'\left( \nabla' \cdot \nabla' \right)^{2}\frac{1}{R} \right] - \frac{1}{120}\nabla' \cdot \left[ \left( \delta' + \delta \right)^{5}\nabla'\left( \nabla' \cdot \nabla' \right)^{2}\frac{1}{R_{2h}} \right] \nonumber\\
&&  + O(\zeta^{6}).
\end{eqnarray}
}
\textcolor{black}{where $\delta_{\pm}=\delta' \pm \delta$.} Applying Fourier transform to Eq.(\ref{eq:bi-b} $\sim$ \ref{eq:bi-b_end}) and utilizing Eq.(\ref{eq:hilbert}), the following equation can be obtained as (refer to Appendix \ref{appB}):
\begin{equation}
\widetilde{\varphi} = \sum_{j=1}^6 \widetilde{\varphi}^{(j)} + I^{(B)},
\label{eq:con-phi}
\end{equation}

\noindent where \(\widetilde{\varphi}^{(j)}\) denotes the quantity in terms of \(j^{\text{th}}\) convolution, and \(I^{(B)}\) is the remaining integration term. 

\subsection {Successive approximations}\label{sec:success}
It is noted that both Eqs.(\ref{eq:con-V}) and (\ref{eq:con-phi}) are implicit, meaning that the evaluation of \(V\) and \(\widetilde{\varphi}\) depends on themselves. Though they can be approximated via an iterative approach \citep{fructus2005,wang2015}, the speed of the convergence largely depends on the strength of wave non-linearity. For very large wave steepness (\textcolor{black}{e.g., Stoke waves of }\(\varepsilon > 0.3\)), it often requires more than three iterations to achieve acceptable convergence. To improve the computational efficiency, this section proposes a successive method to approximate \(V\) and \(\widetilde{\varphi}\). To do so, the convolution parts are rewritten as:
%Collecting terms of the same order in \(\varepsilon\), \(\zeta\) and their product, the following equations are proposed:
\begin{equation}
V = \sum_{l=1}^{7} V_l,%+ I^{(S)}
\label{eq:app-V}
\end{equation}
%\noindent and
\begin{equation}
\widetilde{\varphi} = \sum_{l=1}^6 \widetilde{\varphi}_l,% + I^{(B)}
\label{eq:app-phi}
\end{equation}

\noindent where \(V_l\sim O(\varepsilon^m\zeta^{l-m})\) and \(\widetilde{\varphi}_l\sim O(\varepsilon^m\zeta^{l-m})\) with \(m \in [1,l-1]\). Substituting Eqs.(\ref{eq:app-V}) and (\ref{eq:app-phi}) into Eqs.(\ref{eq:con-V}) and (\ref{eq:con-phi}), and collecting terms of the same order in \(\varepsilon\), \(\zeta\) and their products, the formulations for \(V_l\) and \(\widetilde{\varphi}_l\) at each order can be derived successively. 
The procedures \textcolor{black}{are initiated with:}
%\begin {enumerate}
%\item Step \#1. Evaluate the first-order term \(\sim O(\varepsilon)\) by using:
\begin{equation}
\mathcal{F}\left\{ V_{1} \right\} = k\tanh (kh) \mathcal{F}\left\{ \widetilde{\phi} \right\} ,
%= kC_{h}D_{h}^{- 1}\mathcal{F}\left\{ \widetilde{\phi} \right\}
\end{equation}
\begin{eqnarray}
\mathcal{F}\left\{ {\widetilde{\varphi}}_{1} \right\} = \text{sech} {(kh)} \mathcal{F}\left\{ \widetilde{\phi} \right\}. %= 2\frac{1}{k}D_{h}\sqrt{e_{h}}\mathcal{F}\left\{ V_{1} \right\}
\end{eqnarray}

\noindent \textcolor{black}{Herein, it is noted that the Fourier multipliers $k\tanh (kh)$ and $\text{sech} {(kh)}$ arise naturally, and the exact forms can also be derived from the recursion formulas based on the Dirichlet-Neumann operator approach \citep{guyenne2008high}.}

\textcolor{black}{It is followed by evaluating \(V_l\) and \(\widetilde{\varphi}_l\) at each order based on the formulations as below ($2 \le l \le 7$): }
\textcolor{black}{
\begin{eqnarray}\label{eq:app-V-scs}
\mathcal{F}\left\{ V_{l} \right\} = && \sum_{j=1}^{l-1} \frac{k^j}{j!} \bigg[ \langle j/2 \rangle  \text{sech} (kh) \frac{\mathrm{i}\bm{k}}{k} \cdot\mathcal{F}\left\{ \delta^j \nabla{\widetilde{\varphi}}_{l-j} \right\} -\tanh^{\langle j/2 \rangle}{(kh)} \mathcal{F}\left\{ \eta^jV_{l-j} \right\}  \bigg]\nonumber\\
&& -\tanh{(kh)} \frac{k^{l-2}}{(l-1)!} \mathrm{i}\bm{k} \cdot \mathcal{F}\left\{ \eta^{l-1}\nabla\widetilde{\phi}  \right\} ,
\end{eqnarray}
%\begin{eqnarray}\label{eq:app-phi-scs}
%\mathcal{F}\left\{ \widetilde{\varphi}_{l} \right\} =  \sum_{j=1}^{l-1} \frac{k^{j-1}}{j!} \bigg[ &-&\langle j/2 \rangle \text{sech} (kh) \mathcal{F}\left\{ \eta^jV_{l-j} \right\} \nonumber\\
%&-&  \tanh^{\langle j/2 \rangle}{(kh)} \frac{\mathrm{i}\bm{k}}{k} \cdot  \mathcal{F}\left\{ \delta^j \nabla{\widetilde{\varphi}}_{l-j} \right\}  \bigg],
%\end{eqnarray}
\begin{eqnarray}\label{eq:app-phi-scs}
\mathcal{F}\left\{\widetilde{\varphi}_l\right\} =&&\sum_{j=1}^{l-1}\frac{k^{j-1}}{j!} \bigg[ -\langle j / 2\rangle\operatorname{sech}(k h)   \mathcal{F}\left\{\eta^j V_{l-j}\right\}  + (-1)^j\tanh { }^{(j/ 2\rangle}(k h) \frac{\mathrm{i} \boldsymbol{k}}{k}  \cdot\mathcal{F}\left\{\delta^j \nabla \widetilde{\varphi}_{l-j}\right\} \bigg] \nonumber  \nonumber\\
&&-\langle l/ 2\rangle \mathrm{i} \boldsymbol{k} \cdot \frac{k^{l-3}}{(l-1)!}\operatorname{sech}(k h) \mathcal{F}\left\{\eta^{l-1} \nabla \tilde{\phi}\right\}
\end{eqnarray}
}
\noindent \textcolor{black}{where $\langle j/2 \rangle$ is the reminder of $j$ divided by 2. Their explicit forms are presented in Eqs. (\ref{eq:v2}) $\sim$ (\ref{eq:v7}).}

After all, the overall velocity \(V\) can be estimated based on Eq.(\ref{eq:app-V}), \textcolor{black}{and the numerical scheme is fast \textcolor{black}{owing} to the usage of Fast Fourier Transform. The integration term being omitted is of order \(\sim O\left( \varepsilon^{m}\zeta^{8 - m} \right)\), which renders the present model featuring a highly nonlinear scheme. }
\textcolor{black}{Unlike the HOS model using a recursive formalism and the order $M$ can be specified arbitrarily, the present model can only be applied to 7th order convolution or below. Accounting for the 3rd order convolution \citep{wang2016}, the present model becomes similar with the HOS model ($M=3$) adopted in \cite{xiao2013rogue} with the emphasis on investigating the role of modulational instability on rogue wave occurences,}
\textcolor{black}{or the formalism in \cite{fructus2007} for wave transformation over stationary bathymetry. Nevertheless, the fully nonlinear scheme can recover from the closed formulations for \(V\) and \(\widetilde{\varphi}\) with the inclusions of integration terms given by Eqs. (\ref{app:iphis} \(\sim\) \ref{app:ivarphis}) and (\ref{app:iphib} \(\sim\)\ref{app:ivarphib}). Following \citet{wang2015}, the integration term will only be considered when surface slope is sufficiently large. Yet, it should be noted that the closed form will be required in the cases where local wave front becomes extremely sharp and breaking instabilities become evident. In such cases, the simulation usually breaks down, unless dissipation model or artificial low-pass filter is employed to remove the spectral energy confined in high wavenumber range. }

Regarding this pseudo-spectrum method, the \(2/(l+1)\)-rule is used here for anti-aliasing treatment, which is equivalent to the zero-padding method \citep{canuto1987spectral}. A detailed comparison on the effectiveness between different anti-aliasing schemes can be found in \cite{wang2015}, hence it is not repeated hereby. Meanwhile, the computational efficiency of this successive scheme will be further discussed in Section \ref{sec:dis}. \textcolor{black}{It is emphasized that no smoothing technique is required, and the present model is very stable for the cases without appearance of breaking waves.}

\section {Model verification and validation}\label{sec:verify-validate}
To demonstrate the capability of the proposed model for simulating \textcolor{black}{highly} nonlinear waves, this section will carry out a comprehensive verification and validation test covering a variety of scenarios in variable depth as well as in deep water. The cases are selected with the emphasis on demonstrating the capability of the new model for reproducing the nonlinear wave-seabed and wave-wave interactions. 

\subsection {Solitary wave propagation on constant and plane beach}
The ESBI model is firstly tested for simulating a solitary wave on constant depth. The initial condition for the simulation is set up based on the Tanaka's numerical solution \citep{tanaka1986stability}. The initial wave amplitude over water depth is \(A/h\) = 0.6. The effective domain is about 800\(h\) and is descritized into \textcolor{black}{2048, 4096 and 8192 points in case \#1, \#2 and \#3, respectively. Based on that, within one characteristic solitary wavelength ($=4.24h/\sqrt{A/h}$, containing 95\% of the volume \citep{dean1991water}), there are 14, 28 and 56 points, respectively. } \(\delta\) is set to zero for a constant depth configuration. The ESBI model simulates the free propagation of the solitary wave towards x-direction for a duration of 200\(\sqrt{h/g}\), followed by a backward propagation. The purpose of this test is to verify the accuracy of the model for reproducing a steady solitary wave maintaining its permanent shape during the propagation. 

\begin{figure}
\centering
\begin{subfigure}{.5\textwidth}
\centerline{\includegraphics[width=60mm]{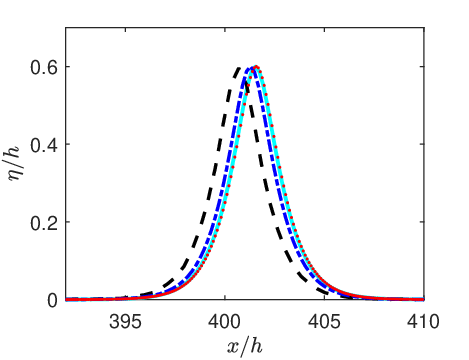}} 
  \caption{Simulated surface after back-propagation. '\textcolor{cyan}{---}': $t=0$; '\(--\)': case \#1; '\textcolor{blue}{--\(\cdot\)--}': case \#2; '\textcolor{red}{\(\cdot \cdot \cdot\)}': case \#3.  }
  \label{fig:sub1}
\end{subfigure}%
\begin{subfigure}{.5\textwidth}
\centerline{\includegraphics[width=60mm]{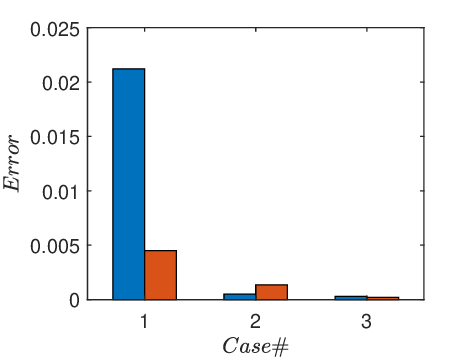}}
  \caption{Error of simulated solitary wave amplitude (\textcolor{blue}{$\blacksquare$}) and phase speed/celerity (\textcolor{red}{$\blacksquare$}) versus the simulation case \#. }
  \label{fig:sub2}
\end{subfigure}

\begin{subfigure}[b]{0.75\textwidth}
   \includegraphics[width=1\linewidth]{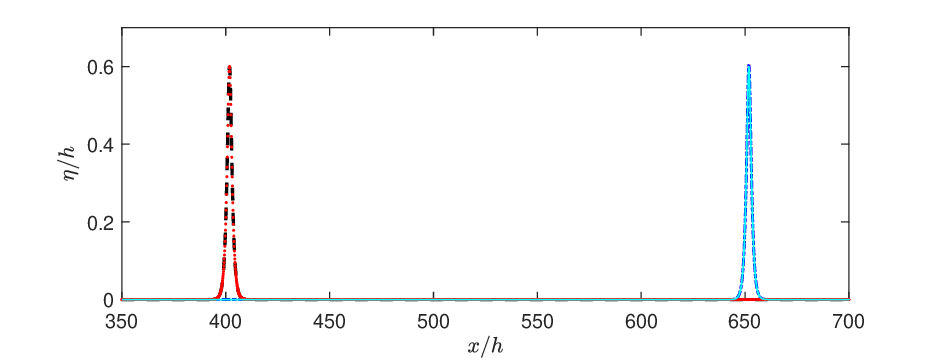}
   \caption{Free surface profiles.'\(--\)': Initial surface at \(t/\sqrt{h/g}=0\); '\textcolor{blue}{--\(\cdot\)--}': Simulated surface at \(t/\sqrt{h/g}=200\); '\textcolor{red}{\(\cdots\)}': Simulated surface at \(t/\sqrt{h/g}=400\);  '\textcolor{cyan}{---}': Tanaka's solution at \(t/\sqrt{h/g}=200\).}
   \label{fig:1} 
\end{subfigure}

\begin{subfigure}[b]{0.75\textwidth}
   \includegraphics[width=1\linewidth]{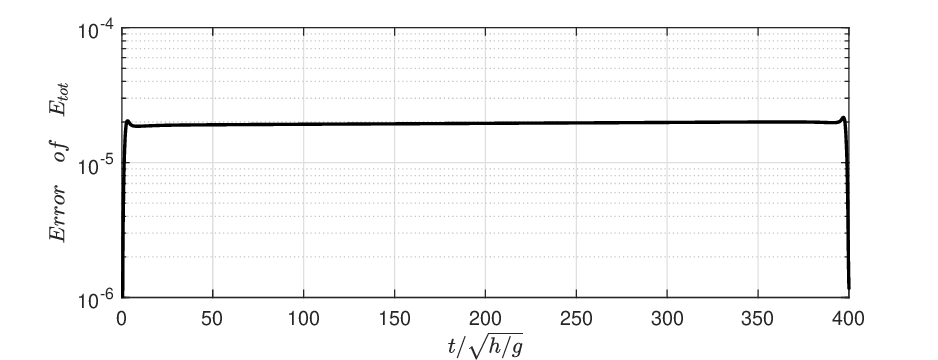}
   \caption{Error of the total energy.}
   \label{fig:2}
\end{subfigure}

\caption{\textcolor{black}{Simulation results of a solitary wave on constant water depth.} }
\label{fig:solitary}
\end{figure}

\textcolor{black}{Firstly, to check the convergence with respect to spatial discretization, the simulated wave surfaces after back-propagation are presented in figure \ref{fig:solitary}(a). It shows that the shapes of the solitary wave are well preserved throughout the propagation in all testing cases, despite that a noticeable phase shift can be observed in case \#1, indicating that its phase speed/celerity may not be sufficiently accurate in this case. The celerities are then calculated (propagating distance over time) and compared with Tanake's solution for quantifying the errors. In addition, the errors of the wave amplitude (changes after back-propagation) are also calculated, and the results are displayed in figure \ref{fig:solitary}(b). It implies that case \#1 not only unsatisfactorily captured the celerity, but also poorly predicted the wave amplitude. The results can be significantly improved in case \#2 with doubled resolution, and further increasing the resolution in case \#3 leads to a precise prediction on both celerity and amplitude. Therefore, it can be said that case \#3 (56 points per characteristic wavelength) produces sufficiently accurate results, which will be used for analysis hereafter. } 

The simulated free surface elevations at different time instants are displayed in figure \ref{fig:solitary}\textcolor{black}{(c)}. It shows that the simulated wave surface coincides with the theory after propagating for a long time and distance. Moreover, it overlaps with the initial surface after a back-propagating simulation, indicating that the present model is very robust. 

In addition, to examine whether the total wave energy (or Hamiltonian) is conserved during the simulation, the following formula is used at each time step:
\begin{equation}\label{eq:energy}
    E_{tot}=\frac{1}{2}\iint_{S_0} \left[V(\bm{x},t) \widetilde{\phi}(\bm{x},t) + \eta(\bm{x},t)^2 \right] \mathrm{d}\bm{x},
\end{equation}

\noindent where the first and second part within the bracket yields the \textcolor{black}{kinetic} and potential energy, respectively. Then the error of total energy conservation is estimated by \(|E_{tot} - E_{tot}(t=0)|/E_{tot}(t=0)\), and it is presented in figure \ref{fig:solitary}\textcolor{black}{(d)}. It is noted that the error of the total energy \textcolor{black}{quickly builds up at the beginning of the simulation, and it stabilized at a constant level ($\sim 10^{-5}$) with no sign of growth in time, even after the propagating direction was changed at \(t=200\sqrt{h/g}\). The conservation of total energy implies} that the present model has achieved good accuracy in simulating a \textcolor{black}{highly} nonlinear solitary wave on constant water depth.  

\begin{figure}
\centering
\begin{subfigure}[b]{0.75\textwidth}
   \includegraphics[width=1\linewidth]{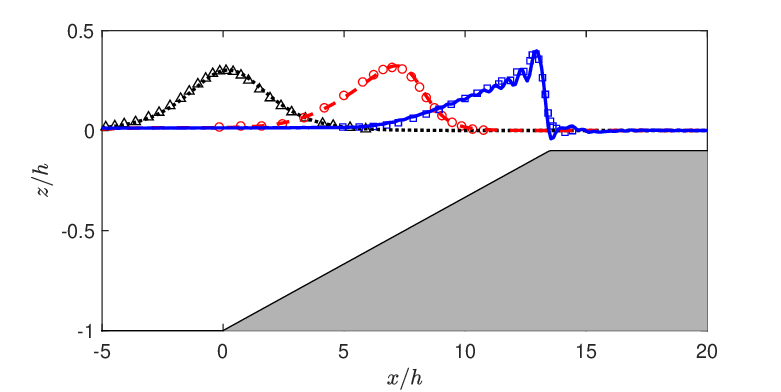}
   \caption{Simulated wave surface. '\(\triangle\)' denotes \(t/\sqrt{h/g}=0\), '\textcolor{red}{\(\circ\)}' denotes \(t/\sqrt{h/g}=6\) and '\textcolor{blue}{\(\square\)}' denotes \(t/\sqrt{h/g}=11\) in \cite{fructus2007} (digitalized from the paper); '\(\cdots\)' denotes \(t/\sqrt{h/g}=0\), '\textcolor{red}{\(--\)}' denotes \(t/\sqrt{h/g}=6\) and '\textcolor{blue}{---}' denotes \(t/\sqrt{h/g}=11\) in present simulation. }
   \label{fig:1} 
\end{subfigure}

\begin{subfigure}[b]{0.75\textwidth}
   \includegraphics[width=1\linewidth]{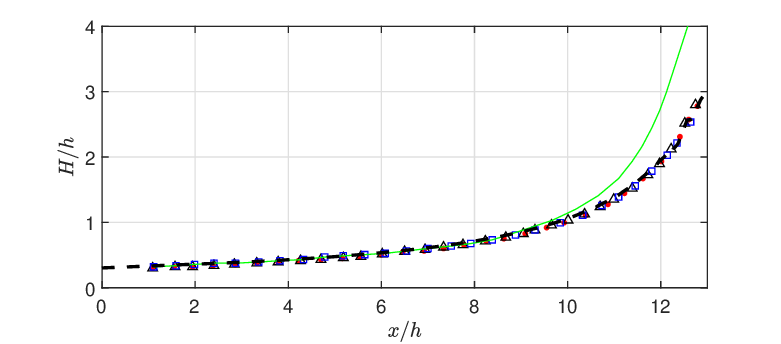}
   \caption{Maximum crest height in space. '\textcolor{blue}{\(\square\)}': \cite{grilli1989efficient}; '\textcolor{green}{---}':\cite{nwogu1993alternative}; '\(\triangle\)': \cite{Wei1995}; '\textcolor{red}{\(\cdots\)}': \cite{fructus2007}; '\(--\)': Present model.}
   \label{fig:2}
\end{subfigure}
\caption{Simulation results of the shoaling of a solitary wave over a plane beach.}
\label{fig:solitaryShoal}
\end{figure}

Next, the ESBI model is verified for simulating the shoaling process of a solitary wave. The computational domain covers 400\(h\) and is descritized into 4096 points. The slope of the plane beach is 1:15 and its foot origins at \(x\)=0. The solitary wave of steepness \(A/h=0.3\) is initiated with its centre located at \(x\)=0 based on Tanaka's numerical solution \citep{tanaka1986stability}. The snapshots of the simulated free surface profiles are extracted at different time instants, which are displayed in figure \ref{fig:solitaryShoal}(a). The results are also compared with the fully nonlinear pseudo-spectrum method by \cite{fructus2007}, where a good agreement between the two simulations is observed. The steepening of the wavefront and saw tooth on the back are captured when the wave reached the top of the slope. 
\textcolor{black}{Note that unlike the numerical instability \citep{longuet1976deformation}, the saw-tooth shape is not artificial in present simulation, but is often observed at the initiation of wave breaking, as an indication of physical breaking instability. Particularly, this saw-tooth shape profile is sometimes challenging to be produced in Boussinesq-type of models, because to obtain a saw-tooth profile for the wave in the surf-zone it is necessary to have full non-linearity at least up to the order of dispersion that is retained \citep{veeramony1998boussinesq}. Nevertheless, more  detailed  analysis  of  the  wave  breaking instability  is  out  of  the  scope  of  the  present  paper, thus it will not be further discussed herein.}
%After this point, the simulation is terminated due to  high wavenumber. is independent of the numerical resolution, and thus models physical steepening towards physical breaking. 

Meanwhile, the localized maximum wave peak height over water depth ratio (\(H/h\)) is extracted in space, and is shown in figure \ref{fig:solitaryShoal}(b). The ratios obtained based on the weakly nonlinear Boussinesq model \citep{nwogu1993alternative}, fully nonlinear Boussinesq model \citep{Wei1995}, fully nonlinear BEM \citep{grilli1989efficient}, and fully nonlinear pseudo-spectrum method \citep{fructus2007} are also presented in this figure for the same simulation. It shows that the values of localized \(H/h\) obtained by using the present ESBI model increases along the propagation with reduced water depth, which is consistent with the other three fully nonlinear numerical models \citep{Wei1995,grilli1989efficient,fructus2007}, indicating that the shoaling process of the solitary wave is successfully simulated. Therefore, it can be said that the present ESBI model can accurately reproduce the \textcolor{black}{highly} nonlinear solitary wave on constant and variable depth in numerical simulations.

\subsection {Wave transformations over a submerged bar}\label{sec:submerged}
\begin{figure}
\centering
\begin{subfigure}[b]{0.75\textwidth}
   \includegraphics[width=1\linewidth]{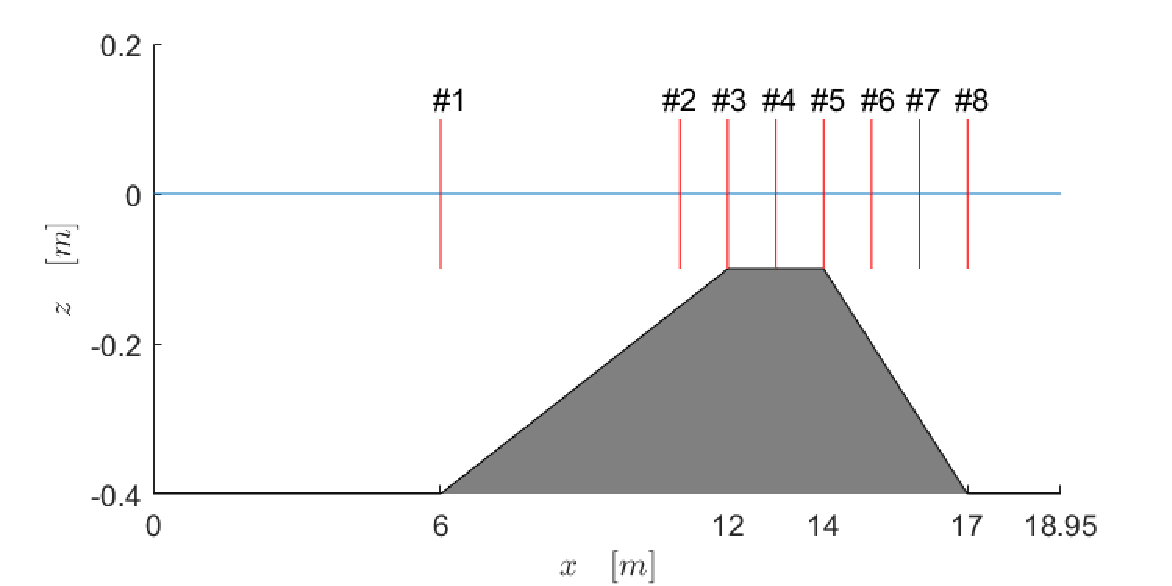}
   \caption{}
   \label{fig:1} 
\end{subfigure}

\begin{subfigure}[b]{0.75\textwidth}
   \includegraphics[width=1\linewidth]{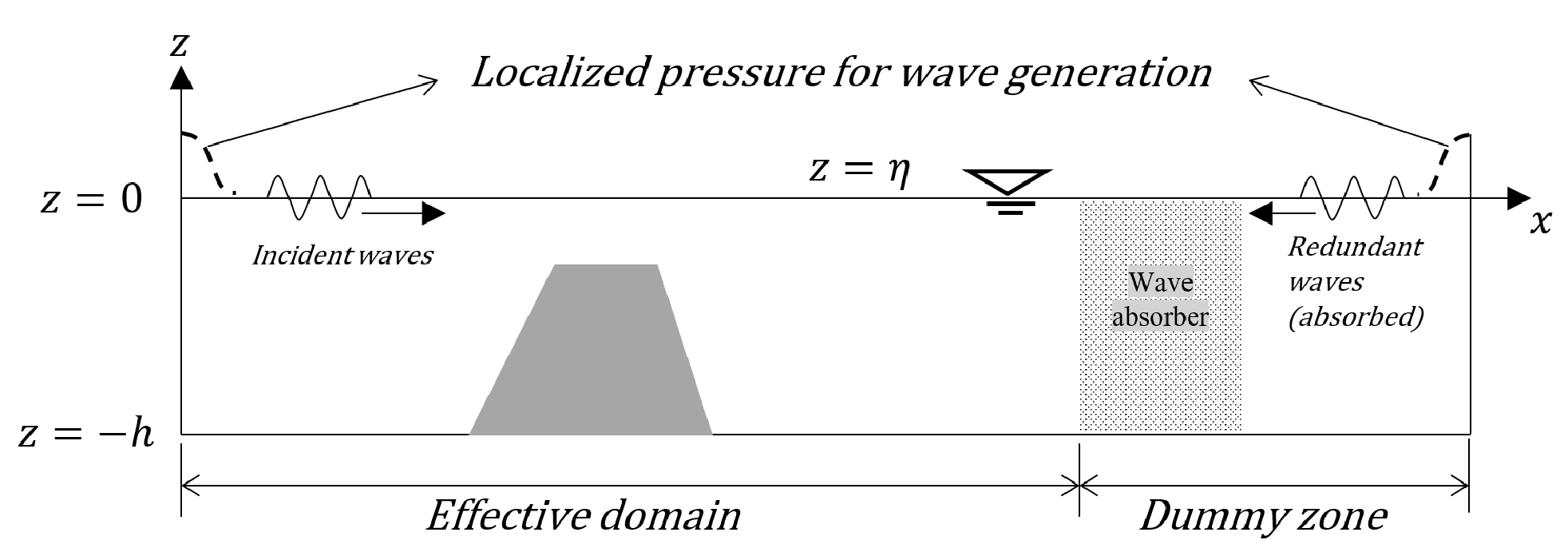}
   \caption{}
   \label{fig:2}
\end{subfigure}
\caption{Illustration of the computational domain for simulating wave propagation over submerged bar.}
\label{fig:beji-domain}
\end{figure}

This section demonstrates the capability of the present ESBI model for reproducing the classic experiments of a plane wave propagation over a submerged bar by \cite{beji1993}. \textcolor{black}{This is a classical experiment that has been extensively used for validations of various numerical models. To be noted, lower-order models are sufficient for these benchmark tests, and it is not the focus to shown that the present model will produced improved results. Instead, the validations here are merely used to demonstrate the capability of the present model for simulating nonlinear wave transformations over variable depth, as the model is originally designed for modelling waves in deep water.} The effective part of the computational domain is shown in figure \ref{fig:beji-domain}(a), where the entire domain is 64 m in length and is descritized into 1024 points. It means that the waves longer than 0.25 m or frequency lower than 2.5 Hz are alias free in the simulations. The water depth before the beach is 0.4 m, and the plane beach facing the incident wave is in a slope of 1:20. The submerged bar is 0.1 m in depth at the flat top of 2 m in length, and the slope of the beach behind the bar is 1:10. For validation purpose, four different wave conditions are selected to test the accuracy of the present model for simulating uniform waves as well as random waves, which are summarized in table \ref{tab:beji}. \textcolor{black}{Regarding the relatively short waves with frequency of 1 Hz, their wavelength will be reduced to about 0.92 m at the top of the bar after the shoaling taking place. The current spatial resolution allows 15 points within one wavelength over the bar, and they are free of aliasing errors (wavelength less than 0.25 m).}

To mimic the wavemaker in laboratory settings, the waves in the numerical simulations are generated through applying a localized oscillating pressure on the surface, i.e., the so-called pneumatic \textcolor{black}{wavemaker \citep{clamond2005efficient}. The pressure can be described mathematically as:
\begin{equation}
    p_g = \sum_{j=1}^{J} a_j \sqrt{\frac{e}{2\pi}}\left[1+\frac{2k_jh}{\sinh{(2k_jh)}}\right] e^{-\frac{(k_jx)^2}{2}} \sin{(\omega_jt+\theta_j)},
\end{equation}
\noindent where $\theta_j$ is the constant phase of the $j^{\text{th}}$ component. It is of a bell shape, which can be split in halves and allocated symmetrically at both ends of the domain (see figure \ref{fig:beji-domain}(b)), without the requirement to treat the singularity at the intersection between the interfaces as in the conventional wave generation technique \citep{grilli1997numerical}. By doing so, the periodic boundary condition (due to the usage of FFT) is not violated}, and the generated waves will \textcolor{black}{radiate from the generation zone symmetrically and} propagate towards the domain centre. \textcolor{black}{In this study, o}nly the domain on the left is effectively used for simulating the wave transformations. Herein, the numerical simulations utilize a artificial damping zone to absorb transmitted waves behind the submerged bar, as well as the back-propagating waves in the dummy zone. 
%whereas \cite{beji1993} used a long mild beach to serve as a wave absorber in the laboratory. 
\textcolor{black}{Within the absorption zone, the present model adopts the sponge layer technique via a damping pressure of the form \citep{clamond2005efficient}:
\begin{equation}\label{eq:pd}
    p_d = \mathcal{F}^{-1}\left\{ \frac{\mathrm{i}\bm{k}}{-k^2} \cdot \mathcal{F} \left\{ e^{-\left(\frac{x-x_0}{4\pi}\right)^2} \nabla \widetilde{\phi} \right\}   \right\}
\end{equation}
\noindent where $x_0$ is the centre position of the damping pressure and is 7/8 total domain length in the present study. In contrast to classical sponge layers, the wave absorption technique of form Eq. (\ref{eq:pd}) is highly effective. More details are omitted here but can be found in section 5 in \cite{clamond2005efficient}.   }

\textcolor{black}{To better understand the configuration of wave generation and absorption, an illustration of the computational domain} is presented in figure \ref{fig:beji-domain}(b). Due to the utilization of periodic boundary condition and the pneumatic wavemaker in ESBI model, the back-scattered waves can propagate through the wave generation zone and be fully absorbed in the dummy zone. Hence, the influence of re-reflected waves on the effective simulation domain is negligible. %The details about the wave generation and absorption technique employed in the ESBI model can be found in \cite{clamond2005efficient}, which is not discussed further in this study. 

\begin{figure}
\centering
\begin{subfigure}{.5\textwidth}
%\centerline{\includegraphics[width=70mm]{figs/elev_mono_long.eps}}
\centerline{\includegraphics[width=70mm]{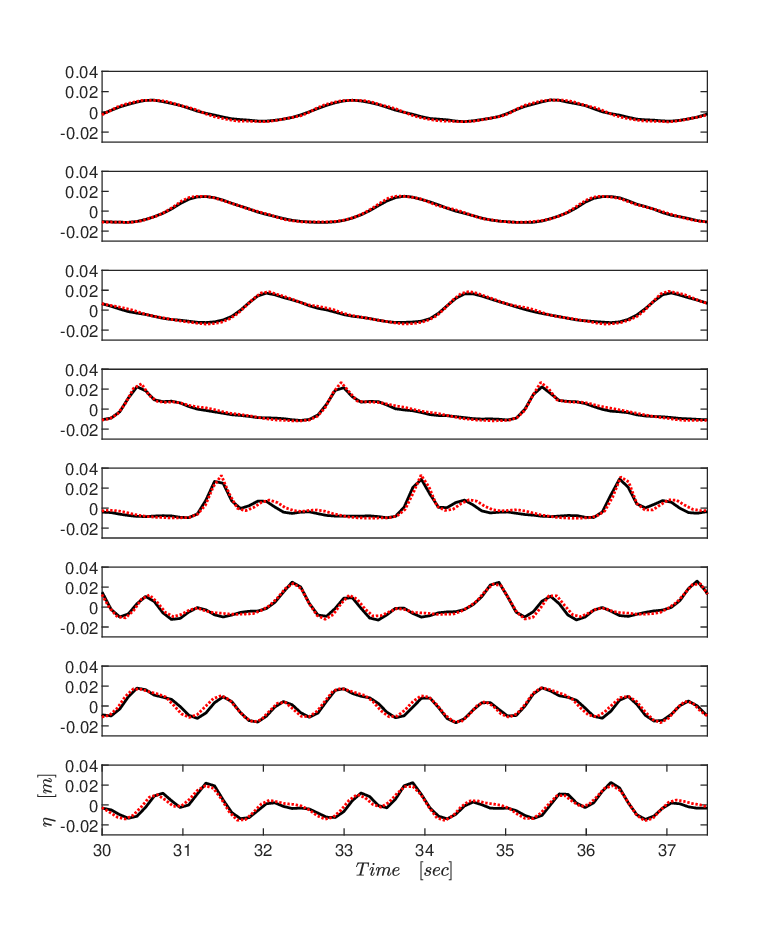}} 
%  \centering
% \includegraphics[width=1.2\linewidth]{figs/elev_mono_long.eps}
  \caption{\textcolor{black}{Case \#1 (long waves)}}
  \label{fig:sub1}
\end{subfigure}%
\begin{subfigure}{.5\textwidth}
%\centerline{\includegraphics[width=70mm]{figs/elev_mono_short.eps}}
\centerline{\includegraphics[width=70mm]{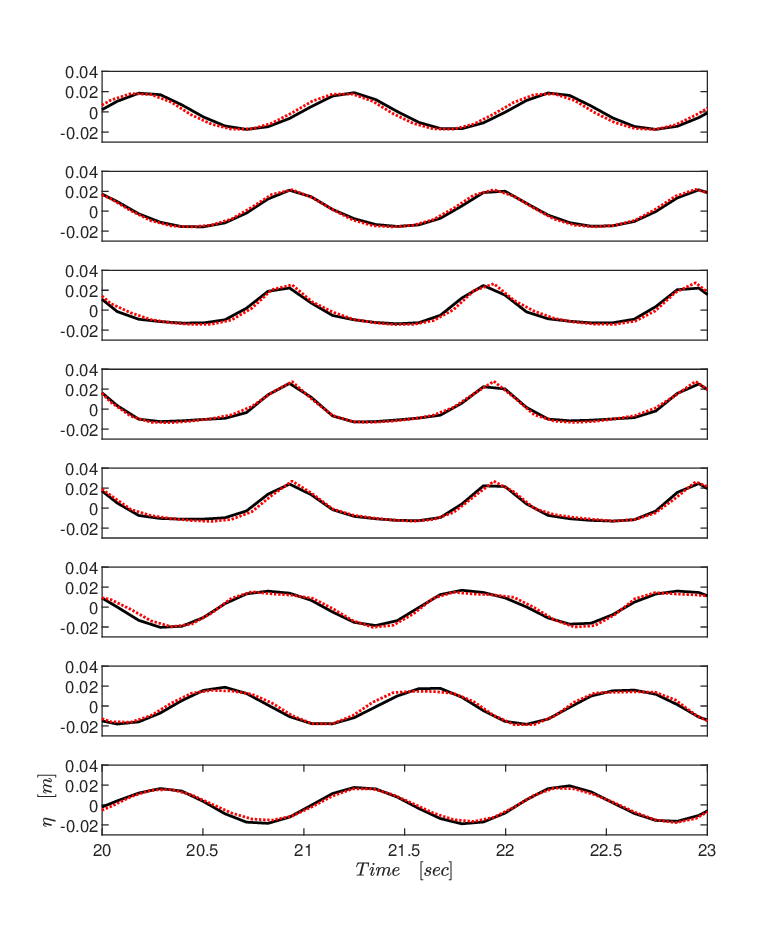}}
%  \centering
%  \includegraphics[width=.4\linewidth]{figs/elev_mono_short.eps}
  \caption{\textcolor{black}{Case \#2 (short waves)}}
  \label{fig:sub2}
\end{subfigure}
\caption{Comparison of the free surface elevation for monochromatic input waves. Plots of gauges \#1 $\sim$ \#8 are arranged from top to bottom. '---': Present model; '\textcolor{red}{\(\cdots\)}': Experiment.}
\label{fig:beji-surf-mono}
\end{figure}

\begin{figure}
\centering
\begin{subfigure}{.5\textwidth}
\centerline{\includegraphics[width=70mm]{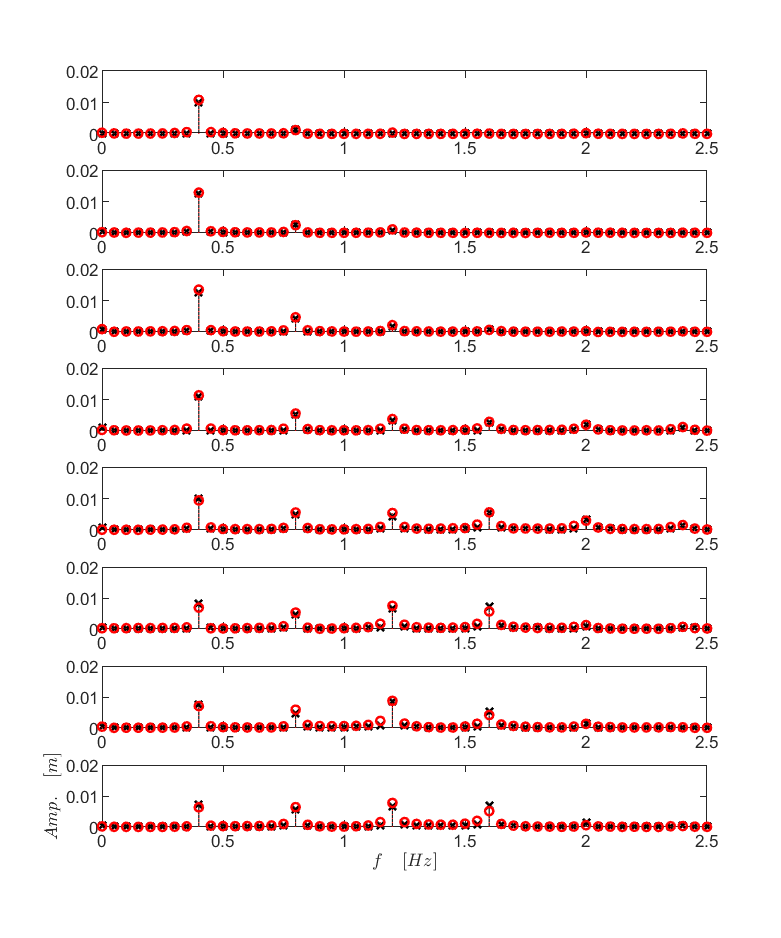}}
  \caption{Case \#1 (long waves)}
  \label{fig:sub1}
\end{subfigure}%
\begin{subfigure}{.5\textwidth}
\centerline{\includegraphics[width=70mm]{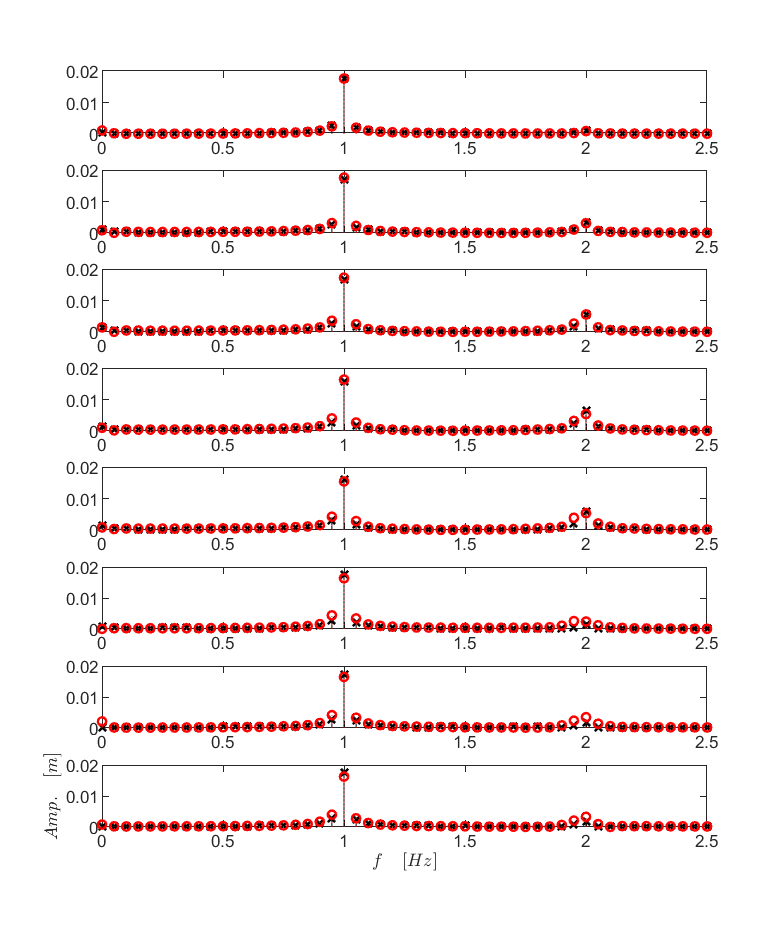}}
  \caption{Case \#2 (short waves)}
  \label{fig:sub2}
\end{subfigure}
\caption{Comparison of the amplitude spectra for monochromatic input waves. Plots of gauges \#1 $\sim$ \#8 are arranged from top to bottom. '\(\times\)': Present model; '\textcolor{red}{\(\circ\)}': Experiment.}
\label{fig:beji-amp-mono}
\end{figure}

For uniform monochromatic incident waves (case 1 and 2), the time signals of the simulated free surface are collected at several locations along the slope (their locations are indicated in figure \ref{fig:beji-domain}) and are compared with those obtained from the laboratory measurements as shown in figure \ref{fig:beji-surf-mono}. Based on the simulation results, it is noted that the shape of long waves will subject to a substantial transformation after propagating over the submerged bar, with a forward leaning front created at the top of the slope. Afterwards, the long waves break up into several short waves behind the bar. Whereas for short waves \textcolor{black}{($k_p|\delta| \approx 1.2$ so the model is in the range of applicability referring to section \ref{sec:limit})}, it is found that the wave profiles exhibit cnoidal wave features with a vertically asymmetric profile (sharper crest and flatter trough) observed at the top of the slope. Then this asymmetry disappears after the bar and waves recover to a vertically symmetric form. Moreover, it is found that the simulated wave surfaces coincide with the laboratory measurements across all gauges for both long and short waves, indicating that the ESBI successfully captured their nonlinear transformations over the submerged bar.    
The asymmetry of the wave profiles over the varying bottom implies that the wave-seabed interactions have induced spectral energy increase at super-harmonics. It is of interests to validate if the present model can precisely reproduce the spectra along the slope. Hence, the amplitude spectra can be extracted from the surface elevation time signal using FFT, and the comparisons between simulations and experiments are presented in figure \ref{fig:beji-amp-mono}. It shows that the energy cascade from peak frequency to super-harmonics is evident in both long and short waves, and the simulations and experiments are consistent. For short waves, the amplitude spectrum after the bar recovers to that before bar, whereas this recovery is not observed for long waves. Meanwhile, good agreement is found between the simulations and laboratory measurements with respect to the harmonic amplitude across all gauges, and the maximum errors are 1\% and 1.5\% for long and short wave simulations, respectively. Therefore, it can be said that the ESBI model can accurately reproduce nonlinear transformations of monochromatic waves over submerged bar in simulations. 

\begin{table}
  \begin{center}
\def~{\hphantom{0}}
  \begin{tabular}{cccc}
      Case No. &  Wave height [cm]   & Peak frequency [Hz]  & Input spectrum    \\[3pt]
%      No.  &  (unit:cm)      & (unit:Hz)        & spectrum     \\[3pt]
      1    &   2.9           & 0.4              & Monochrome \\
      2    &   4.1           & 1.0              & Monochrome \\
      3    &   2.9           & 0.4              & JONSWAP \\
      4    &   4.1           & 1.0              & JONSWAP \\
  \end{tabular}
  \caption{Selected wave conditions for simulating wave propagation over submerged bar.}
  \label{tab:beji}
  \end{center}
\end{table}

%\begin{figure}
%  \centerline{\includegraphics[width=100mm]{figs/beji-domain.eps}}
%  \caption{Illustration of the computational domain for simulating wave propagation over submerged bar.}
%\label{fig:beji-domain}
%\end{figure}

In cases 3 and 4, the transformations of random waves are simulated by using ESBI model. To reproduce the random waves in simulations, the wave amplitude and phase of each spectral component are derived first based on the surface elevation signal recorded on gauge \#1 in the laboratory. A shift in distance is then applied to the phase in order to obtain those at the wavemaker, which are used for wave generation in simulations. 
\textcolor{black}{Notably, wave breaking may occur over the bar in these irregular wave tests. Like many other FPNF models, the present model requires special treatment in order to stabilize the simulation with a good representation of the energy dissipation due to wave breaking. This has become an active research area leading to many important development of breaker models, and various energy dissipation models in physically meaningful approaches have been proposed such as those employed in HOS model \citep{seiffert2018simulation} and Whispers3D \citep{papoutsellis2019modelling,simon2019comparing}, as well as a unified model adopted in the BE model \citep{grilli2020fully,mohanlal2023unified}. 
Nevertheless, discussions on the dissipation model is out of the scope of the present study. For simplicity, a simple low-pass filter is appied to the free surface and velocity potential at every time step \citep{xiao2013rogue}:
\begin{equation}\label{eq:filter}
    \Lambda = e^{-\left|\frac{k}{\beta_1} \right|^{\beta_2} }
\end{equation}
\noindent where $\beta_1=8$ and $\beta_2=30$ can well represent the energy dissipation for a variety of breakers over a broad range of wave steepness \citep{xiao2013rogue}.}

\begin{figure}
\centering
\begin{subfigure}{.5\textwidth}
%\centerline{\includegraphics[width=70mm]{figs/elev_rand_long.eps}}
\centerline{\includegraphics[width=70mm]{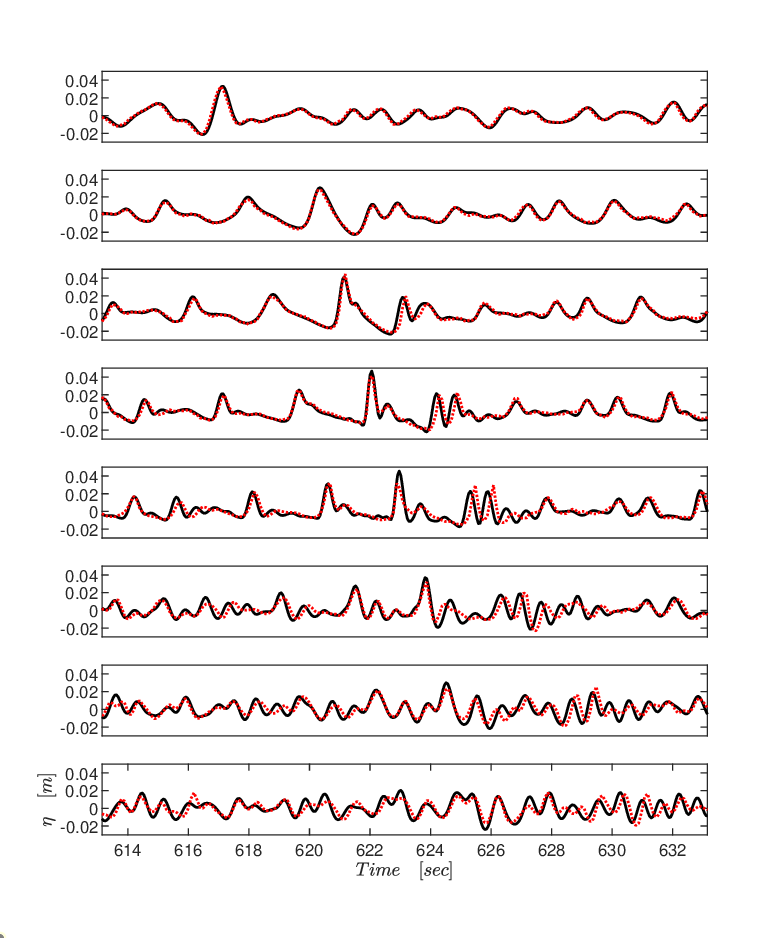}}
  \caption{\textcolor{black}{Case \#3 (long waves)}}
  \label{fig:sub1}
\end{subfigure}%
\begin{subfigure}{.5\textwidth}
%\centerline{\includegraphics[width=70mm]{figs/elev_rand_short.eps}}
\centerline{\includegraphics[width=70mm]{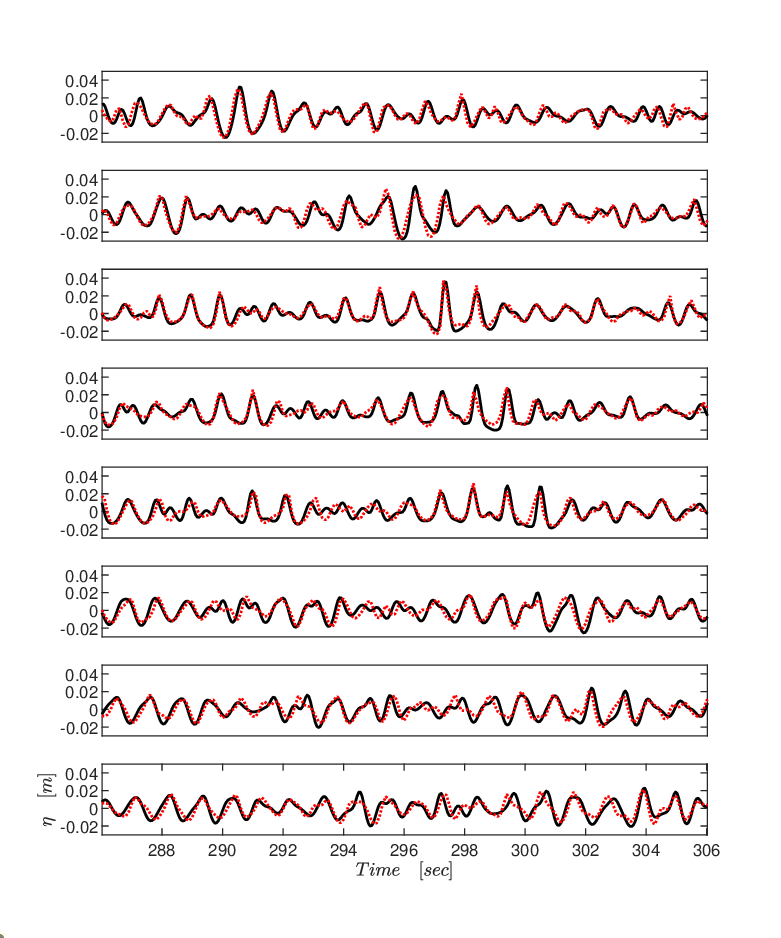}}
  \caption{\textcolor{black}{Case \#4 (short waves)}}
  \label{fig:sub2}
\end{subfigure}
\caption{Comparison of the free surface elevation for random waves. '---': Present model; '\textcolor{red}{\(\cdots\)}': Experiment.}
\label{fig:beji-surf-rand}
\end{figure}

Similar with previous, the simulated free surfaces are extracted at the gauges deployed along the slope and they are plotted in figure \ref{fig:beji-surf-rand}. It shows that the simulated surface elevation agrees generally well with the laboratory measurements, except that a phase shift in time is observed at the last few gauges over the dropping slope in short wave simulations. \textcolor{black}{In the midst, there is also a general reduction of the height of the largest waves along the dropping slope, where the numerical model consistently over-predicted their heights (see figure \ref{fig:beji-surf-rand}(a)). This reduction may be associated with the bottom friction and flow separation at the top of the bar, which are not represented in the present model. However, with the explicit calculation of the velocity potential on the bottom surface, the tangential velocity can be easily derived in the present model. This enables the calculation of the bottom friction using a simple quadratic equation, which can be incorporated into the present model in a future study. }

\begin{figure}
\centering
\begin{subfigure}{.5\textwidth}
\centerline{\includegraphics[width=70mm]{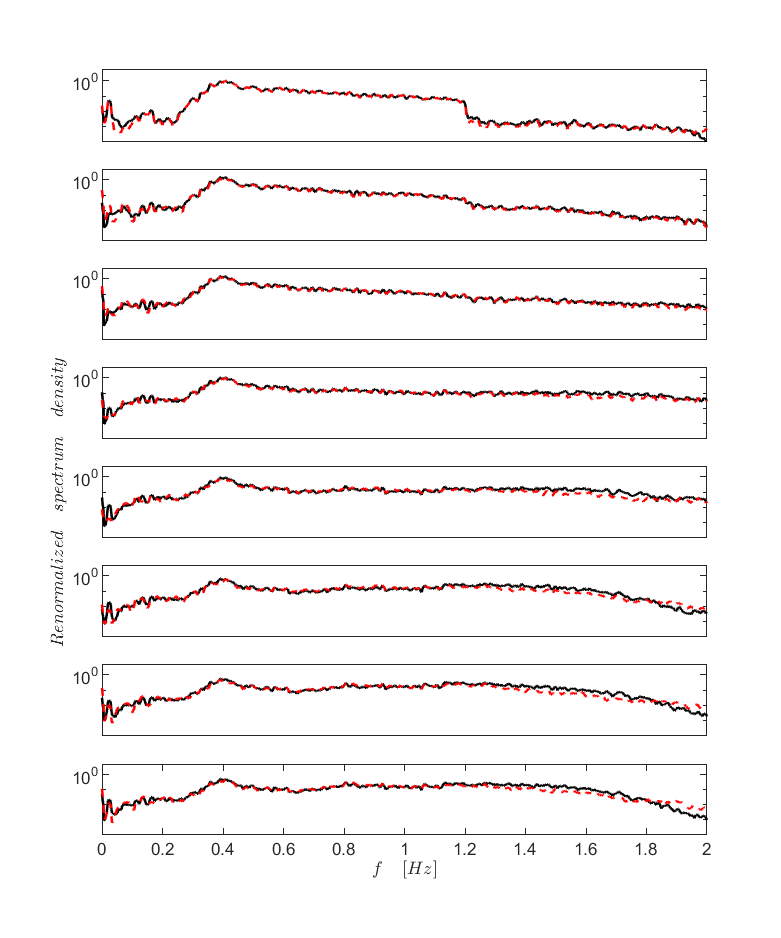}}
  \caption{Case \#3 (long waves)}
  \label{fig:sub1}
\end{subfigure}%
\begin{subfigure}{.5\textwidth}
\centerline{\includegraphics[width=70mm]{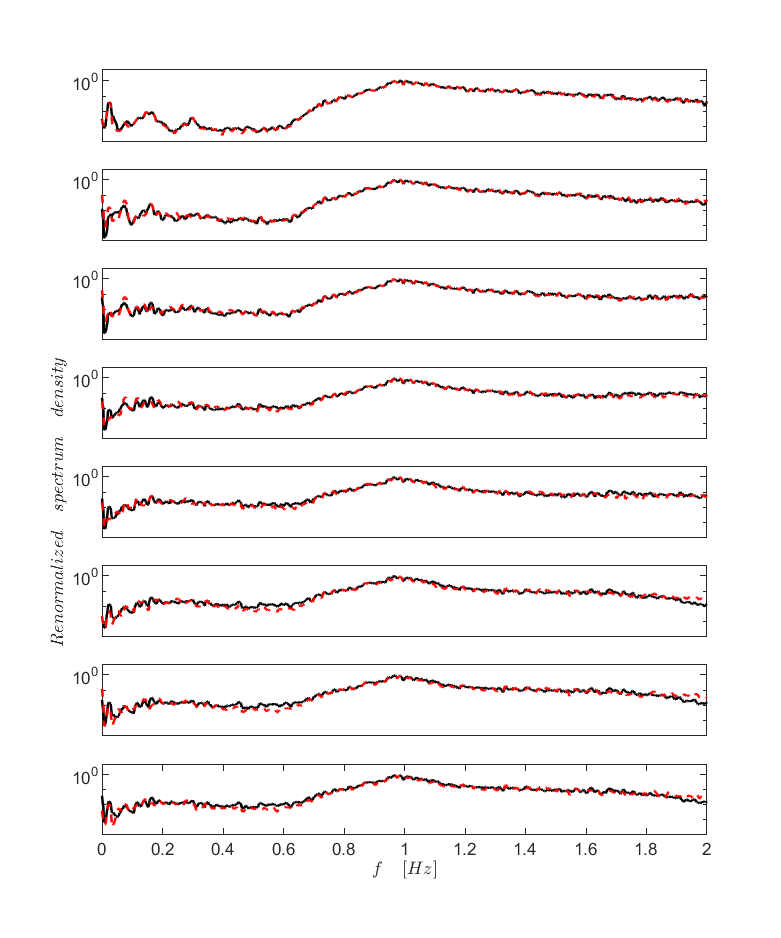}}
  \caption{Case \#4 (short waves)}
  \label{fig:sub2}
\end{subfigure}
\caption{Comparison of the free surface elevation for random waves. '---': Present model; '\textcolor{red}{\(--\)}': Experiment.}
\label{fig:beji-spec-rand}
\end{figure}

In the meantime, to examine the energy cascade among spectral components for random waves, the free surface elevation is converted to spectrum density using FFT, and the comparisons between simulations and experiments are displayed in figure \ref{fig:beji-spec-rand}. 
It is deduced from the spectrum at gauge \#1 in figure \ref{fig:beji-spec-rand}(a) that a cut-off frequency of 1.2 Hz was applied for wave generation in the laboratory. However, the spectral energy in the range beyond cut-off frequency quickly builds up along the propagation due to the nonlinear wave-wave and wave-seabed interactions, as shown in the spectra on the following gauges. \textcolor{black}{It is also interesting to note that the present model over-predicted the energy on the gauges over the dropping slope. This is consistent with the over-prediction of the largest wave heights as seen in figure \ref{fig:beji-surf-rand}(a).}
Meanwhile, figure \ref{fig:beji-spec-rand}(b) shows that the energy cascade in the short wave simulations is not as evident as in long wave simulations. Overall, the spectra based on the simulations are in good agreement with those obtained in the laboratory, and the errors of the total spectral energy (zeroth moment) at the last gauge are 9\% and 3\% in long and short wave simulations, respectively. Therefore, it can be said that the ESBI model accurately reproduces the transformations of random waves over the submerged bar, as well as the energy cascade among spectral components. 

However, it should be addressed that, as shown in figure \ref{fig:beji-spec-rand}, long (or low-frequency) waves were excited and recorded in the laboratory as well as in simulations. The emergence of such long waves is attributed to two main sources. One is owing to the difference interactions between two components of proximate frequencies \citep{longuet1962radiation}. Such long waves are produced immediately after waves are generated from the wavemaker and bounded to the surface envelope. The other source is due to the scattering of free long waves at the edge of the slope \citep{mei1984long,liu1989note}. The short wave group induced bound and free long waves are studied recently through a coupled phase-averaged and phase-resolving approach \citep{wang2022}. However, another set of the evolution equation for the random phase is required in order to simulate random short waves induced long waves, which is not available so far. Regarding this purpose, the present model can be adopted to describe short and long waves simultaneously in one phase-resolving simulation, which however is not the focus here and will be investigated systematically in another study.

\subsection {Bragg reflection over submerged sinusoidal bars}
\begin{figure}
  \centerline{\includegraphics[width=100mm]{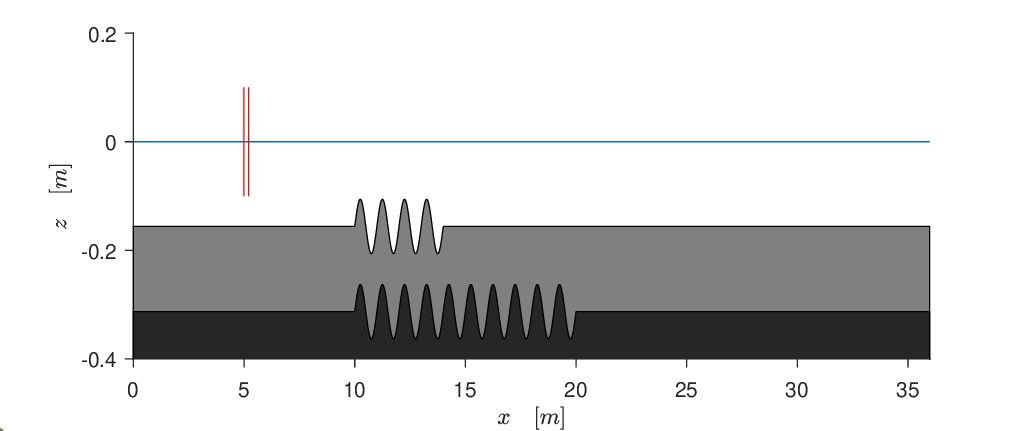}}
  \caption{Illustration of the computational domain for simulating Bragg reflection over submerged sinusoidal bars.Grey and black patches denote the bathemetry for Test 1 and 2, respectively. The location of the wave gauges are marked by the red vertical lines, which are used for recording the surface elevation to calculate the reflection coefficients.}
\label{fig:bragg-domain}
\end{figure}

This section reports the validation results of using ESBI for simulating the Bragg reflection phenomenon. It is well known that periodic undulating topographies (such as sandbars) can interact with incident waves. And wave reflections will be amplified when the wavelength is approximately twice of that corresponding to the bottom undulations. Hence, this phenomenon is also often referred to as Bragg resonance. To validate the efficacy of the present ESBI model, this study selects and reproduces two cases from the experiments conducted by  \cite{davies1984surface}, which are classical experiments for benchmarking numerical models. \textcolor{black}{Though it is noted that it is sufficient to predict the Bragg reflection with lower-order models, it is still an interesting experiment for validation since the ESBI is used to simulate Bragg reflection for the first time.}
The computational domain is illustrated in figure \ref{fig:bragg-domain}, and the seabed conditions are summarized in table \ref{tab:bragg-waveCon}, namely, a relatively deep water case with longer undulating bottom and a shallower depth with shorter undulating bottom. The domain is descritized into 2048 points, and waves are generated and absorbed using the same technique as described in Section \ref{sec:submerged}. For each case, 110 simulations of uniform incident waves are being carried out with different peak periods, leading to a variety of bar length to wavelength ratios (\(2S/L\)) ranging from 0.5 \(\sim\) 1.6. For comparing the simulation results with the theoretical models \citep{liu2019analytical,hsu2007application}, the wave height is fixed as 1 cm for wave generation in the simulations. This is to minimize the influences of the wave nonlinearities on quantifying the wave reflection. After the simulations, the time history of free surface elevations is collected on two wave gauges deployed in front of the sinusoidal bars (see figure \ref{fig:bragg-domain}), which are used to estimate the reflection coefficients based on Goda \& Suzuki's method (\citeyear{goda1976estimation}). 

\begin{figure}
\centering
\begin{subfigure}{.45\textwidth}
\centerline{\includegraphics[width=65mm]{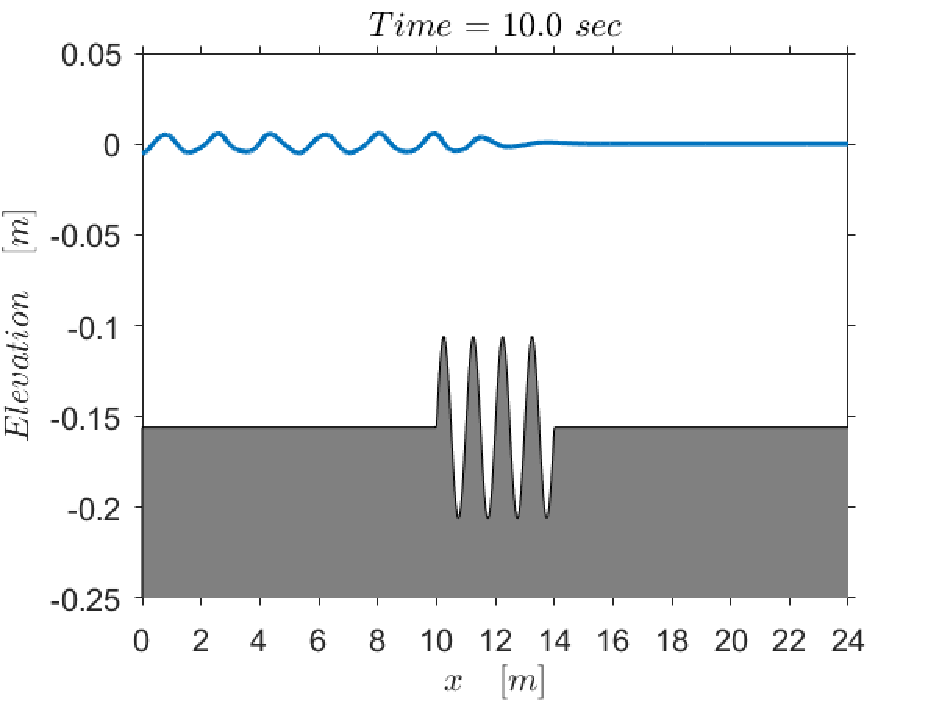}}
  \caption{}
  \label{fig:sub1}
\end{subfigure}%
\begin{subfigure}{.45\textwidth}
\centerline{\includegraphics[width=65mm]{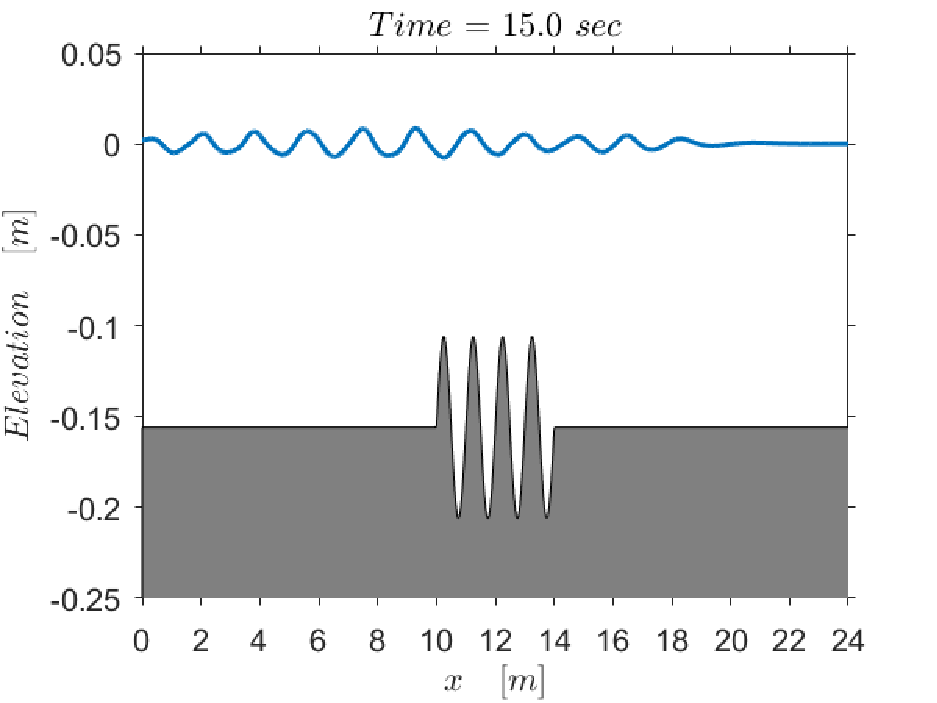}}
  \caption{}
  \label{fig:sub2}
\end{subfigure}
\begin{subfigure}{.45\textwidth}
\centerline{\includegraphics[width=65mm]{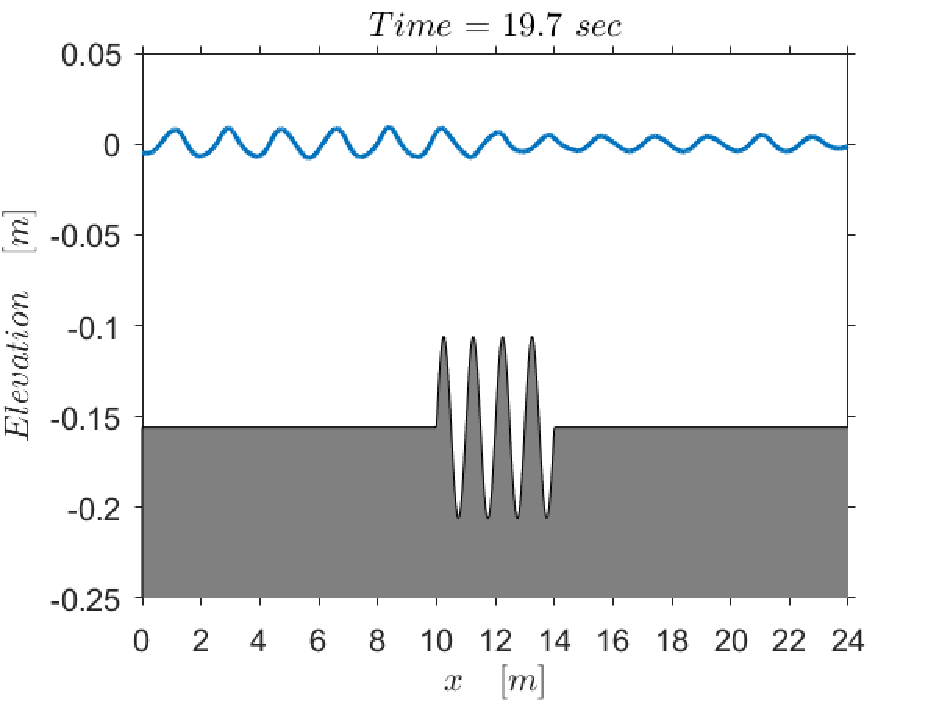}}
  \caption{}
  \label{fig:sub3}
\end{subfigure}
\begin{subfigure}{.45\textwidth}
\centerline{\includegraphics[width=65mm]{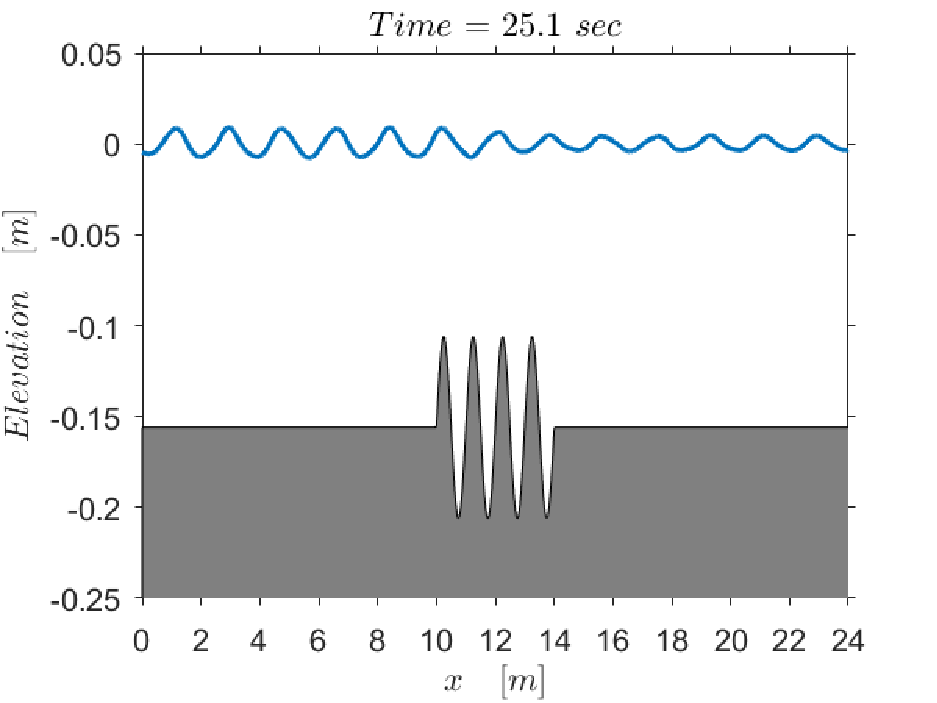}}
  \caption{}
  \label{fig:sub4}
\end{subfigure}
\caption{Snapshots of the surface elevation at different time for case \(2S/L\) = 1.1 and water depth 0.156 m.}
\label{fig:bragg-surf}
\end{figure}

For clarity, only the wave surface profiles in one of the simulations of \(2S/L\) = 1.1 in test \#1 are displayed in figure \ref{fig:bragg-surf}. A uniform wave train is generated and the leading waves arrive at the undulating bars after 10 seconds as shown in figure \ref{fig:bragg-surf}(a). 
%Note that the waves are not perfectly uniform, which may be due to the large wave length to depth ratio, implying a cnoidal wave theory should be employed.  
The resonance is then triggered with scattered waves being excited when the wave train propagates over the undulations (figure \ref{fig:bragg-surf}(b)). The scattered waves will propagate towards both directions and superimpose with the incident waves, resulting in a standing wave train as shown in figure \ref{fig:bragg-surf}(c). It can be seen that the wave height becomes amplified prior to the undulations, whereas it reduces after the periodic bars. The standing waves will persist if the incident waves are not interrupted (see figure \ref{fig:bragg-surf}(d)).     

\begin{figure}
\centering
\begin{subfigure}{.5\textwidth}
\centerline{\includegraphics[width=70mm]{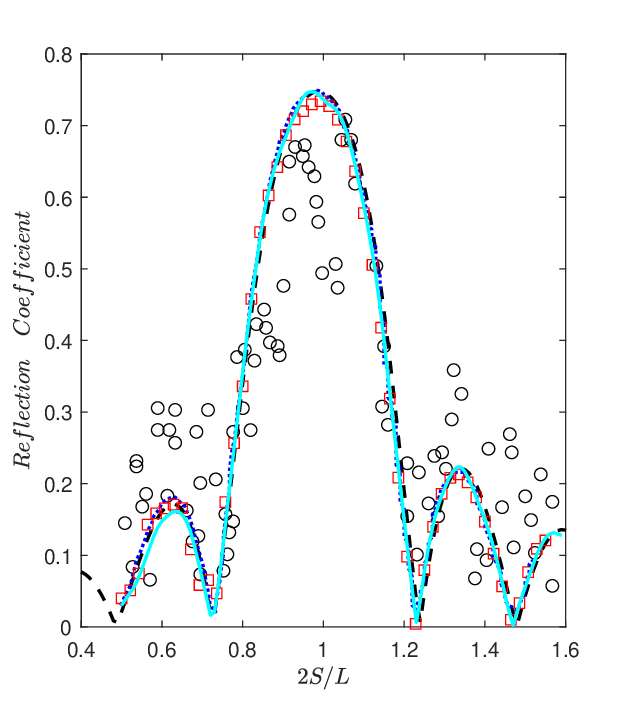}}
  \caption{Water depth: 0.156 m}
  \label{fig:sub1}
\end{subfigure}%
\begin{subfigure}{.5\textwidth}
\centerline{\includegraphics[width=70mm]{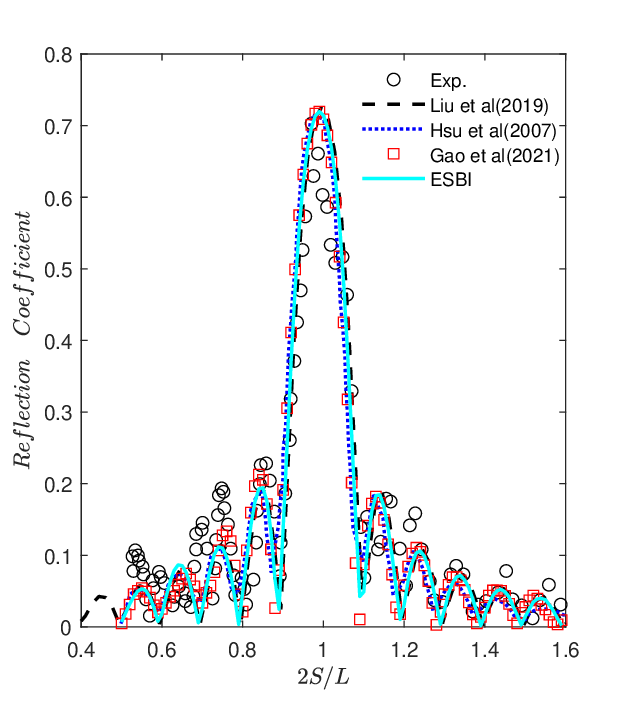}}
  \caption{Water depth: 0.313 m}
  \label{fig:sub2}
\end{subfigure}
\caption{Comparisons of the reflection coefficients for Bragg reflection simulations. '\(\circ\)': Experiment; '\textcolor{blue}{\(\cdots\)}': \cite{hsu2007application}; '\(--\)': \cite{liu2019analytical};  '\textcolor{red}{\(\square\)}': \cite{gao2021investigation}; '\textcolor{cyan}{---}': Present model.}
\label{fig:bragg-coe}
\end{figure}

Next, the reflection coefficients obtained in the numerical simulations based on the ESBI model are compared with those obtained from the experiments \citep{davies1984surface} as shown in figure \ref{fig:bragg-coe}. The theoretical predictions based on the MS equation \citep{liu2019analytical} and Boussinesq equation \citep{hsu2007application}, as well as those obtained from numerical simulations based on the fully nonlinear Boussinesq model FUNWAVE-TVD \citep{gao2021investigation} are also presented for comparisons. A good agreement between the simulation results based on the ESBI and the theories \citep{liu2019analytical,hsu2007application} in both test cases is observed, all of which are generally consistent with the coefficients estimated from the laboratory measurements. The simulation results based on the present model reconfirm that the maximum reflection occurs at the ratio \(2S/L\) = 1, and decays when it increases or decreases. It is also noted that the present ESBI model yields slightly better predictions of the reflection coefficients than the FUNWAVE-TVD \citep{gao2021investigation} in relatively deep water (test \#2) as shown in figure \ref{fig:bragg-coe}(b), when compared with the two theoretical models \citep{liu2019analytical,hsu2007application}. 
\textcolor{black}{A possible explanation may be associated with the reduced accuracy of the FUNWAVE-TVD model on the wave dispersion in relatively deep water. For reader's interests, a direct comparison between the ESBI and FUNWAVE-TVD on simulating focusing waves with the emphasis on accuracy of dispersion was reported in \cite{wang2015numerical}. }
Afterall, it can be summarized that the present ESBI model can be reliably used to simulate the Bragg reflection.  

\begin{table}
  \begin{center}
\def~{\hphantom{0}}
  \begin{tabular}{ccccc}
      Test No.  & ~Number of bars       & ~Bar length [m]~      & Bar height [m]~        &   Water depth [m] \\[3pt]
%      No.   & ~of bars      &   (unit:m)        &   (unit:m)         &   (unit:m)  \\[3pt]
       1        & ~4               &     1           &     0.05             & 0.156 \\
       2        & 10               &     1           &     0.05             & 0.313 \\
  \end{tabular}
  \caption{Selected bathymetry conditions for simulating Bragg reflection over submerged sinusoidal bars.}
  \label{tab:bragg-waveCon}
  \end{center}
\end{table}

\subsection {Nonlinear evolution of a Peregrine breather}\label{sec:peregrine}
\textcolor{black}{Modulational} instability of water waves has recently received significantly scientific interests, which provide an explanation to possible mechanisms of rogue wave occurrences. It is noted that the models originally developed for simulating waves in shallow or finite water depth is rarely used to study wave instabilities in deep water scenarios, even though substantial improvements have been made, e.g., the accuracy on wave dispersion of the Boussinesq models has been significantly improved \citep[and so on]{Lynett2004,Lynett2004b,Chen2006,Shi2012,Liu2018}. In contrast, the ESBI is originally developed for deep water, while with the new improvements incorporated, its application is not restricted by water depth. To demonstrate that, this section will conduct a verification and validation case by simulating a nonlinear breather wave phenomenon that can only occur in relatively deep water.

The Peregrine breather is one of the theoretical solutions of the nonlinear Schrödinger equation, which describes the unidirectional dynamics of Stokes wave train subject to \textcolor{black}{modulational} instability \citep{peregrine1983water}. Under specific circumstances, an initially disturbed uniform wave train will develop into a large wave and recovers to uniformity. Because of this feature, it can be used to explain the occurrences of rogue wave extremity in deep seas \citep{akhmediev2009waves,shrira2010makes}. Such a case can be used to validate the present ESBI model for its capability of modeling the \textcolor{black}{modulational} instability. Regarding this purpose, the experiment in \cite{chabchoub2011rogue} is reproduced herein. 
The simulation domain is 35 m in length and is descritized into 4096 points.
%(64 points per peak wavelength). 
The bottom is flat so that \(\delta\) is set to zero in the model. Similar with previous, the pneumatic wavemaker is specified to generate a target wave based on the analytical solution \citep{peregrine1983water}:
\begin{equation}\label{eq:PB}
    \eta(\xi,\tau)=\mathrm{Re}\left\{\varepsilon_0 \left\{1+\frac{4(-1+\mathrm{i}\varepsilon^2\tau)}{1+[2\sqrt{2}\varepsilon_0\left(\xi-c_g\tau\right)]^2+\varepsilon_0^4\tau^2}\right\}e^{\mathrm{i}[\xi-(1+\varepsilon_0^2/2)\tau]}\right\}.
\end{equation}

\noindent where \(\mathrm{Re}\{*\}\) denotes the real part, \(\varepsilon_0\) is the wave steepness of the carrier waves (uniform part), \(c_g\) is the wave group speed, \(\xi = x - x_f\) and \(\tau = t - t_f\) with \(x_f\) and \(t_f\) being the specified focusing location and time, respectively. To reproduce the experiment configuration, the peak wavenumber in the simulation is set to \(k_p = 11.63 \) m\(^{-1}\), with some of the dimensionless variables given as \(h = 11.63\), \(\varepsilon_0 = 0.1163\), \(x_f/L_p = 11\), \(t_f/T_p = 68\). 

\begin{figure}
\centering
\begin{subfigure}[b]{0.75\textwidth}
   \includegraphics[width=1\linewidth]{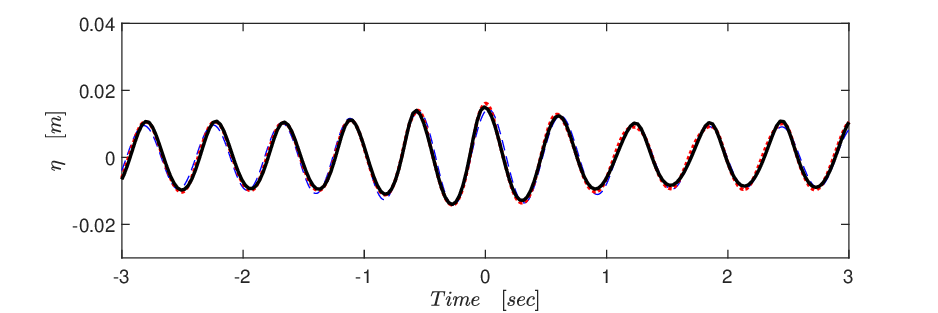}
   \caption{\(x\) = 1.1 m}
   \label{fig:Ng1} 
\end{subfigure}

\begin{subfigure}[b]{0.75\textwidth}
   \includegraphics[width=1\linewidth]{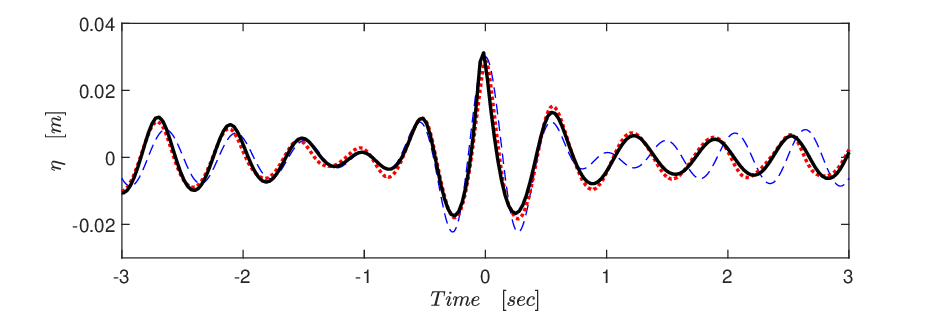}
   \caption{\(x\) = 9.1 m}
   \label{fig:Ng2}
\end{subfigure}
\caption{Comparison of the free surface elevation for Peregrine breather simulation. '\textcolor{blue}{\(--\)}': Theoretial solution based on Eq. (\ref{eq:PB}); '\textcolor{red}{\(\cdots\)}': Experiment by \cite{chabchoub2011rogue}; '---': Present model.}
\label{fig:PB}
\end{figure}

The wave surface elevation is recorded at two locations, i.e., 1.1 m and 9.1 m away from the wavemaker in both simulation and laboratory, and they are displayed in figure \ref{fig:PB}. The theoretically predicted surface profile can also be estimated by using Eq. (\ref{eq:PB}), which is also presented in the figure. The wave profiles are re-centred so that the peaks appear at time equal to zero. 
It shows that the simulated free surface agrees very well with the measurement in laboratory near the wavemaker and at the focusing point. The development of a nearly uniform wave train into an extremely large wave is well captured in the numerical simulation. The amplification of the largest crest height appears to be in good agreement with the theory, i.e., 3 times of the carrier wave amplitude. Meanwhile, both the focused breathers at 9.1 m in simulation and experiment show discrepancies from the theory. Specifically, they both exhibit narrower crest and elevated trough compared with the analytical solution. The simulated breather is slightly sharper with a more elevated following trough comparing with the laboratory measurement. Nevertheless, the difference of the maximum wave height at focusing point between the simulation result and experiment data is only 3.2\%. Therefore, it can be concluded that the present ESBI model can be reliably used for simulating breather-type extreme waves due to \textcolor{black}{modulational} instability in deep water. 

\subsection {Instability of crossing random seas}\label{sec:crossing}
%Crossing seas have been identified as an indicator linked to increased probability of extreme waves.
The bimodal condition of crossing seas featuring two spectral peaks is shown to be linked to increased probability of extreme waves \citep{mcallister2019laboratory}. One of the possible mechanisms is due to the \textcolor{black}{modulational} instability, which can be represented by using a crossing breather model \citep{onorato2006modulational,shukla2006instability}. And the probability of extreme waves is associated with the aperture angle between the two wave systems. 
Therefore, this section validates the ESBI model for simulating crossing random seas to test its capability of modeling wave instability in directional seas. The experiment to be reproduced in simulation refers to \cite{toffoli2011extreme}. The computational domain covers 40 \(\times\) 40 peak wavelengths, and is resolved into 1024 \(\times\) 512 collocation points in x (along-wave) and y (cross-wave) directions, respectively. The selected domain size and resolution in space ensure that the Fourier modes up to 7- and 3-times peak wavenumber in the x- and y- directions, respectively, are aliasing free. 
The wave condition is based on the JONSWAP spectrum with peakedness \(\gamma = 6\) and wave steepness \(k_pH_s = 0.28\). And the two wave \textcolor{black}{systems cross} at an aperture angle of 40\textdegree, which corresponds to the most hazardous angle leading to highest probability distribution tail. Several wave probes are deployed every 4 peak wavelengths along the centre of the domain in x-direction. The wave generation zone is deployed along \(x=0\) and they are absorbed at a distance near the other end. Each simulation lasts for 1000 peak periods (equivalent to a typical 3-hour sea state). 

\begin{figure}
\centering
\begin{subfigure}[b]{0.75\textwidth}
   \includegraphics[width=1\linewidth]{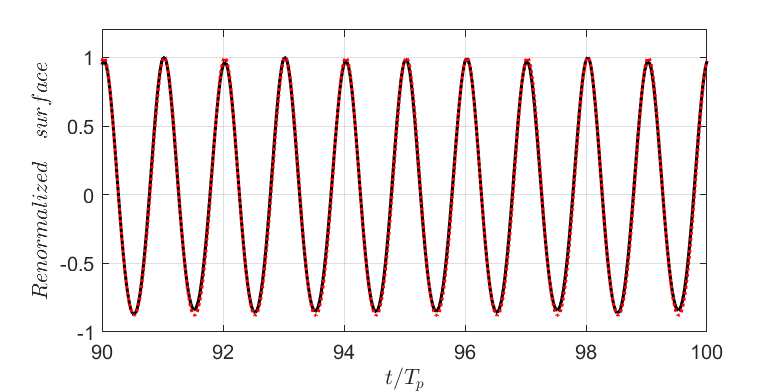}
   \caption{}
%   \label{fig:Ng1} 
\end{subfigure}

\begin{subfigure}[b]{0.75\textwidth}
   \includegraphics[width=1\linewidth]{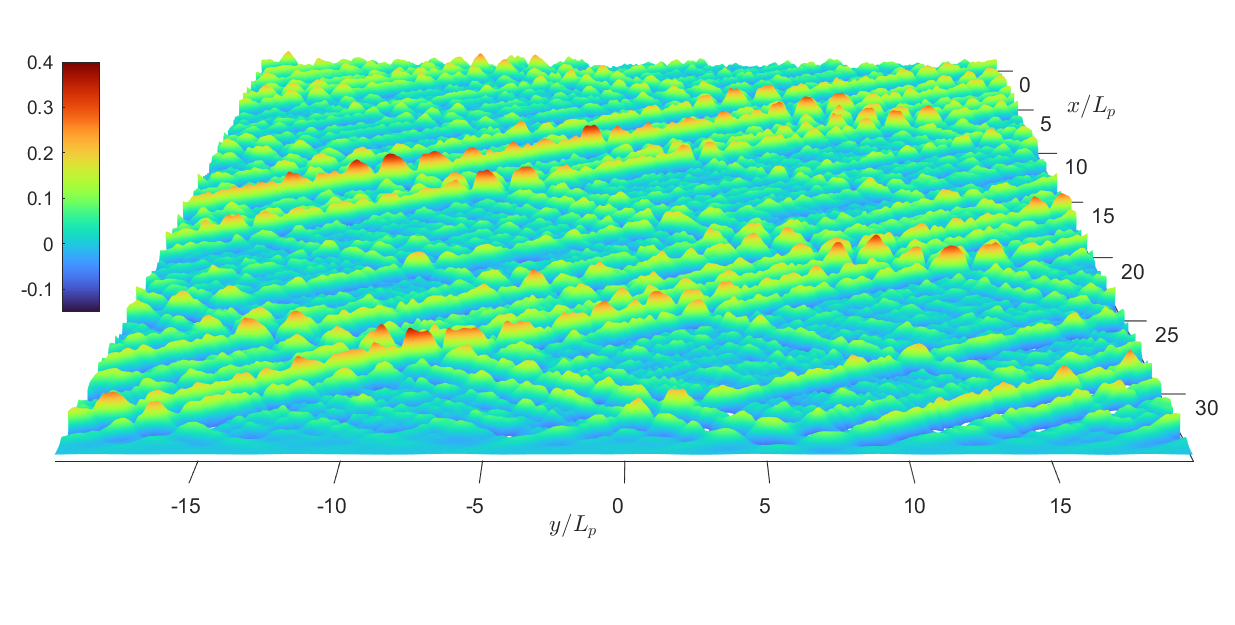}
   \caption{}
%   \label{fig:Ng2}
\end{subfigure}
\caption{\textcolor{black}{Time history of simulated oblique wave train surface (a) and snapshot of the simulated crossing sea surface (b).} In (a): '$--$': ESBI results;  '\textcolor{red}{\(\cdots\)}': Fenton's numerical solution. In (b): colour bar indicates the magnitude of \(\eta\). }
\label{fig:cross-surf}
\end{figure}

\textcolor{black}{Before proceed to simulating crossing seas, the present model is verified for reproducing an obliquely propagating uniform wave train of steepness $\varepsilon = 0.14$ at the direction of 20\textdegree\ with respect to $y=0$. The surface elevation time signal is collected at the location $x/L_p = 32$, and compared with the Fenton's numerical solution \citeyearpar{fenton1988numerical}, as shown in figure \ref{fig:cross-surf}(a). A good agreement between the simulation and Fenton's solution can be observed, despite that minor irregularities can be found for the simulated free surface. This may be due to the shortage of the wave generation technique based on linear wave theory. Nevertheless, the present model can be said to accurately simulate propagating waves in oblique direction. Next, crossing random seas are simulated by using the present model, and a} snapshot of the simulated free surface is shown in figure \ref{fig:cross-surf}\textcolor{black}{(b)}, where several groups of large waves can be seen in space at the time being. 
\textcolor{black}{It is worth noting that wave breaking is inevitable under such wave conditions.} 
%As is well known, the FNPF models become unstable when waves begin to break and overturn, and the present model is no exception. To stabilize the simulation, various energy dissipation models in physically meaningful approaches can be employed such as in HOS model \citep{seiffert2018simulation}, Whispers3D \citep{papoutsellis2019modelling,simon2019comparing}, and BE model \citep{grilli2020fully,mohanlal2023unified}. }
\textcolor{black}{To stabilize the simulation, the low-pass filter of Eq. (\ref{eq:filter}) \citep{xiao2013rogue} is employed again to suppress the breaking. %This filter is shown to well represent the energy dissipation quantitatively over a broad range of wave steepness, breaker types and directional spreading. Hence it is adopted which is shown to be sufficient regarding the purpose of modelling evolution of kurtosis in crossing seas. 
It is reiterated here that this study is not aimed to justify the superiority of various dissipation models. This low-pass filter is later shown to be sufficient regarding the purpose of modelling evolution of kurtosis in crossing seas.}

\begin{figure}
\centering
\begin{subfigure}{.5\textwidth}
\centerline{\includegraphics[width=70mm]{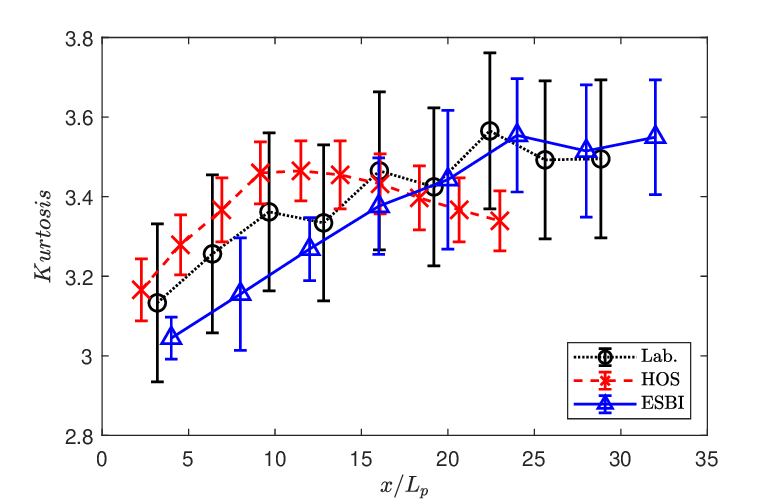}}
  \caption{}
  \label{fig:sub1}
\end{subfigure}%
\begin{subfigure}{.5\textwidth}
\centerline{\includegraphics[width=70mm]{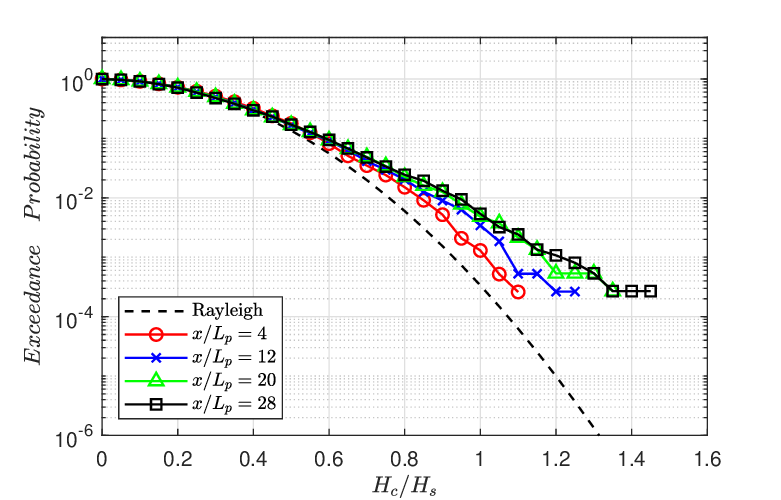}}
  \caption{}
  \label{fig:sub2}
\end{subfigure}
\caption{\textcolor{black}{Comparison of the kurtosis between the numerical simulation and experiment for crossing random seas (a) and wave crest exceedance probability based on simulation results (b).} }
\label{fig:cross-kurtosis}
\end{figure}

It is noted that a quantitative and direct comparison of free surface elevation is difficult due to the lack of information of random phases for wave generation in present study to reproduce exactly the same wave surface in laboratory. Hence, the kurtosis is selected for making comparisons between numerical simulations and experiment. Its magnitude is often adopted as an indicator of rogue wave probability, which can infer the accuracy of the ESBI model for reproducing the associated nonlinear quasi-resonance mechanisms. Note that the kurtosis requires large number of data to minimize the statistical uncertainty, therefore, a total of 4 realizations are performed herein.
At each probe, the kurtosis is then estimated from the simulated surface elevation time history and averaged over 4 realizations, and the values are compared with those obtained in laboratory as shown in figure \ref{fig:cross-kurtosis}\textcolor{black}{(a)}. The figure also displays the simulation results obtained by using the HOS model \citep{toffoli2011extreme}. \textcolor{black}{It should be pointed out that the configurations of the HOS model is substantially different. Firstly, the domain size is much smaller, i.e., 10 \(\times\) 10 peak wavelengths, which is resolved by using 256 \(\times\) 256 grid points. Aliasing errors are eliminated by removing Fourier modes higher than 6 times peak wavenumber in both along and cross wave directions. The simulation duration is 60 peak periods and a fourth-order Runge-Kutta method with a time step size of 1/100 peak period is used for time integration. A total number of 100 realizations are performed to obtain sufficiently long dataset comparable with those collected in the laboratory. Due to the differences in setups, discrepancies are expected between the results from HOS and ESBI simulations. The results of HOS model are still presented here rather to give a qualitative description of the role of modulational instability in the nonlinear evolution of crossing sea states.  }

The curve obtained based on laboratory measurements shows that the kurtosis increases rapidly along the mean wave direction, and reaches the peak of 3.6 at the distance about 22.4 peak wavelengths away from the wavemaker. After this point, its value drops slightly and becomes stabilized at 3.5.
Such a variation of kurtosis in space is attributed to the regain of spectral equilibrium \citep{barratt2022impact}. During the readjustment of the wave spectrum along the propagating direction, the nonlinear physics can lead to occurrence of extreme waves, though the detailed mechanisms are not well understood. 
The simulation results based on ESBI model report the same trend with the kurtosis increases along the mean wave direction. And the slope of this increasing trend is very similar with that derived from laboratory. However, the values near wavemaker is slightly smaller than the experiment, which may be due to the difference in wave generation techniques. The kurtosis is peaked at 24 peak wavelengths away from the wavemaker and then becomes stabilized afterwards. The stabilized kurtosis in simulation is generally consistent with that reported in laboratory. 
\textcolor{black}{To interpret the increased kurtosis along the mean wave direction, the wave crest exceedance probabilities at selected gauges are also estimated and presented in \ref{fig:cross-kurtosis}(b), where $H_c$ stands for crest height (distance from peak to $z=0$). Note that $H_c/H_s>1.2$ is often used for rogue wave criterion along with the condition of overall height greater than $2H_s$ \citep{kharif2008rogue}. The figure shows that with the increased distance from the wavemaker, the enhanced probability of wave crests greater than $1.2H_s$ is observed, indicating more frequent appearances of extremes in the crossing seas. It is seen that the growth of the distribution tail slows down near the end of the domain, which is in accordance with the stabilized kurtosis near the end of the tank. While the probability distributions obtained from the simulations are consistently higher than the Rayleigh distribution, implying that the Rayleigh distribution may not well represent the wave crest probability and their enhanced probability may be due to nonlinear wave effects.}

\textcolor{black}{Interestingly, as shown in figure \ref{fig:cross-kurtosis}(a),} the simulations based on the HOS model report a quite different variation of kurtosis along the mean wave direction, with the value of the kurtosis dropping constantly after it reached the peak of 3.5. One of the reasons is that the HOS model initiated the simulations and utilized the periodic boundary condition, which is not fully consistent with the laboratory settings. And the kurtosis is calculated at fixed time from a surface spatial distribution in the HOS computations, while its value as a function of time is converted into kurtosis as a function of space using the group velocity normal to the wavemaker \citep{toffoli2010evolution}. Apart from that, the simulation results based on the ESBI model agree generally well with that derived from laboratory. The stabilized kurtosis near the end of the tank is accurately predicted in the simulations based on the ESBI model, which implies that the formations of extreme waves due to the instability of two crossing nonlinear wave trains are successfully captured. \textcolor{black}{It is reiterated here that it is not intended to compare the HOS and ESBI model in this section, as the two models use completely different configurations. A fair comparison of the diagnostic procedures between the two models will be reported in the next. }

%\begin{figure}
%  \centerline{\includegraphics[width=80mm]{figs/cross-kurtosis_ka0.25.eps}}
%  \caption{Comparison of the kurtosis between the numerical simulation and experiment for crossing random seas. '\(\circ\)': Experiment; '\textcolor{red}{\(\times\)}': HOS model; '\textcolor{blue}{\(\triangle\)}': Present model.}
%\label{fig:cross-kurtosis}
%\end{figure}

\section {Discussions on the diagnostic procedure: a comparison between ESBI and HOS} \label{sec:dis} 
The HOS method employs the perturbation expansion and suggests that the solution of the velocity potential satisfying Eq.(\ref{eq:laplace}) can be written as:
\begin{equation}\label{eq:HOS}
    \phi(\bm{x},z=\eta,t)=\sum_{m=1}^{M}\sum_{n=0}^{M-m}\frac{\eta^n}{n!}\frac{\partial^n}{\partial z^n} \phi^{(m)}(\bm{x},z=0,t),
\end{equation}

\noindent where \(M\) is the given order in perturbation series, which becomes a user-specified parameter in the numerical modeling. The unknown at each order can be obtained from a successive evaluation by using:
\begin{equation}\label{eq:HOSM}
    \phi^{(m)}(\bm{x},z=0,t)=-\sum_{n=1}^{m-1}\frac{\eta^n}{n!}\frac{\partial^n}{\partial z^n} \phi^{(m-n)}(\bm{x},z=0,t), \quad m=2,3,...M,
\end{equation}

\noindent with initiation \(\phi^{(1)}=\Tilde{\phi}\). For each \(\phi^{(m)}\), the time- and space-dependent variables can be separated as:
\begin{equation}\label{eq:HOSE}
    \phi^{(m)}(\bm{x},z,t)=\sum_{k=1}^{K} \phi^{(m)}_k (t) \Phi_k (\bm{x},z).
\end{equation}

\noindent Substituting Eq.(\ref{eq:HOSE}) into Eq.(\ref{eq:HOSM}) and applying appropriate eigenfunctions in deep or finite waters, the amplitude of each modal component can be determined. The surface vertical velocity can then be calculated by using:
\begin{equation}\label{eq:HOSZ}
    \frac{\partial \phi}{\partial z}(\bm{x},z=\eta,t)=\sum_{m=1}^{M}\sum_{n=0}^{M-m}\frac{\eta^n}{n!} \sum_{k=1}^{K} \phi^{(m)}_k (t) \frac{\partial^{n+1}}{\partial z^{n+1}}\Phi_k (\bm{x},z=0).
\end{equation}

In the case of free surface and velocity potential specified by using Stokes wave theory, the perturbation in HOS converges exponentially fast as \(M\) increases. But the convergence ceases beyond critical steepness of \(\varepsilon \approx 0.35\) owing to the singularities in the analytical continuation of the velocity potential \citep{dommermuth1987high}. 
For a direct comparison between HOS and ESBI model, Fenton's fully nonlinear numerical solution \citep{fenton1988numerical} of Stokes waves is used here, and different wave steepness is selected between \(\varepsilon =\) 0.1 \(\sim\) 0.35. A Fortran program is made based on Eqs. (\ref{eq:HOS} \(\sim\) \ref{eq:HOSZ}), which will be used for comparisons with the present model. 
\textcolor{black}{In deep water, the recurrence formula for estimating $V_l$ in ESBI model, i.e., Eq.(\ref{eq:app-V-scs}), reduces to:
\begin{eqnarray}
\mathcal{F}\left\{ V_{l} \right\} &=&  \sum_{j=1}^{l-1}   - \frac{k^j}{j!}\mathcal{F}\left\{ \eta^jV_{l-j} \right\} - \frac{k^{l-2}}{(l-1)!} \mathrm{i}\bm{k} \cdot \mathcal{F}\left\{ \eta^{l-1}\nabla\widetilde{\phi}  \right\} ,    
\end{eqnarray}
\noindent for $2\le l \le 7$. It can be seen that the present model substantially simplifies the estimation of vertical velocity in comparison with the HOS model that requires two separate steps using Eqs. (\ref{eq:HOSM}) and (\ref{eq:HOSZ}). }

\begin{figure}
\centering
\begin{subfigure}{.33\textwidth}
\centerline{\includegraphics[width=45mm]{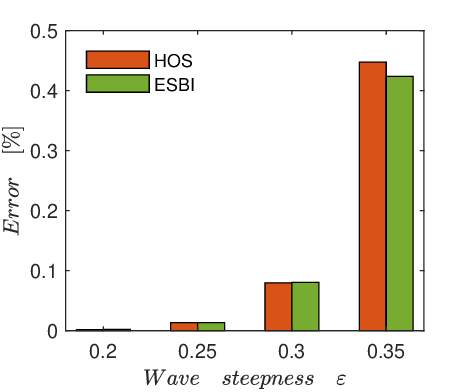}}
  \caption{}
  \label{fig:sub1}
\end{subfigure}%
\begin{subfigure}{.33\textwidth}
\centerline{\includegraphics[width=45mm]{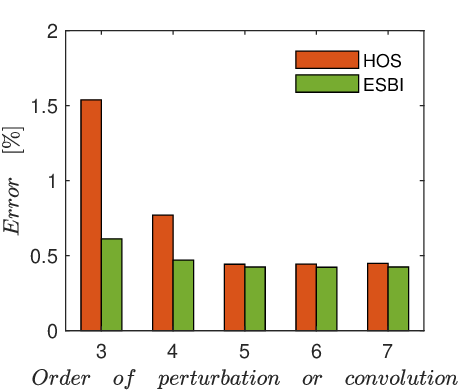}}
  \caption{}
  \label{fig:sub2}
\end{subfigure}
\begin{subfigure}{.33\textwidth}
%\centerline{\includegraphics[width=45mm]{figs/hos-fft.eps}}
\centerline{\includegraphics[width=45mm]{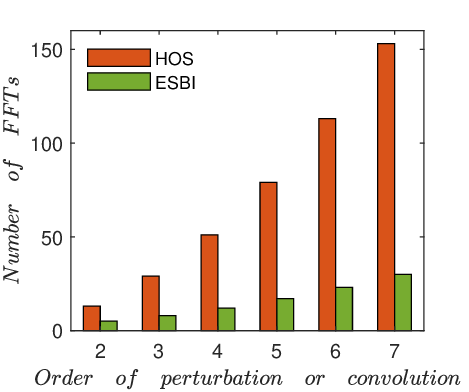}}
  \caption{}
  \label{fig:sub1}
\end{subfigure}%

\caption{Comparisons between HOS and ESBI for performing the diagnose procedure. (a) Error of estimating vertical velocity of a Stokes wave of various steepness at 7th order perturbation or convolution. (b) Error of estimating vertical velocity of a Stokes wave due to the truncation to perturbation or convolution. (c) Number of FFT operations required in HOS and ESBI. }
\label{fig:hos-fft}
\end{figure}

Both models cover a domain of one wavelength and is descritized with 64 points. Thus, the first 16 Fourier modes are aliasing free, which are sufficient to minimize dependence of the accuracy of the diagnostic procedure on the aliasing treatment scheme. Both models are then used to calculate the vertical surface velocity \(V\) for various wave steepness at different perturbation or convolution orders. The error of each model is then estimated by comparing with Fenton's numerical solution \citep{fenton1988numerical}, and the results are summarized in figure \ref{fig:hos-fft} (a) and (b). Figure \ref{fig:hos-fft}(a) shows the errors of estimating the vertical velocity at 7th order perturbation in HOS and 7th order convolution in ESBI. Both models are sufficiently accurate, and the maximum errors are below 0.5\% regarding the largest wave steepness being considered, and the errors quickly diminish for reduced wave steepness. The ESBI marginally outperforms the HOS for the largest wave steepness case.      

\begin{figure}
\centering
\centerline{\includegraphics[width=90mm]{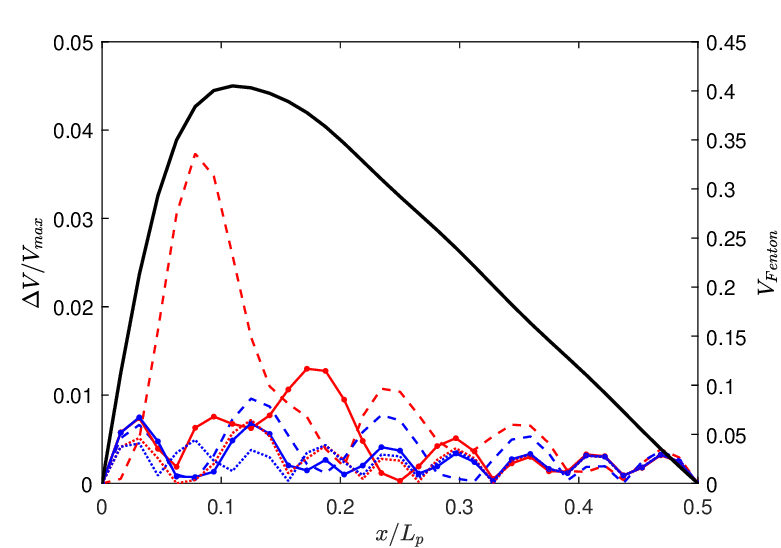}}
\caption{\textcolor{black}{Vertical velocity difference between model results and Fenton's solution for a Stokes wave of steepness \(\varepsilon = 0.35\). Only half of the velocity profile is presented for simplicity.} '---': Fenton's numerical solution (\citeyear{fenton1988numerical}) (right ordinate); '\textcolor{red}{\(---\)}': $\Delta V_\text{hos}$ with M=3; '\textcolor{blue}{\(---\)}': $\Delta V_\text{esbi}$ at 3r order; '\textcolor{red}{\(-\bullet-\)}': $\Delta V_\text{hos}$ with M=4; '\textcolor{blue}{\(-\bullet-\)}': $\Delta V_\text{esbi}$ at 4th order; '\textcolor{red}{\(\cdots\)}':$\Delta V_\text{hos}$ with M=5; '\textcolor{blue}{\(\cdots\)}':$\Delta V_\text{esbi}$ at 5th order.}
\label{fig:hos-v}
\end{figure}

To examine the speed of the convergence of the perturbation or convolution, the wave steepness is then fixed to 0.35 and the errors of using HOS and ESBI for calculating the vertical velocity are then evaluated at different order of perturbation and convolution. The errors are presented in figure \ref{fig:hos-fft}(b), which shows that the ESBI produces relatively smaller error than the HOS at the same order of perturbation or convolution. Above 5th order, the errors of both models do not reduce further. For an acceptable tolerance of 0.5\%, the ESBI model would have converged at 4th order convolution, while the HOS model requires at least 5th order perturbation. For clarity, \textcolor{black}{the profile of \(V\) obtained by using Fenton's solution (ordinate on the right), and its difference from those estimated from HOS model ($\Delta V_\text{hos}$) and ESBI model ($\Delta V_\text{esbi}$) at 3rd, 4th and 5th order perturbation or convolution (ordinate on the left) are displayed in figure \ref{fig:hos-v}. 
%At the 3rd order, the maximum of $\Delta V_\text{hos}$ (or $\Delta V_\text{esbi}$) appears at the wave front (or peak), whereas the difference is much smaller thoughout the entire profile for the ESBI model. At the 4th order, $\Delta V_\text{hos}$ reduces significantly with its maxima shifted to the leeside of the peak. Meanwhile, $\Delta V_\text{esbi}$ reduces slightly and is generally smaller than $\Delta V_\text{hos}$. At the fifth order, it is found that the maxima of $\Delta V_\text{hos}$ appears underneath the peak, 
The changing profiles of $\Delta V_\text{hos}$ with increasing order of perturbation indicates that the HOS model improves the accuracy of the solution, and the approximation converges rapidly until 5th order perturbation. In general, the magnitude of $\Delta V_\text{esbi}$ is smaller than $\Delta V_\text{hos}$ at the same order of convolution, and the overall magnitude of $\Delta V_\text{hos}$ and $\Delta V_\text{esbi}$ become comparable at 5th order.
} This finding does not alter if the estimation is made for evaluating the Eulerian speed \(\partial \phi / \partial z\) at the free surface, while the results are omitted here for simplicity.   

\textcolor{black}{Typically, the complexity of a numerical model is often indicated by using a function of the total number of discretization points. Roughly spreaking, due to the use of FFT, the number of operations required at each order in both HOS and ESBI is $O(N \ln{N})$, where $N$ is the total number of grid points. For perturbations or convolutions up to order $M$, the operation count is $O(MN\ln{N})$ within one diagnostic procedure in both models. Therefore, to be more precise,} it is also of interests to know the number of FFT operations demanded in each model, which determines the computational efficiency of pseudo-spectrum methods as both models relies on FFT for repeatedly converting quantities in physical and Fourier space. It is noted that the total operation of FFTs required to complete this diagnose procedure in HOS \textcolor{black}{(calculate ${\partial \phi / \partial z}$ from given $\Tilde{\phi}$)} is \(3M^2+M-1\). Meanwhile, the number of FFTs required for the diagnotic procedure for ESBI \textcolor{black}{(calculate $V$ from given $\Tilde{\phi}$)} is counted\textcolor{black}{, which is $M^2/2+M/2+2$ ($M$ is used to denote the order of convolution). The total number of FFT operations for calculating $\partial \phi / \partial z$ in HOS and $V$ in ESBI from given $\Tilde{\phi}$ are compared and presented} in figure \ref{fig:hos-fft}(c). In general, it is found that the ESBI requires significantly less FFT operations for fixed order of perturbation and convolution. At the 7th order of perturbation and convolution, the number of FFT operations required in HOS (153 operations) can be 5 times of that in ESBI (\textcolor{black}{30} operations). Meanwhile, referring to the Stokes wave case and if 0.5\% tolerance is chosen, the ESBI needs 14 FFT operations at 4th order convolution, whereas the HOS requires 79 FFT operations at 5th order perturbation. Therefore, it can be said that the present EBI model is more computationally efficient than the HOS model. 

\textcolor{black}{It is worthy noting that other fast models are also available for simulating free propagating waves over variable depth, such as the FMM \citep{wang1995efficient,harris2022nonlinear} and REEF3D::FNPF \citep{wang2022flexible}, etc., which also incorporate parallelized algorithm to ensure fast simulations on large domains. Such models are not subject to limitations of small $k_p|\delta|$ assumption (which will be further discussed in the next). However, comparison with such models will require a rigorous benchmark test on both the diagonistic and prognostic procedure, which is out of the scope of this study. }

\textcolor{black}{Lastly, it should be mentioned that parallelization of the present model is implemented with the OpenMP (Open Multi-Processing) technique for performing the FFTs. The results reported in section \ref{sec:crossing} are obtained based on the parallel computations on a workstation equipped with Intel(R) Xeon(R) Gold 5220R CPU @ 2.2 GHz. The total elapsed time with 50 computational threads is about 35 hours per realization. On an even larger spatiotemporal scale, the present model is also ready with the FFTW3 library \citep{frigo2005design} on MPI (Message Passing Interface) platform and cuFFT library for computing FFTs on an NVIDIA(R) GPU. Such applications will be reported in future studies.  }

%\begin{figure}
%  \centerline{\includegraphics[width=100mm]{figs/hos-fft.eps}}
%  \caption{Number of FFTs required in HOS and ESBI for diagnose procedure.}
%\label{fig:hos-fft}
%\end{figure}

%\begin{figure}
%\centering
%\begin{subfigure}{.33\textwidth}
%\centerline{\includegraphics[width=45mm]{figs/V_m3.eps}}
%  \caption{3rd order}
%  \label{fig:sub1}
%\end{subfigure}%
%\begin{subfigure}{.33\textwidth}
%\centerline{\includegraphics[width=45mm]{figs/V_m4.eps}}
%  \caption{4th order}
%  \label{fig:sub1}
%\end{subfigure}%
%\begin{subfigure}{.33\textwidth}
%\centerline{\includegraphics[width=45mm]{figs/V_m5.eps}}
%  \caption{5th order}
%  \label{fig:sub2}
%\end{subfigure}
%\caption{Comparisons of the vertical velocity profiles for a Stokes wave of steepness \(\varepsilon = 0.35\). Only the velocity on the wave crest is presented for simplicity. '\(\cdots\)': Fenton's numerical solution (\citeyear{fenton1988numerical}); '\textcolor{red}{\(--\)}': HOS; '\textcolor{blue}{---}': ESBI. }
%\label{fig:hos-v}
%\end{figure}

\section {\textcolor{black}{Discussions on the limitations of the present model and a quantitative criterion}} \label{sec:limit}
\textcolor{black}{
It is shown that the present model can be very computationally efficient and promising for the applications of modelling nonlinear waves propagation over large spatiotemporal scales. In this section, the limitations of the present model will be discussed. Like the other models of same type, the present model cannot simulate breaking waves. To overcome this restriction, breaker models should be incorporated \citep[and so on]{svendsen1978wave,schaffer1993boussinesq,kennedy2000boussinesq,guignard2001modeling,grilli2020fully,mohanlal2023unified}. Besides, due to the usage of Taylor expansions, the accuracy of the present model is also subject to the magnitude of the bathymetry variation, which will be emphasized in this section. 
It should be pointed out that the validity of the Taylor expansions adopted in Eqs. (\ref{eq:bi-s_end}, \ref{eq:bi-b_start} $\sim$ \ref{eq:bi-b_end}) relies on the magnitude of $\zeta=k_p|\delta|$. Therefore, the approximations become less accurate when $\zeta$ increases, leading to imprecise predictions of $V$. Hence, it is desirable to define a quantitative criterion for using the present model. 
}

\textcolor{black}{
Considering shoreward-propagating waves, $h$ is the water depth (reference level) far away from the shore, and $\delta$ is the variation above the reference level. In deep water, the hyperbolic functions reduce to $\tanh{(kh)}=1$ and $\text{sech } {(kh)} = 0$, thus the present model becomes independent of $h$ and $\delta$, which is equivalent with the deep water model by \cite{wang2015}. On the other hand, over a long shallow shelf, $h$ is small and so is $\delta$, the Taylor expansions are valid and the present model is precise. Otherwise, in finite depth and $h$ is not a small quantity, the convergence of the Taylor expansions and the accuracy of the present model will rely on the magnitude of $\delta$. 
}

\begin{figure}
\centering
\centerline{\includegraphics[width=90mm]{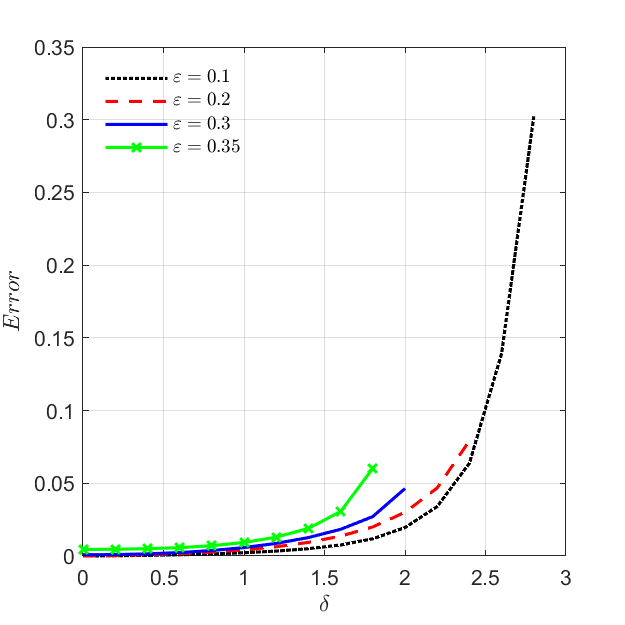}}
\caption{\textcolor{black}{Error of vertical velocity of using present model for Stokes wave of various steepness. }}
\label{fig:rmse}
\end{figure}

\textcolor{black}{
To quantify the limit of $\delta$, the present model is used to approximate $V$ of a Stokes wave in finite water depth, and its error is calculated through comparing with Fenton's numerical solution \citeyearpar{fenton1988numerical}. To do that, one wavelength is resolved into 64 points in the computational domain, meaning that the first 16 Fourier modes are free of alias errors. The water depth is fixed to $h=\pi$, with $\delta$ being assigned different constant values in each computation, so that velocities $V$ corresponding to various $d=h-\delta$ can be calculated by using the present model. Then the numerical results obtained from the present model are compared with Fenton's solutions, and the root mean square errors are computed, which are displayed in figure \ref{fig:rmse}. It shows that the error of the present model gradually grows with increasing $\delta$, and the growth is faster for larger wave steepness (up to the limit of the Fenton's solver at a specific depth $d$). It is also found that the error of the present model are well below 10\% if $\delta$ is less than 2.4, regardless of the wave steepness. 
}

\begin{figure}
\centering
\begin{subfigure}{.33\textwidth}
\centerline{\includegraphics[width=45mm]{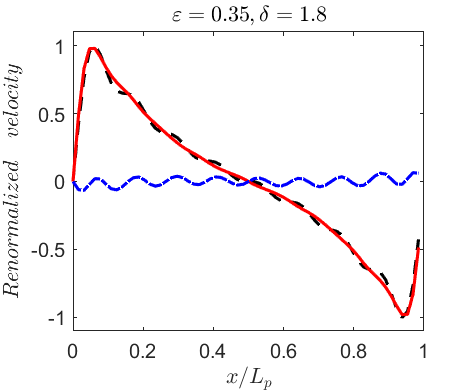}}
  \caption{}
  \label{fig:sub1}
\end{subfigure}%
\begin{subfigure}{.33\textwidth}
\centerline{\includegraphics[width=45mm]{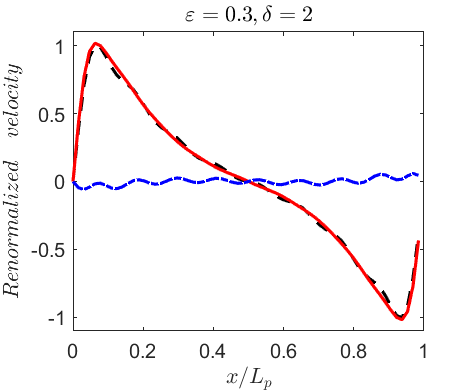}}
  \caption{}
  \label{fig:sub2}
\end{subfigure}
\begin{subfigure}{.33\textwidth}
\centerline{\includegraphics[width=45mm]{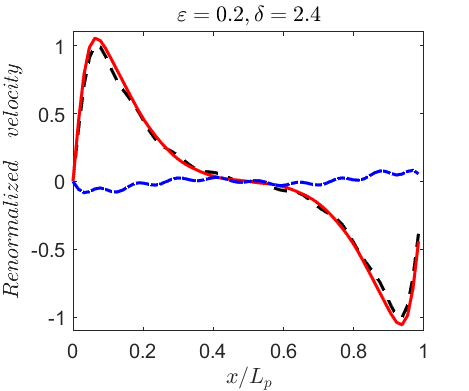}}
  \caption{}
  \label{fig:sub3}
\end{subfigure}%

\begin{subfigure}{.33\textwidth}
\centerline{\includegraphics[width=45mm]{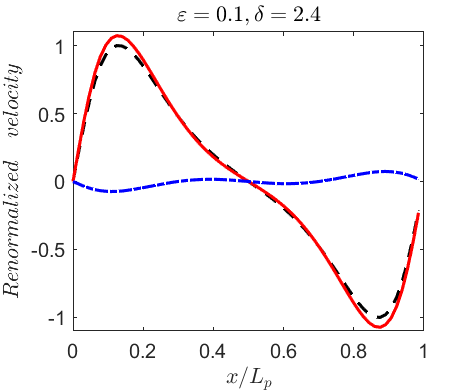}}
  \caption{}
  \label{fig:sub4}
\end{subfigure}%
\begin{subfigure}{.33\textwidth}
\centerline{\includegraphics[width=45mm]{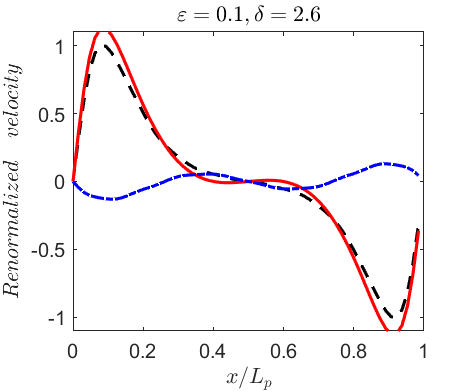}}
  \caption{}
  \label{fig:sub5}
\end{subfigure}

\caption{Comparisons of profiles of $V$ (renormalized) between Fenton's solutions and ESBI predictions. '$---$': Fenton's numerical solution \citeyearpar{fenton1988numerical}; '\textcolor{red}{---}': ESBI results; '\textcolor{blue}{- -- -}': Their differences. } 
\label{fig:limit-v}
\end{figure}

\begin{figure}
\centering
\begin{subfigure}{.33\textwidth}
\centerline{\includegraphics[width=45mm]{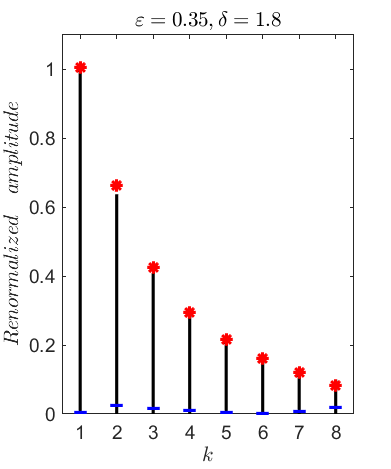}}
  \caption{}
  \label{fig:sub1}
\end{subfigure}%
\begin{subfigure}{.33\textwidth}
\centerline{\includegraphics[width=45mm]{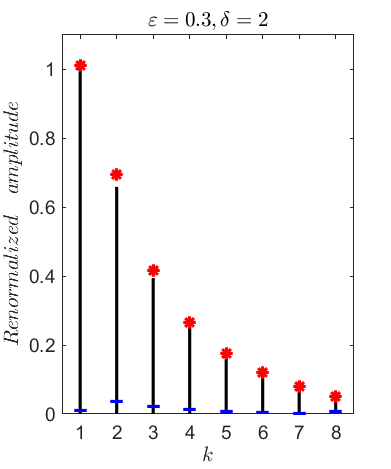}}
  \caption{}
  \label{fig:sub2}
\end{subfigure}
\begin{subfigure}{.33\textwidth}
\centerline{\includegraphics[width=45mm]{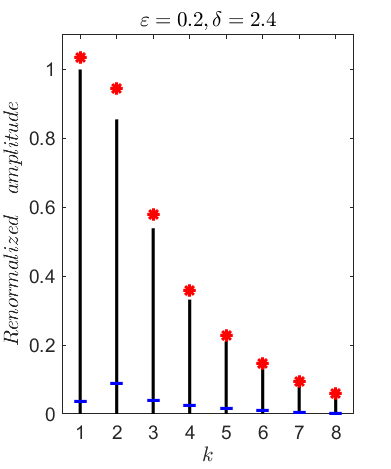}}
  \caption{}
  \label{fig:sub3}
\end{subfigure}%

\begin{subfigure}{.33\textwidth}
\centerline{\includegraphics[width=45mm]{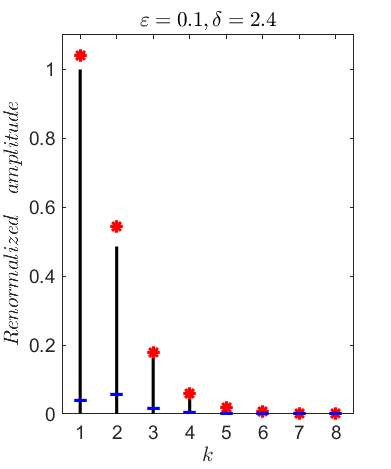}}
  \caption{}
  \label{fig:sub4}
\end{subfigure}%
\begin{subfigure}{.33\textwidth}
\centerline{\includegraphics[width=45mm]{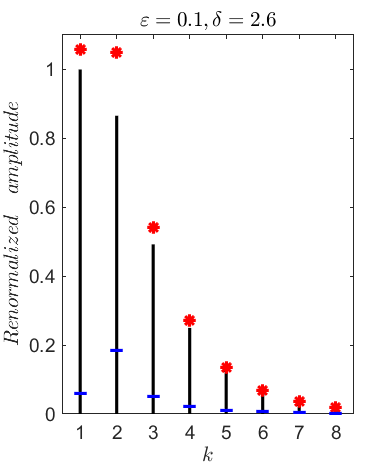}}
  \caption{}
  \label{fig:sub5}
\end{subfigure}

\caption{Comparisons of amplitude spectra (renormalized) of $V$ between Fenton's solution and ESBI prediction. '$|$': Fenton's numerical solution \citeyearpar{fenton1988numerical}; '\textcolor{red}{*}': ESBI results; '\textcolor{blue}{--}': Their differences.  }
\label{fig:limit-vk}
\end{figure}

\textcolor{black}{
To visualize the error for the limiting cases, the profiles of the velocities are extracted corresponding to $\varepsilon=0.35$ $\&$ $\delta=1.8$, $\varepsilon=0.3$ $\&$ $\delta=2$, $\varepsilon=0.2$ $\&$ $\delta=2.4$ and $\varepsilon=0.1$ $\&$ $\delta=2.4$ and plotted in figure \ref{fig:limit-v}, and their amplitude spectra are presented in figure \ref{fig:limit-vk}. In addition, the case leading to error larger than 10\%, i.e., $\varepsilon=0.1$ $\&$ $\delta=2.6$ are also included in the figures. Figure \ref{fig:limit-v} (a-d) indicate that the present model reasonably predicted the velocity $V$ with very small differences concentrated near the peaks of $V$. The inspection of the amplitude spectra in figure \ref{fig:limit-vk} (a-d) implies that the present model can well reproduced the amplitude of the first harmonic but slightly overpredicted the amplitude of the second harmonic. This overprediction is the cause that downgrades the overall precision of the present model. In the case with very large $\delta$, it can be seen that the velocity profile obtained from the present model deviates from Fenton's solution quite significantly, as shown in figure \ref{fig:limit-v}(e). The amplitude spectra in figure \ref{fig:limit-vk}(e) reveals that the present model overpredicted the amplitudes of the first three harmonics.   
}

\textcolor{black}{
Therefore, it is suggested that the present model should be used with extremely caution in finite water depth when $\zeta=k_p|\delta|>2.4$. Note that this quantitative criterion is derived based on the case of Stoke waves, whereas the waves in reality are composed of a variety of components of different frequencies (or wavenumber). In other words, the spectral components of higher frequencies (or wavenumber) may subject to relatively large errors comparing with the peak component. Therefore, for random waves with broad-banded spectrum, the suggested quantitative criterion should be adopted with $k_p$ being replaced by the highest wavenumber to check if the considered bandwidth is valid. 
%The present model should be comprehensively benchmarked if being employed with $\zeta$ close to the suggested limit.    
}

\section {Concluding remarks} \label{sec:con}
This study presents a new pseudo-spectrum numerical model based on the \textcolor{black}{highly} nonlinear potential flow theory for simulating wave propagation on variable water depth. A set of new formulas for estimating the vertical velocity on free surface based on the boundary integral equation is obtained. A successive approximation scheme is also proposed and the calculation of the vertical velocity can be fast \textcolor{black}{owing} to the usage of Fast Fourier Transform (FFT). 

The new model is verified and validated for simulating a variety of nonlinear wave cases on variable depth as well as in deep water, including free propagating solitary wave, wave transformations over submerged bar, Bragg reflection over undulating bars, nonlinear evolution of Peregrine breather, \textcolor{black}{obliquely propagating uniform waves} and extreme waves in crossing random seas. Good agreement is achieved between the numerical simulation results and laboratory measurements. A comparison of the diagnostic procedure is made between the present model and the Higher-Order Spectrum (HOS) method. It is found that the present model outperforms the HOS in the speed of convergence for estimating the vertical velocity of a Stoke wave. It is also noted that the present model requires significantly less number of FFTs than the HOS method considering the same order in perturbation and convolution. For the same level of accuracy with acceptable tolerance as 0.5\%, the present model can be about 5 times faster (or 5 times less FFT operations) than the HOS method for estimating the vertical velocity of a Stoke wave at the limiting wave steepness.    

It is highlighted that the present model provides a computationally efficient approach for simulating \textcolor{black}{highly} nonlinear water waves in arbitrary depth. This enables the study of nonlinear ocean waves on a large spatiotemporal scale in phase-resolving simulations. Nevertheless, the limitation of the present model is the incapability of modeling wave propagation over a sharp step. \textcolor{black}{Though it is suggested that the present model should be comfortably used when $\zeta=k_p|\delta|<2.4$ in finite water depth, one should opt for other FNPF models without such limitations, e.g., BE or FE models, in the case with presence of steep bottom steps beyond the range of applicability of present model.} Another limitation is the modeling of breaking waves due to the intrinsic restrictions of potential theory assuming inviscid and irrotational flow. One widely adopted remedy is to use artificial filtering technique as outlined in many potential flow models such as in \cite{xiao2013rogue}, \textcolor{black}{or physically meaningful approaches  \citep{seiffert2018simulation,papoutsellis2019modelling,simon2019comparing,grilli2020fully,mohanlal2023unified}.} Alternatively, hybrid modeling approach can be employed by coupling with the CFD models based on the Navier-Stokes equation or Lattice-Boltzmann method to simulate localized breaking phenomenon. 

%\backsection[Supplementary data]{\label{SupMat}Supplementary material and movies are available at \\https://doi.org/10.1017/jfm.2019...}

%\backsection[Acknowledgements]{The author appreciate Dr. Zhengtong Yang at The Technology Centre for Offshore and Marine, Singapore, and Prof. Junliang Gao at Jiangsu University of Science and Technology, China for providing experiment data for model validations.}

%\backsection[Funding]{The author shows gratitude to the sponsorship provided by University Grants Committee, Hong Kong (P0039692), Research Institute for Sustainable Urban Development at PolyU (P0042840, P0043578), Department of Science and Technology of Guangdong Province, China (22202206050000278). 
%Please provide details of the sources of financial support for all authors, including grant numbers. Where no specific funding has been provided for research, please provide the following statement: "This research received no specific grant from any funding agency, commercial or not-for-profit sectors." 
%}

%\backsection[Declaration of interests]{The authors report no conflict of interest.}

%\backsection[Data availability statement]{The data that support the findings of this study are openly available in [repository name] at http://doi.org/[doi], reference number [reference number]. See JFM's \href{https://www.cambridge.org/core/journals/journal-of-fluid-mechanics/information/journal-policies/research-transparency}{research transparency policy} for more information}

%\backsection[Author ORCIDs]{J. Wang, https://orcid.org/0000-0003-1821-2362}

%\backsection[Author contributions]{Authors may include details of the contributions made by each author to the manuscript'}

\appendix

\section{}\label{appA}
The formulations for \(V^{(j)}\) at each order are obtained as:
%\begin{equation}
%V = S_{\phi} + S_{V} + S_{\varphi} + I_{S}
%\end{equation}
\begin{eqnarray}
V^{(1)} = k\tanh{(kh)}\mathcal{F}\left\{ \widetilde{\phi} \right\},
\end{eqnarray}

\begin{eqnarray}
V^{(2)} &=& - i\bm{k} \cdot \mathcal{F}\left\{ \eta\nabla\widetilde{\phi} \right\} + C_{h}k \Big(- \mathcal{F}\left\{ \eta \mathcal{F}^{- 1}\left\{ D_{h}^{- 1} k\mathcal{F}\left\{ \widetilde{\phi} \right\} \right\} \right\} \nonumber\\
&&+  e_{h}\mathcal{F}\left\{ \eta V \right\} + \mathcal{F}\left\{ \eta \mathcal{F}^{- 1}\left\{ e_{h}\mathcal{F}\left\{ V \right\} \right\} \right\} +2\sqrt{e_{h}} k^{-1} i\bm{k} \cdot \mathcal{F}\left\{ \delta\nabla\widetilde{\varphi} \right\} \Big),
\end{eqnarray}

\begin{eqnarray}
V^{(3)} &=& C_{h}k \bigg( \frac{1}{2}e_{h}i\bm{k} \cdot \mathcal{F}\left\{ \eta^{2}\nabla\widetilde{\phi} \right\} + \mathcal{F}\left\{ \eta \mathcal{F}^{- 1}\left\{ e_{h}i\bm{k} \cdot \mathcal{F}\left\{ \eta\nabla\widetilde{\phi} \right\} \right\} \right\} \nonumber\\
&&-\frac{1}{2}\mathcal{F}\left\{ \eta^{2}\mathcal{F}^{- 1}\left\{ e_{h}k^{2}\mathcal{F}\left\{ \widetilde{\phi} \right\} \right\} \right\} - \frac{1}{2}C_{h}^{- 1} k\mathcal{F} \left\{ \eta^{2}V \right\} \nonumber\\
&&+\mathcal{F}\left\{ \eta \mathcal{F}^{- 1}\left\{ D_{h}^{- 1} k\mathcal{F} \left\{ \eta V \right\} \right\} \right\} - \frac{1}{2}\mathcal{F}\left\{ \eta^{2}\mathcal{F}^{- 1}\left\{ C_{h}^{- 1} k\mathcal{F} \left\{ V \right\} \right\} \right\} \nonumber\\
&& - 2\mathcal{F}\left\{ \eta \mathcal{F}^{- 1}\left\{ \sqrt{e_h}i\bm{k} \cdot \mathcal{F}\left\{ \delta\nabla\widetilde{\varphi} \right\} \right\} \right\}
\bigg),
\end{eqnarray}

\begin{eqnarray}
V^{(4)} &=& C_{h}k \bigg( - \frac{1}{6}C_{h}^{- 1}ki\bm{k} \cdot \mathcal{F}\left\{ \eta^{3}\nabla\widetilde{\phi} \right\} +\frac{1}{2}\mathcal{F}\left\{ \eta \mathcal{F}^{- 1}\left\{ D_{h}^{- 1}ki\bm{k} \cdot \mathcal{F}\left\{ \eta^{2}\nabla\widetilde{\phi} \right\} \right\} \right\} \nonumber\\
&&- \frac{1}{2}\mathcal{F}\left\{ \eta^{2}\mathcal{F}^{- 1}\left\{ C_{h}^{- 1}ki\bm{k} \cdot \mathcal{F}\left\{ \eta\nabla\widetilde{\phi} \right\} \right\} \right\} -\frac{1}{6}\mathcal{F}\left\{ \eta^{3}\mathcal{F}^{- 1}\left\{ D_{h}^{- 1}k^{3}\mathcal{F}\left\{ \widetilde{\phi} \right\} \right\} \right\} \nonumber\\
&&+\frac{1}{6}e_{h}k^{2}\mathcal{F}\left\{ \eta^{3}V \right\} + \frac{1}{2}\mathcal{F}\left\{ \eta^{2}\mathcal{F}^{- 1}\left\{ e_{h}k^{2}\mathcal{F}\left\{ \eta V \right\} \right\} \right\} \nonumber\\
&&+\frac{1}{2}\mathcal{F}\left\{ \eta \mathcal{F}^{- 1}\left\{ e_{h}k^{2}\mathcal{F}\left\{ \eta^{2}V \right\} \right\} \right\} + \frac{1}{6}\mathcal{F}\left\{ \eta^{3}\mathcal{F}^{- 1}\left\{ e_{h}k^{2}\mathcal{F}\left\{ V \right\} \right\} \right\} \nonumber\\
&& +\mathcal{F}\left\{ \eta^{2}\mathcal{F}^{- 1}\left\{ \sqrt{e_h}ki\bm{k} \cdot \mathcal{F}\left\{ \delta\nabla\widetilde{\varphi} \right\} \right\} \right\}  + \frac{1}{3}\sqrt{e_{h}}ki\bm{k} \cdot \mathcal{F}\left\{ \delta^{3}\nabla\widetilde{\varphi} \right\} 
\bigg),
\end{eqnarray}

\begin{eqnarray}
V^{(5)} &=& C_{h}k \bigg( \frac{1}{24}e_{h}k^{2}i\bm{k} \cdot \mathcal{F}\left\{ \eta^{4}\nabla\widetilde{\phi} \right\} +\frac{1}{6}\mathcal{F}\left\{ \eta \mathcal{F}^{- 1}\left\{ e_{h}k^{2}i\bm{k} \cdot \mathcal{F}\left\{ \eta^{3}\nabla\widetilde{\phi} \right\} \right\} \right\} \nonumber\\
&&+ \frac{1}{4}\mathcal{F}\left\{ \eta^{2}\mathcal{F}^{- 1}\left\{ e_{h}k^{2}i\bm{k} \cdot \mathcal{F}\left\{ \eta^{2}\nabla\widetilde{\phi} \right\} \right\} \right\} +\frac{1}{6}\mathcal{F}\left\{ \eta^{3}\mathcal{F}^{- 1}\left\{ e_{h}k^{2}i\bm{k} \cdot \mathcal{F}\left\{ \eta\nabla\widetilde{\phi} \right\} \right\} \right\} \nonumber\\
&&- \frac{1}{24}\mathcal{F}\left\{ \eta^{4}\mathcal{F}^{- 1}\left\{ e_{h}k^{4}\mathcal{F}\left\{ \widetilde{\phi} \right\} \right\} \right\} -\frac{1}{24}C_{h}^{- 1}k^{3}\mathcal{F}\left\{ \eta^{4}V \right\} \nonumber\\
&& + \frac{1}{6}\mathcal{F}\left\{ \eta \mathcal{F}^{- 1}\left\{ D_{h}^{- 1}k^{3}\mathcal{F}\left\{ \eta^{3}V \right\} \right\} \right\} -\frac{1}{4}\mathcal{F}\left\{ \eta^{2}\mathcal{F}^{- 1}\left\{ C_{h}^{- 1}k^{3}\mathcal{F}\left\{ \eta^{2}V \right\} \right\} \right\} \nonumber\\
&&+ \frac{1}{6}\mathcal{F}\left\{ \eta^{3}\mathcal{F}^{- 1}\left\{ D_{h}^{- 1}k^{3}\mathcal{F}\left\{ \eta V \right\} \right\} \right\} -\frac{1}{24}\mathcal{F}\left\{ \eta^{4}\mathcal{F}^{- 1}\left\{ C_{h}^{- 1}k^{3}\mathcal{F}\left\{ V \right\} \right\} \right\} \nonumber\\
&&-  \frac{1}{3}\mathcal{F}\left\{\eta \mathcal{F}^{- 1}\left\{ \sqrt{e_h}k^{2}i\bm{k} \cdot \mathcal{F}\left\{ \delta^{3}\nabla\widetilde{\varphi} \right\} \right\} \right\} 
-\frac{1}{3}\mathcal{F}\left\{ \eta^{3}\mathcal{F}^{- 1}\left\{ \sqrt{e_h}k^{2}i\bm{k} \cdot \mathcal{F}\left\{ \delta\nabla\widetilde{\varphi} \right\} \right\} \right\}
\bigg),\nonumber\\
\end{eqnarray}

\begin{eqnarray}
V^{(6)} &=& C_{h}k \bigg( -\frac{1}{120}C_{h}^{- 1}k^{3}i\bm{k} \cdot \mathcal{F}\left\{ \eta^{5}\nabla\widetilde{\phi} \right\} + \frac{1}{24}\mathcal{F}\left\{ \eta \mathcal{F}^{- 1}\left\{ D_{h}^{- 1}k^{3}i\bm{k} \cdot \mathcal{F}\left\{ \eta^{4}\nabla\widetilde{\phi} \right\} \right\} \right\} \nonumber\\
&&-\frac{1}{12}\mathcal{F}\left\{ \eta^{2}\mathcal{F}^{- 1}\left\{ C_{h}^{- 1}k^{3}i\bm{k} \cdot \mathcal{F}\left\{ \eta^{3}\nabla\widetilde{\phi} \right\} \right\} \right\} + \frac{1}{12}\mathcal{F}\left\{ \eta^{3}\mathcal{F}^{- 1}\left\{ D_{h}^{- 1}k^{3}i\bm{k} \cdot \mathcal{F}\left\{ \eta^{2}\nabla\widetilde{\phi} \right\} \right\} \right\} \nonumber\\
&&-\frac{1}{24}\mathcal{F}\left\{ \eta^{4}\mathcal{F}^{- 1}\left\{ C_{h}^{- 1}k^{3}i\bm{k} \cdot \mathcal{F}\left\{ \eta\nabla\widetilde{\phi} \right\} \right\} \right\} + \frac{1}{120}\mathcal{F}\left\{ \eta^{5}\mathcal{F}^{- 1}\left\{ D_{h}^{- 1}k^{3}i\bm{k} \cdot \mathcal{F}\left\{ \nabla\widetilde{\phi} \right\} \right\} \right\}\nonumber\\
&& + \frac{1}{120}e_{h}k^{4}\mathcal{F}\left\{ \eta^{5}V \right\} +\frac{1}{24}\mathcal{F}\left\{ \eta \mathcal{F}^{- 1}\left\{ e_{h}k^{4}\mathcal{F}\left\{ \eta^{4}V \right\} \right\} \right\} + \frac{1}{12}\mathcal{F}\left\{ \eta^{2}\mathcal{F}^{- 1}\left\{ e_{h}k^{4}\mathcal{F}\left\{ \eta^{3}V \right\} \right\} \right\} \nonumber\\
&& +\frac{1}{12}\mathcal{F}\left\{ \eta^{3}\mathcal{F}^{- 1}\left\{ e_{h}k^{4}\mathcal{F}\left\{ \eta^{2}V \right\} \right\} \right\} + \frac{1}{24}\mathcal{F}\left\{ \eta^{4}\mathcal{F}^{- 1}\left\{ e_{h}k^{4}\mathcal{F}\left\{ \eta V \right\} \right\} \right\} \nonumber\\
&&  +\frac{1}{120}\mathcal{F}\left\{ \eta^{5}\mathcal{F}^{- 1}\left\{ e_{h}k^{4}\mathcal{F}\left\{ V \right\} \right\} \right\} +\frac{1}{60}\sqrt{e_{h}}k^{3}i\bm{k} \cdot \mathcal{F}\left\{ \delta^{5}\nabla\widetilde{\varphi} \right\} \nonumber\\
&& + \frac{1}{6}\mathcal{F}\left\{ \eta^{2}\mathcal{F}^{- 1}\left\{ \sqrt{e_{h}}k^{3}i\bm{k} \cdot \mathcal{F}\left\{ \delta^{3}\nabla\widetilde{\varphi} \right\} \right\} \right\} +\frac{1}{12}\mathcal{F}\left\{ \eta^{4}\mathcal{F}^{- 1}\left\{ \sqrt{e_{h}}k^{3}i\bm{k} \cdot \mathcal{F}\left\{ \delta\nabla\widetilde{\varphi} \right\} \right\} \right\}
\bigg), \nonumber\\
\end{eqnarray}
\noindent and
\begin{eqnarray}
V^{(7)} &=& C_{h}k \bigg( \frac{1}{720}e_{h}k^{4}i\bm{k} \cdot \mathcal{F}\left\{ \eta^{6}\nabla\widetilde{\phi} \right\} + \frac{1}{120}\mathcal{F}\left\{ \eta \mathcal{F}^{- 1}\left\{ e_{h}k^{4}i\bm{k} \cdot \mathcal{F}\left\{ \eta^{5}\nabla\widetilde{\phi} \right\} \right\} \right\} \nonumber\\
&& +\frac{1}{48}\mathcal{F}\left\{ \eta^{2}\mathcal{F}^{- 1}\left\{ e_{h}k^{4}i\bm{k} \cdot \mathcal{F}\left\{ \eta^{4}\nabla\widetilde{\phi} \right\} \right\} \right\} + \frac{1}{36}\mathcal{F}\left\{ \eta^{3}\mathcal{F}^{- 1}\left\{ e_{h}k^{4}i\bm{k} \cdot \mathcal{F}\left\{ \eta^{3}\nabla\widetilde{\phi} \right\} \right\} \right\} \nonumber\\
&& +\frac{1}{48}\mathcal{F}\left\{ \eta^{4}\mathcal{F}^{- 1}\left\{ e_{h}k^{4}i\bm{k} \cdot \mathcal{F}\left\{ \eta^{2}\nabla\widetilde{\phi} \right\} \right\} \right\} + \frac{1}{120}\mathcal{F}\left\{ \eta^{5}\mathcal{F}^{- 1}\left\{ e_{h}k^{4}i\bm{k} \cdot \mathcal{F}\left\{ \eta\nabla\widetilde{\phi} \right\} \right\} \right\} \nonumber\\
&& +\frac{1}{720}\mathcal{F}\left\{ \eta^{6}\mathcal{F}^{- 1}\left\{ e_{h}k^{4}i\bm{k} \cdot \mathcal{F}\left\{ \nabla\widetilde{\phi} \right\} \right\} \right\} \nonumber - \frac{1}{720}C_{h}^{- 1}k^{5}\mathcal{F}\left\{ \eta^{6}V \right\} \nonumber\\
&& +\frac{1}{120}\mathcal{F}\left\{ \eta \mathcal{F}^{- 1}\left\{ D_{h}^{- 1}k^{5}\mathcal{F}\left\{ \eta^{5}V \right\} \right\} \right\} - \frac{1}{48}\mathcal{F}\left\{ \eta^{2}\mathcal{F}^{- 1}\left\{ C_{h}^{- 1}k^{5}\mathcal{F}\left\{ \eta^{4}V \right\} \right\} \right\} \nonumber\\
&& +\frac{1}{36}\mathcal{F}\left\{ \eta^{3}\mathcal{F}^{- 1}\left\{ D_{h}^{- 1}k^{5}\mathcal{F}\left\{ \eta^{3}V \right\} \right\} \right\} - \frac{1}{48}\mathcal{F}\left\{ \eta^{4}\mathcal{F}^{- 1}\left\{ C_{h}^{- 1}k^{5}\mathcal{F}\left\{ \eta^{2}V \right\} \right\} \right\} \nonumber\\
&& +\frac{1}{120}\mathcal{F}\left\{ \eta^{5}\mathcal{F}^{- 1}\left\{ D_{h}^{- 1}k^{5}\mathcal{F}\left\{ \eta V \right\} \right\} \right\} - \frac{1}{720}\mathcal{F}\left\{ \eta^{6}\mathcal{F}^{- 1}\left\{ C_{h}^{- 1}k^{5}\mathcal{F}\left\{ V \right\} \right\} \right\} \nonumber\\
&&  - \frac{1}{60}\mathcal{F}\left\{ \eta \mathcal{F}^{- 1}\left\{ \sqrt{e_{h}}k^{4}i\bm{k} \cdot \mathcal{F}\left\{ \delta^{5}\nabla\widetilde{\varphi} \right\} \right\} \right\} -\frac{1}{18}\mathcal{F}\left\{ \eta^{3}\mathcal{F}^{- 1}\left\{  \sqrt{e_{h}}k^{4}i\bm{k} \cdot \mathcal{F}\left\{ \delta^{3}\nabla\widetilde{\varphi} \right\} \right\} \right\} \nonumber\\
&& -\frac{1}{60}\mathcal{F}\left\{ \eta^{5}\mathcal{F}^{- 1}\left\{ \sqrt{e_{h}}k^{4}i\bm{k} \cdot \mathcal{F}\left\{ \delta\nabla\widetilde{\varphi} \right\} \right\} \right\}
\bigg).
\end{eqnarray}

\noindent where  \(e_{h} = e^{- 2kh}\), \(C_{h} = \left( 1 + e_{h} \right)^{- 1}\), \(D_{h} = \left( 1 - e_{h} \right)^{- 1}\). 
The remaining integration part consists of three parts, i.e., \(I^{(S)} = I_{\phi}^{(S)}+I_V^{(S)}+I_{\varphi}^{(S)}\), and each part is given by:

\begin{eqnarray}\label{app:iphis}
I_{\phi}^{(S)} &=& \frac{1}{2\pi}\iint_{S_{0}}^{} {\widetilde{\phi}}' \Bigg\{ -\frac{ \bm{R} \cdot \nabla'\eta' - \left( \eta' + \eta \right) }{\left[R^{2} + \left( \eta' - \eta \right)^{2}\right]^{3/2}} -\frac{ \bm{R} \cdot \nabla'\eta' - \left( \eta' + \eta \right) - 2h }{\left[R^{2} + \left( \eta' + \eta + 2h \right)^{2}\right]^{3/2}}  \nonumber\\
&&+ \frac{\partial}{\partial(2h)}\frac{1}{R_{2h}} - \nabla' \cdot \left[ \left( \eta' - \eta \right)\nabla'\frac{1}{R} \right] - \nabla' \cdot \left[ \left( \eta' + \eta \right)\nabla'\frac{1}{R_{2h}} \right] \nonumber\\
 &&- \nabla' \cdot \left[ \frac{1}{2}\left( \eta' + \eta \right)^{2}\nabla'\frac{\partial}{\partial(2h)}\frac{1}{R_{2h}} \right] + \frac{1}{6}\nabla' \cdot \left[ \left( \eta' - \eta \right)^{3}\nabla'\nabla' \cdot \nabla'\frac{1}{R} \right] \nonumber\\
&&+ \frac{1}{6}\nabla' \cdot \left[ \left( \eta' + \eta \right)^{3}\nabla'\nabla' \cdot \nabla'\frac{1}{R_{2h}} \right] + \frac{1}{24}\nabla' \cdot \left[ \left( \eta' + \eta \right)^{4}\nabla'\nabla' \cdot \nabla'\frac{\partial}{\partial(2h)}\frac{1}{R_{2h}} \right] \nonumber\\
&& -\frac{1}{120}\nabla' \cdot \left[ \left( \eta' - \eta \right)^{5}\nabla'\left( \nabla' \cdot \nabla' \right)^{2}\frac{1}{R} \right] - \frac{1}{120}\nabla' \cdot \left[ \left( \eta' + \eta \right)^{5}\nabla'\left( \nabla' \cdot \nabla' \right)^{2}\frac{1}{R_{2h}} \right] \nonumber\\
&& -\frac{1}{720}\nabla' \cdot \left[ \left( \eta' + \eta \right)^{6}\nabla'\left( \nabla' \cdot \nabla' \right)^{2}\frac{\partial}{\partial(2h)}\frac{1}{R_{2h}} \right] \Bigg\} 
\mathrm{d} \bm{x}',
\end{eqnarray}

\begin{eqnarray}\label{app:ivs}
I_V^{(S)} &=&  -\frac{1}{2\pi}\iint_{S_{0}}V'\Bigg\{\frac{1}{\left[R^{2} + \left( \eta' - \eta \right)^{2}\right]^{1/2}}+ \frac{1}{\left[R^{2} + \left( \eta' + \eta + 2h \right)^{2}\right]^{1/2}} \nonumber\\
&&-\frac{1}{R} - \frac{1}{R_{2h}} - \left( \eta' + \eta \right)\frac{\partial}{\partial(2h)}\frac{1}{R_{2h}} + \frac{1}{2}\left( \eta' - \eta \right)^{2}\nabla' \cdot \nabla'\frac{1}{R} \nonumber\\
&& +\frac{1}{2}\left( \eta' + \eta \right)^{2}\nabla' \cdot \nabla'\frac{1}{R_{2h}} + \frac{1}{6}\left( \eta' + \eta \right)^{3}\nabla' \cdot \nabla'\frac{\partial}{\partial(2h)}\frac{1}{R_{2h}} \nonumber\\
&& -\frac{1}{24}\left( \eta' - \eta \right)^{4}\left( \nabla' \cdot \nabla' \right)^{2}\frac{1}{R} - \frac{1}{24}\left( \eta' + \eta \right)^{4}\left( \nabla' \cdot \nabla' \right)^{2}\frac{1}{R_{2h}} \nonumber\\
&& -\frac{1}{120}\left( \eta' + \eta \right)^{5}\left( \nabla' \cdot \nabla' \right)^{2}\frac{\partial}{\partial(2h)}\frac{1}{R_{2h}} + \frac{1}{720}\left( \eta' - \eta \right)^{6}\left( \nabla' \cdot \nabla' \right)^{3}\frac{1}{R} \nonumber\\
&& +\frac{1}{720}\left( \eta' + \eta \right)^{6}\left( \nabla' \cdot \nabla' \right)^{3}\frac{1}{R_{2h}} \Bigg\}  \mathrm{d} \bm{x}' ,
\end{eqnarray}
\noindent and
\begin{eqnarray}\label{app:ivarphis}
I_{\varphi}^{(S)} &=& -\frac{1}{2\pi}\iint_{B_{0}}^{} {\widetilde{\varphi}}'\Bigg\{ \frac{\bm{R} \cdot \nabla'\delta' + \delta' - \eta - h}{(R^{2} + \left( \delta' - \eta - h \right)^{2})^{3/2}} + \frac{\bm{R} \cdot \nabla'\delta' + \delta' + \eta + h}{(R^{2} + \left( \delta' + \eta + h \right)^{2})^{3/2}} \nonumber\\
&& -2\nabla' \cdot \left( \delta'\nabla'\frac{1}{R_{h}} \right) - 2\eta\nabla' \cdot \left( \delta'\nabla'\frac{\partial}{\partial h}\frac{1}{R_{h}} \right) +\frac{1}{3}\nabla' \cdot \left( \delta'3\nabla'\nabla' \cdot \nabla'\frac{1}{R_{h}} \right) \nonumber\\
&&  + \eta^{2}\nabla' \cdot \left( \delta'\nabla'\nabla' \cdot \nabla'\frac{1}{R_{h}} \right) +\frac{1}{3}\eta\nabla' \cdot \left( \delta'^{3}\nabla'\nabla' \cdot \nabla'\frac{\partial}{\partial h}\frac{1}{R_{h}} \right) \nonumber\\
&&  + \frac{1}{3}\eta^{3}\nabla' \cdot \left( \delta'\nabla'\nabla' \cdot \nabla'\frac{\partial}{\partial h}\frac{1}{R_{h}} \right) -\frac{1}{60}\nabla' \cdot \left[ \delta'^{5}\nabla'\left( \nabla' \cdot \nabla' \right)^{2}\frac{1}{R_{h}} \right] \nonumber\\
&&  - \frac{1}{6}\eta^{2}\nabla' \cdot \left[ \delta'^{3}\nabla'\left( \nabla' \cdot \nabla' \right)^{2}\frac{1}{R_{h}} \right] -\frac{1}{12}\eta^{4}\nabla' \cdot \left[ \delta'\nabla'\left( \nabla' \cdot \nabla' \right)^{2}\frac{1}{R_{h}} \right] \nonumber\\
&&  - \frac{1}{60}\eta\nabla' \cdot \left[ \delta'^{5}\nabla'\left( \nabla' \cdot \nabla' \right)^{2}\frac{\partial}{\partial h}\frac{1}{R_{h}} \right]-\frac{1}{18}\eta^{3}\nabla' \cdot \left[ \delta'^{3}\nabla'\left( \nabla' \cdot \nabla' \right)^{2}\frac{\partial}{\partial h}\frac{1}{R_{h}} \right]  \nonumber\\
&& - \frac{1}{60}\eta^{5}\nabla' \cdot \left[ \delta'\nabla'\left( \nabla' \cdot \nabla' \right)^{2}\frac{\partial}{\partial h}\frac{1}{R_{h}} \right] \Bigg\}  \mathrm{d} \bm{x}'.
\end{eqnarray}

\section{}\label{appB}
 The formulations for \(\widetilde{\varphi}^{(j)}\) at each order are obtained as:
\begin{eqnarray}
\widetilde{\varphi}^{(1)} =  2 D_{h} \sqrt{e_{h}}k^{-1} \mathcal{F}\left\{ V \right\},
\end{eqnarray}

\begin{eqnarray}
\widetilde{\varphi}^{(2)} &=& D_{h} \Big( 2\mathcal{F}\left\{ \delta \mathcal{F}^{- 1}\left\{ \sqrt{e_{h}}\mathcal{F}\left\{ V \right\} \right\} \right\} +2\sqrt{e_{h}} k^{-1}  i \bm{k} \cdot \mathcal{F}\left\{ \eta\nabla\widetilde{\phi} \right\} \nonumber\\
&& -C_{h}^{- 1}k^{-1} i\bm{k} \cdot \mathcal{F}\left\{ \delta\nabla\widetilde{\varphi} \right\} - \mathcal{F}\left\{ \delta \mathcal{F}^{- 1}\left\{ D_{h}^{- 1}k\mathcal{F}\left\{ \widetilde{\varphi} \right\} \right\} \right\}
\Big),
\end{eqnarray}

\begin{eqnarray}
\widetilde{\varphi}^{(3)} &=& D_{h} \bigg( \sqrt{e_{h}}k\mathcal{F}\left\{ \eta^{2}V \right\} +\mathcal{F}\left\{ \delta^{2}\mathcal{F}^{- 1}\left\{ \sqrt{e_{h}}k\mathcal{F}\left\{ V \right\} \right\} \right\} \nonumber\\
&& + 2\mathcal{F}\left\{ \delta \mathcal{F}^{- 1}\left\{ \sqrt{e_{h}}i\bm{k} \cdot \mathcal{F}\left\{ \eta\nabla\widetilde{\phi} \right\} \right\} \right\} + \frac{1}{2}D_{h}^{- 1}i\bm{k} \cdot \mathcal{F}\left\{ \delta^{2}\nabla\widetilde{\varphi} \right\} \nonumber\\
&& -\mathcal{F}\left\{ \delta \mathcal{F}^{- 1}\left\{ C_{h}^{- 1}i\bm{k} \cdot \mathcal{F}\left\{ \delta\nabla\widetilde{\varphi} \right\} \right\} \right\} - \frac{1}{2}\mathcal{F}\left\{ \delta^{2}\mathcal{F}^{- 1}\left\{ D_{h}^{- 1}k^{2}\mathcal{F}\left\{ \widetilde{\varphi} \right\} \right\} \right\} 
\bigg),
\end{eqnarray}

\begin{eqnarray}
\widetilde{\varphi}^{(4)} &=& D_{h} \bigg(\frac{1}{3}\mathcal{F}\left\{ \delta^{3}\mathcal{F}^{- 1}\left\{ \sqrt{e_{h}} k^{2}\mathcal{F}\left\{ V \right\} \right\} \right\} +\mathcal{F}\left\{ \delta \mathcal{F}^{- 1}\left\{ \sqrt{e_{h}} k^{2}\mathcal{F}\left\{ \eta^{2}V \right\} \right\} \right\} \nonumber\\
&& +\frac{1}{3}\sqrt{e_{h}} ki\bm{k} \cdot \mathcal{F}\left\{ \eta^{3}\nabla\widetilde{\phi} \right\} + \mathcal{F}\left\{ \delta^{2}\mathcal{F}^{- 1}\left\{ \sqrt{e_{h}} ki \bm{k} \cdot \mathcal{F}\left\{ \eta\nabla\widetilde{\phi} \right\} \right\} \right\} \nonumber\\
&& -\frac{1}{6}C_{h}^{- 1}ki\bm{k} \cdot \mathcal{F}\left\{ \delta^{3}\nabla\widetilde{\varphi} \right\} - \frac{1}{6}\mathcal{F}\left\{ \delta^{3}\mathcal{F}^{- 1}\left\{ D_{h}^{- 1}k^{3}\mathcal{F}\left\{ \widetilde{\varphi} \right\} \right\} \right\} \nonumber\\
&& +\frac{1}{2}\mathcal{F}\left\{ \delta \mathcal{F}^{- 1}\left\{ D_{h}^{- 1}ki\bm{k} \cdot \mathcal{F}\left\{ \delta^{2}\nabla\widetilde{\varphi} \right\} \right\} \right\} - \frac{1}{2}\mathcal{F}\left\{ \delta^{2}\mathcal{F}^{- 1}\left\{ C_{h}^{- 1}ki\bm{k} \cdot \mathcal{F}\left\{ \delta\nabla\widetilde{\varphi} \right\} \right\} \right\} \bigg), \nonumber\\
\end{eqnarray}

\begin{eqnarray}
\widetilde{\varphi}^{(5)} &=& D_{h} \bigg( \frac{1}{12}\sqrt{e_{h}}k^{3}\mathcal{F}\left\{ \eta^{4}V \right\} +\frac{1}{2}\mathcal{F}\left\{ \delta^{2}\mathcal{F}^{- 1}\left\{ \sqrt{e_{h}}k^{3}\mathcal{F}\left\{ \eta^{2}V \right\} \right\} \right\} \nonumber\\
&&+ \frac{1}{12}\mathcal{F}\left\{ \delta^{4}\mathcal{F}^{- 1}\left\{ \sqrt{e_{h}}k^{3}\mathcal{F}\left\{ V \right\} \right\} \right\} +\frac{1}{3}\mathcal{F}\left\{ \delta \mathcal{F}^{- 1}\left\{ \sqrt{e_{h}}k^{2}i\bm{k} \cdot \mathcal{F}\left\{ \eta^{3}\nabla\widetilde{\phi} \right\} \right\} \right\} \nonumber\\
&&  + \frac{1}{3}\mathcal{F}\left\{ \delta^{3}\mathcal{F}^{- 1}\left\{ \sqrt{e_{h}}k^{2}i\bm{k} \cdot \mathcal{F}\left\{ \eta\nabla\widetilde{\phi} \right\} \right\} \right\} -\frac{1}{24}e_{h}k^{2} i\bm{k} \cdot \mathcal{F}\left\{  \delta^{4}\nabla\widetilde{\varphi}  \right\} \nonumber\\
&& - \frac{1}{6}\mathcal{F}\left\{ \delta \mathcal{F}^{- 1}\left\{ e_{h}k^{2} i\bm{k} \cdot \mathcal{F}\left\{   \delta^{3}\nabla\widetilde{\varphi} \right\} \right\} \right\} -\frac{1}{4}\mathcal{F}\left\{ \delta^{2}\mathcal{F}^{- 1}\left\{ e_{h}k^{2} i\bm{k} \cdot \mathcal{F}\left\{  \delta^{2}\nabla\widetilde{\varphi} \right\} \right\} \right\} \nonumber\\
&&  - \frac{1}{6}\mathcal{F}\left\{ \delta^{3}\mathcal{F}^{- 1}\left\{ e_{h}k^{2} i\bm{k} \cdot \mathcal{F}\left\{  \delta\nabla\widetilde{\varphi}  \right\} \right\} \right\} +\frac{1}{24}\mathcal{F}\left\{ \delta^{4}\mathcal{F}^{- 1}\left\{ e_{h}k^{4}\mathcal{F}\left\{\widetilde{\varphi} \right\} \right\} \right\} 
\bigg), \nonumber\\
\end{eqnarray}
\noindent and
\begin{eqnarray}
\widetilde{\varphi}^{(6)} &=& D_{h} \bigg( \frac{1}{12}\mathcal{F}\left\{ \delta \mathcal{F}^{- 1}\left\{ \sqrt{e_{h}}k^{4}\mathcal{F}\left\{ \eta^{4}V \right\} \right\} \right\} + \frac{1}{6}\mathcal{F}\left\{ \delta^{3}\mathcal{F}^{- 1}\left\{ \sqrt{e_{h}}k^{4}\mathcal{F}\left\{ \eta^{2}V \right\} \right\} \right\} \nonumber\\
&& +\frac{1}{60}\mathcal{F}\left\{ \delta^{5}\mathcal{F}^{- 1}\left\{ \sqrt{e_{h}}k^{4}\mathcal{F}\left\{ V \right\} \right\} \right\} +\frac{1}{60}\sqrt{e_{h}}k^{3}i\bm{k} \cdot \mathcal{F}\left\{ \eta^{5}\nabla\widetilde{\phi} \right\} \nonumber\\
&&  + \frac{1}{6}\mathcal{F}\left\{ \delta^{2}\mathcal{F}^{- 1}\left\{ \sqrt{e_{h}}k^{3}i\bm{k} \cdot \mathcal{F}\left\{ \eta^{3}\nabla\widetilde{\phi} \right\} \right\} \right\}  +\frac{1}{12}\mathcal{F}\left\{ \delta^{4}\mathcal{F}^{- 1}\left\{ \sqrt{e_{h}}k^{3}i\bm{k} \cdot \mathcal{F}\left\{ \eta\nabla\widetilde{\phi} \right\} \right\} \right\} \nonumber\\
&&  - \frac{1}{120} C_{h}^{- 1}k^{3}i\bm{k} \cdot \mathcal{F}\left\{ \delta^{5}\nabla\widetilde{\varphi} \right\} + \frac{1}{24}\mathcal{F}\left\{ \delta \mathcal{F}^{- 1}\left\{ D_{h}^{- 1}k^{3}i\bm{k} \cdot \mathcal{F}\left\{ \delta^{4}\nabla\widetilde{\varphi} \right\} \right\} \right\} \nonumber\\
&& - \frac{1}{12} \mathcal{F}\left\{ \delta^{2}\mathcal{F}^{- 1}\left\{ C_{h}^{- 1}k^{3}i\bm{k} \cdot \mathcal{F}\left\{ \delta^{3}\nabla\widetilde{\varphi} \right\} \right\} \right\} +\frac{1}{12}\mathcal{F}\left\{ \delta^{3}\mathcal{F}^{- 1}\left\{ D_{h}^{- 1}k^{3}i\bm{k} \cdot \mathcal{F}\left\{ \delta^{2}\nabla\widetilde{\varphi} \right\} \right\} \right\}\nonumber\\
&&  - \frac{1}{24} \mathcal{F}\left\{ \delta^{4}\mathcal{F}^{- 1}\left\{ C_{h}^{- 1}k^{3}i\bm{k} \cdot \mathcal{F}\left\{ \delta\nabla\widetilde{\varphi} \right\} \right\} \right\} - \frac{1}{120}\mathcal{F}\left\{ \delta^{5}\mathcal{F}^{- 1}\left\{ D_{h}^{- 1}k^{5} \mathcal{F}\left\{ \widetilde{\varphi} \right\} \right\} \right\} 
\bigg). \nonumber\\
\end{eqnarray}

The remaining integration term can be split into \(I^{(B)}=I^{(B)}_{\phi}+I^{(B)}_V+I^{(B)}_{\varphi}\), and each part is given by:

\begin{eqnarray}\label{app:iphib}
I^{(B)}_{\phi} &=& {\widetilde{\phi}}'\Bigg\{ \frac{- \bm{R} \cdot \nabla'\eta' + \eta' - \delta + h}{[R^{2} + \left( \eta' - \delta + h \right)^{2}]^{3/2}} + \frac{- \bm{R} \cdot \nabla'\eta' + \eta' + \delta - h}{[R^{2} + \left( \eta' + \delta - h \right)^{2}]^{3/2}} \nonumber\\
&& -\nabla' \cdot \left( 2\eta'\nabla'\frac{1}{R_{h}} \right) + 2\delta\nabla' \cdot \left( \eta'\nabla'\frac{\partial}{\partial h}\frac{1}{R_{h}} \right) \nonumber\\
&& +\frac{1}{3}\nabla' \cdot \left( \eta'^{3}\nabla'\nabla' \cdot \nabla'\frac{1}{R_{h}} \right) + \delta^{2}\nabla' \cdot \left( \eta'\nabla'\nabla' \cdot \nabla'\frac{1}{R_{h}} \right) \nonumber\\
&& +\delta^{2}\nabla' \cdot \left( \eta'\nabla'\nabla' \cdot \nabla'\frac{1}{R_{h}} \right) - \frac{1}{3}\delta\nabla' \cdot \left( \eta'^{3}\nabla'\nabla' \cdot \nabla'\frac{\partial}{\partial h}\frac{1}{R_{h}} \right) \nonumber\\
&& -\frac{1}{3}\delta^{3}\nabla' \cdot \left( \eta'\nabla'\nabla' \cdot \nabla'\frac{\partial}{\partial h}\frac{1}{R_{h}} \right) - \frac{1}{60}\nabla' \cdot \left[ \eta'^{5}\nabla'\left( \nabla' \cdot \nabla' \right)^{2}\frac{1}{R_{h}} \right] \nonumber\\
&& -\frac{1}{6}\delta^{2}\nabla' \cdot \left[ \eta'^{3}\nabla'\left( \nabla' \cdot \nabla' \right)^{2}\frac{1}{R_{h}} \right] - \frac{1}{12}\delta^{4}\nabla' \cdot \left[ \eta'\nabla'\left( \nabla' \cdot \nabla' \right)^{2}\frac{1}{R_{h}} \right]\Bigg\} \mathrm{d} \bm{x}' ,
\end{eqnarray}

\begin{eqnarray}\label{app:ivb}
I^{(B)}_V &=& \frac{1}{2\pi}\iint_{S_{0}}^{} V'\Bigg\{ \frac{1}{\left[R^{2} + \left( \eta' - \delta + h \right)^{2}\right]^{1/2}} + \frac{1}{\left[R^{2} + \left( \eta' + \delta - h \right)^{2}\right]^{1/2}} \nonumber\\
&& -\frac{2}{R_{h}} + 2\delta\frac{\partial}{\partial h}\frac{1}{R_{h}} + \left( \eta'^{2} + \delta^{2} \right)\nabla' \cdot \nabla'\frac{1}{R_{h}} -\frac{1}{3}\left( \delta^{3} + 3\eta'^{2}\delta \right)\nabla' \cdot \nabla'\frac{\partial}{\partial h}\frac{1}{R_{h}} \nonumber\\
&&  - \frac{1}{24}\left( 2\eta'^{4} + 12\eta'^{2}\delta^{2} + 2\delta^{4} \right)\left( \nabla' \cdot \nabla' \right)^{2}\frac{1}{R_{h}} \nonumber\\
&& -\frac{1}{120}\left( - 10\eta'^{4}\delta - 20\eta'^{2}\delta^{3} - 2\delta^{5} \right)\left( \nabla' \cdot \nabla' \right)^{2}\frac{\partial}{\partial h}\frac{1}{R_{h}}\Bigg\} \mathrm{d} \bm{x}',
\end{eqnarray}

\noindent and

\begin{eqnarray}\label{app:ivarphib}
I^{(B)}_{\varphi} &=& - \frac{1}{2\pi}\iint_{B_{0}}^{}{\widetilde{\varphi}}'\Bigg\{ \frac{- \bm{R} \cdot \nabla'\delta' + \delta' - \delta}{[R^{2} + \left( \delta^{'} - \delta \right)^{2}]^{3/2}} + \frac{- \bm{R} \cdot \nabla'\delta' + \delta' + \delta - 2h}{[R^{2} + \left( \delta^{'} + \delta - 2h \right)^{2}]^{3/2}} \nonumber\\
&& -\nabla' \cdot \left[ \left( \delta' - \delta \right)\nabla'\frac{1}{R} \right] - \nabla' \cdot \left[ \left( \delta' + \delta \right)\nabla'\frac{1}{R_{2h}} \right] - \frac{\partial}{\partial\left( 2h \right)}\frac{1}{R_{2h}} \nonumber\\
&& +\frac{1}{2}\nabla' \cdot \left[ \left( \delta' + \delta \right)^{2}\nabla'\frac{\partial}{\partial\left( 2h \right)}\frac{1}{R_{2h}} \right] + \frac{1}{6}\nabla' \cdot \left[ \left( \delta' - \delta \right)^{3}\nabla'\nabla' \cdot \nabla'\frac{1}{R} \right] \nonumber\\
&& +\frac{1}{6}\nabla' \cdot \left[ \left( \delta' + \delta \right)^{3}\nabla'\nabla' \cdot \nabla'\frac{1}{R_{2h}} \right] - \frac{1}{24}\nabla' \cdot \left[ \left( \delta' + \delta \right)^{4}\nabla'\nabla' \cdot \nabla'\frac{\partial}{\partial(  2h)}\frac{1}{R_{2h}} \right] \nonumber\\
&& -\frac{1}{120}\nabla' \cdot \left[ \left( \delta' - \delta \right)^{5}\nabla'\left( \nabla' \cdot \nabla' \right)^{2}\frac{1}{R} \right] -\frac{1}{120}\nabla' \cdot \left[ \left( \delta' + \delta \right)^{5}\nabla'\left( \nabla' \cdot \nabla' \right)^{2}\frac{1}{R_{2h}} \right]\Bigg\} \mathrm{d} \bm{x}' .\nonumber\\
\end{eqnarray}

\color{black}
\section{}\label{appC}
%\item Step \#2. Evaluate the second-order terms
The second-order terms \(\sim O\left( \varepsilon^{m}\zeta^{2 - m} \right)\) are given as:
\begin{eqnarray}\label{eq:v2}
\mathcal{F}\left\{ V_{2} \right\} 
&= & - k\tanh (kh) \mathcal{F}\left\{ \eta V_{1} \right\} - \mathrm{i}\bm{k} \cdot \left[ \mathcal{F}\left\{ \eta\nabla\widetilde{\phi} \right\} - \text{sech} (kh) \mathcal{F}\left\{ \delta\nabla{\widetilde{\varphi}}_{1} \right\} \right],
\end{eqnarray}
\begin{eqnarray}
\mathcal{F}\left\{ {\widetilde{\varphi}}_{2} \right\} 
&= &- \text{sech}{(kh)}\mathcal{F}\left\{ \eta V_{1} \right\} - k^{-1}\tanh{(kh)} \mathrm{i}\bm{k} \cdot \mathcal{F}\left\{ \delta\nabla{\widetilde{\varphi}}_{1} \right\}.
\end{eqnarray}

%\item Step \#3. Evaluate the third-order terms
\noindent The third-order terms\(\sim O\left( \varepsilon^{m}\zeta^{3 - m} \right)\) are given as:
\begin{eqnarray}
\mathcal{F}\left\{ V_{3} \right\} 
 &= &- k\tanh (kh) \mathcal{F}\left\{ \eta V_{2} \right\} - \frac{1}{2}k^{2}\mathcal{F}\left\{ \eta^{2}V_{1} \right\} \nonumber\\
 && -\mathrm{i}\bm{k} \cdot \left[ \frac{1}{2}k\tanh (kh) \mathcal{F}\left\{ \eta^{2}\nabla\widetilde{\phi} \right\} - \text{sech} (kh) \mathcal{F}\left\{ \delta\nabla{\widetilde{\varphi}}_{2} \right\} \right],
\end{eqnarray}
\begin{eqnarray}
\mathcal{F}\left\{ {\widetilde{\varphi}}_{3} \right\} 
&= &- \text{sech}{(kh)}\mathcal{F}\left\{ \eta V_{2} \right\} + \mathrm{i}\bm{k} \cdot \bigg[ - \frac{1}{2}\text{sech}{(kh)}\mathcal{F}\left\{ \eta^{2}\nabla\widetilde{\phi} \right\} \nonumber\\
&& -k^{-1}\tanh{(kh)}\mathcal{F}\left\{ \delta\nabla{\widetilde{\varphi}}_{2} \right\} + \frac{1}{2}\mathcal{F}\left\{ \delta^{2}\nabla{\widetilde{\varphi}}_{1} \right\} \bigg].
\end{eqnarray}

%\item Step \#4. Evaluate the fourth-order terms
\noindent The fourth-order terms \(\sim O\left( \varepsilon^{m}\zeta^{4 - m} \right)\) are given as:
\begin{eqnarray}
\mathcal{F}\left\{ V_{4} \right\} 
&= & - \tanh (kh) \left( k \mathcal{F}\left\{ \eta V_{3} \right\} + \frac{1}{6}k^{3} \mathcal{F}\left\{ \eta^{3}V_{1} \right\} \right) - \frac{1}{2}k^{2}\mathcal{F}\left\{ \eta^{2}V_{2} \right\} \nonumber\\
&& - \mathrm{i}\bm{k} \cdot \left[ \frac{1}{6}k^{2}\mathcal{F}\left\{ \eta^{3}\nabla\widetilde{\phi} \right\} - \text{sech} (kh) \left( \mathcal{F}\left\{ \delta\nabla{\widetilde{\varphi}}_{3} \right\} + \frac{1}{6}k^{2}\mathcal{F}\left\{ \delta^{3}\nabla{\widetilde{\varphi}}_{1} \right\} \right) \right],
\end{eqnarray}
\begin{eqnarray}
 \mathcal{F}\left\{ {\widetilde{\varphi}}_{4} \right\} 
&= &- \text{sech}{(kh)} \left( \mathcal{F}\left\{ \eta V_{3} \right\} + \frac{1}{6}k^{2}\mathcal{F}\left\{ \eta^{3}V_{1} \right\} \right) + \mathrm{i}\bm{k} \cdot \bigg[ \frac{1}{2}\mathcal{F}\left\{ \delta^{2}\nabla{\widetilde{\varphi}}_{2} \right\} \nonumber\\
&&  - \tanh{(kh)} \left( k^{-1}\mathcal{F}\left\{ \delta\nabla{\widetilde{\varphi}}_{3} \right\} + \frac{1}{6}k\mathcal{F}\left\{ \delta^{3}\nabla{\widetilde{\varphi}}_{1} \right\} \right) \bigg].
\end{eqnarray}

%\item Step \#5. Evaluate the fifth-order terms
\noindent The fifth-order terms \(\sim O\left( \varepsilon^{m}\zeta^{5 - m} \right)\) are given as:
\begin{eqnarray}
\mathcal{F}\left\{ V_{5} \right\} 
&= &- \tanh (kh) \left( k \mathcal{F}\left\{ \eta V_{4} \right\} + \frac{1}{6}k^{3} \mathcal{F}\left\{ \eta^{3}V_{2} \right\} \right) - \frac{1}{2}k^{2}\mathcal{F}\left\{ \eta^{2}V_{3} \right\} \nonumber\\
&& -\frac{1}{24}k^{4}\mathcal{F}\left\{ \eta^{4}V_{1} \right\} + \mathrm{i}\bm{k} \cdot \bigg[ - \frac{1}{24}k^{3}\tanh (kh) \mathcal{F}\left\{ \eta^{4}\nabla \widetilde{\phi} \right\} \nonumber\\
&& +\text{sech} (kh) \left( \mathcal{F}\left\{ \delta\nabla{\widetilde{\varphi}}_{4} \right\} + \frac{1}{6}k^{2}\mathcal{F}\left\{ \delta^{3}\nabla{\widetilde{\varphi}}_{2} \right\} \right) \bigg],
\end{eqnarray}
\begin{eqnarray}
\mathcal{F}\left\{ {\widetilde{\varphi}}_{5} \right\} 
&= &- \text{sech}{(kh)} \left( \mathcal{F}\left\{ \eta V_{4} \right\} + \frac{1}{6}k^{2}\mathcal{F}\left\{ \eta^{3}V_{2} \right\} \right) + \mathrm{i}\bm{k} \cdot \bigg[ - \frac{1}{24}k^{2}\text{sech}{(kh)}\mathcal{F}\left\{ \eta^{4}\nabla \widetilde{\phi} \right\} \nonumber\\
&& - \tanh{(kh)} \left( k^{-1}\mathcal{F}\left\{ \delta\nabla{\widetilde{\varphi}}_{4} \right\} + \frac{1}{6}k\mathcal{F}\left\{ \delta^{3}\nabla{\widetilde{\varphi}}_{2} \right\} \right)   \nonumber\\
&&  + \frac{1}{2}\mathcal{F}\left\{ \delta^{2}\nabla{\widetilde{\varphi}}_{3} \right\} + \frac{1}{24}k^{2}\mathcal{F}\left\{ \delta^{4}\nabla{\widetilde{\varphi}}_{1} \right\} \bigg].
\end{eqnarray}

%\item Step \#6. Evaluate the sixth-order terms
\noindent The sixth-order terms \(\sim O\left( \varepsilon^{m}\zeta^{6 - m} \right)\) are given as: 
\begin{eqnarray}
\mathcal{F}\left\{ V_{6} \right\} 
&= &- \tanh{(kh)} \left( k\mathcal{F}\left\{ \eta V_{5} \right\} + \frac{1}{6}k^{3}\mathcal{F}\left\{ \eta^{3}V_{3} \right\} + \frac{1}{120}k^{5}\mathcal{F}\left\{ \eta^{5}V_{1} \right\} \right) \nonumber\\
&& - \frac{1}{2}k^{2}\mathcal{F}\left\{ \eta^{2}V_{4} \right\} -\frac{1}{24}k^{4}\mathcal{F}\left\{ \eta^{4}V_{2} \right\} + \mathrm{i}\bm{k} \cdot \bigg[ - \frac{1}{120}k^{4}\mathcal{F}\left\{ \eta^{5}\nabla \widetilde{\phi} \right\} \nonumber\\
&& +\text{sech}{(kh)} \left( \mathcal{F}\left\{ \delta\nabla{\widetilde{\varphi}}_{5} \right\} + \frac{1}{6}k^{2}\mathcal{F}\left\{ \delta^{3}\nabla{\widetilde{\varphi}}_{3} \right\} + \frac{1}{120} k^{4}\mathcal{F}\left\{ \delta^{5}\nabla{\widetilde{\varphi}}_{1} \right\} \right) \bigg],
\end{eqnarray}
\begin{eqnarray}
\mathcal{F}\left\{ {\widetilde{\varphi}}_{6} \right\}
&= &- {\text{sech}}{(kh)} \left( \mathcal{F}\left\{ \eta V_{5} \right\} + \frac{1}{6}k^{2}\mathcal{F}\left\{ \eta^{3}V_{3} \right\} + \frac{1}{120}k^{4}\mathcal{F}\left\{ \eta^{5}V_{1} \right\} \right) \nonumber\\
&& + \mathrm{i}\bm{k} \cdot\bigg[ - \tanh{(kh)} \left( k^{-1}\mathcal{F}\left\{ \delta\nabla{\widetilde{\varphi}}_{5} \right\} + \frac{1}{6}k\mathcal{F}\left\{ \delta^{3}\nabla{\widetilde{\varphi}}_{3} \right\} + \frac{1}{120}k^{3}\mathcal{F}\left\{ \delta^{5}\nabla{\widetilde{\varphi}}_{1} \right\}   \right) \nonumber\\
&& + \frac{1}{2}\mathcal{F}\left\{ \delta^{2}\nabla{\widetilde{\varphi}}_{4} \right\} +\frac{1}{24}k^{2}\mathcal{F}\left\{ \delta^{4}\nabla{\widetilde{\varphi}}_{2} \right\}  \bigg].
\end{eqnarray}

%\item Step \#7. Evaluate the seventh-order velocity
\noindent The seventh-order \(\sim O\left( \varepsilon^{m}\zeta^{7 - m} \right)\) velocity is:

\begin{eqnarray}\label{eq:v7}
\mathcal{F}\left\{ V_{7} \right\} %= &&kC_{h}\bigg[ e_{h}\mathcal{F}\left\{ \eta V_{6} \right\} + \mathcal{F}\left\{ \eta \mathcal{F}^{- 1}\left\{ e_{h}\mathcal{F}\left\{ V_{6} \right\} \right\} \right\} - \frac{1}{2}C_{h}^{- 1}{k\mathcal{F}}\left\{ \eta^{2}V_{5} \right\} \nonumber\\
&=&  - \tanh{(kh)} \left( k\mathcal{F}\left\{ \eta V_{6} \right\} \color{black}+ \frac{1}{6}k^{3}\mathcal{F}\left\{ \eta^{3}V_{4} \right\}  +\frac{1}{120}k^{5}\mathcal{F}\left\{ \eta^{5}V_{2} \right\} \right)\nonumber\\
&& - \frac{1}{2}k^{2}\mathcal{F}\left\{ \eta^{2}V_{5} \right\} - \frac{1}{24}k^{4}\mathcal{F}\left\{ \eta^{4}V_{3} \right\} - \frac{1}{720}k^{6}\mathcal{F}\left\{ \eta^{6}V_{1} \right\}  \nonumber\\
&& + \mathrm{i}\bm{k} \cdot \bigg[ - \frac{1}{720}k^{5}\tanh{(kh)}\mathcal{F}\left\{ \eta^{6}\nabla \widetilde{\phi} \right\} +\text{sech}{(kh)} \bigg( \mathcal{F}\left\{ \delta\nabla{\widetilde{\varphi}}_{6} \right\} \nonumber\\
 && + \frac{1}{6}k^{2}\mathcal{F}\left\{ \delta^{3}\nabla{\widetilde{\varphi}}_{4} \right\} + \frac{1}{120}k^{4}\mathcal{F}\left\{ \delta^{5}\nabla{\widetilde{\varphi}}_{2} \right\} \bigg) \bigg].
\end{eqnarray}
%\end {enumerate}
\color{black}

%\printbibliography

%\bibliographystyle{jfm}
\bibliographystyle{plain}
\bibliography{jfm}

\begin{thebibliography}{100}

\bibitem{akhmediev2009waves}
N.~Akhmediev and M.~Ankiewicz, A.and~Taki.
\newblock Waves that appear from nowhere and disappear without a trace.
\newblock {\em Physics Letters A}, 373(6):675--678, 2009.

\bibitem{alpert1993wavelet}
B.~Alpert, G.~Beylkin, R.~Coifman, and V.~Rokhlin.
\newblock Wavelet-like bases for the fast solution of second-kind integral equations.
\newblock {\em SIAM journal on Scientific Computing}, 14(1):159--184, 1993.

\bibitem{barratt2022impact}
D.~Barratt, T.~S. van~den Bremer, and T.~A.~A. Adcock.
\newblock The impact of the spectral tail on the evolution of the kurtosis of random seas.
\newblock {\em Journal of Offshore Mechanics and Arctic Engineering}, 144(6):061702, 2022.

\bibitem{bateman2001efficient}
W.~J.~D. Bateman, C.~Swan, and P.~H. Taylor.
\newblock On the efficient numerical simulation of directionally spread surface water waves.
\newblock {\em Journal of Computational Physics}, 174(1):277--305, 2001.

\bibitem{beji1993}
S.~Beji and J.~A. Battjes.
\newblock Experimental investigation of wave propagation over a bar.
\newblock {\em Coastal engineering}, 19(1-2):151--162, 1993.

\bibitem{bingham2007}
H.~B. Bingham and H.~Zhang.
\newblock On the accuracy of finite-difference solutions for nonlinear water waves.
\newblock {\em Journal of Engineering Mathematics}, 58:211--228, 2007.

\bibitem{canuto1987spectral}
C.~Canuto, M.~Y. Hussaini, A.~Quarteroni, and T.~A. Zang.
\newblock Spectral methods in fluid dynamics springer-verlag.
\newblock {\em New York}, 1987.

\bibitem{castro2019}
M.~J. Castro and M.~Semplice.
\newblock Third-and fourth-order well-balanced schemes for the shallow water equations based on the cweno reconstruction.
\newblock {\em International Journal for Numerical Methods in Fluids}, 89(8):304--325, 2019.

\bibitem{cavaleri2018wave}
L.~Cavaleri, S.~Abdalla, A.~Benetazzo, L.~Bertotti, J.-R. Bidlot, {\O}.~Breivik, S.~Carniel, R.~E. Jensen, J.~Portilla-Yandun, W.~E. Rogers, and A.~Roland.
\newblock Wave modelling in coastal and inner seas.
\newblock {\em Progress in Oceanography}, 167:164--233, 2018.

\bibitem{cavaleri2007wave}
L.~Cavaleri, J.-H. G.~M. Alves, F.~Ardhuin, A.~Babanin, M.~Banner, K.~Belibassakis, M.~Benoit, M.~Donelan, J.~Groeneweg, T.~H.~C. Herbers, and P.~A. E.~M. Hwang.
\newblock Wave modelling--the state of the art.
\newblock {\em Progress in Oceanography}, 75(4):603--674, 2007.

\bibitem{celebi1998}
M.~S. Celebi, M.~H. Kim, and R.~F. Beck.
\newblock {Fully nonlinear 3-D numerical wave tank simulation}.
\newblock {\em Journal of Ship Research}, 42(1):33--45, 1998.

\bibitem{chabchoub2011rogue}
A.~Chabchoub, N.~P. Hoffmann, and N.~Akhmediev.
\newblock Rogue wave observation in a water wave tank.
\newblock {\em Physical Review Letters}, 106(20):204502, 2011.

\bibitem{Chen2006}
Q.~Chen.
\newblock {Fully nonlinear Boussinesq-type equations for waves and currents over porous beds}.
\newblock {\em Journal of Engineering Mechanics}, 132:220--230, 2006.

\bibitem{Chen1995}
Y.~Chen and P.~L.-F. Liu.
\newblock {Modified Boussinesq equations and associated parabolic models for water wave propagation.}
\newblock {\em Journal of Fluid Mechanics}, 288:351--381, 1995.

\bibitem{clamond2007note}
D.~Clamond, D.~Fructus, and J.~Grue.
\newblock A note on time integrators in water-wave simulations.
\newblock {\em Journal of Engineering Mathematics}, 58:149--156, 2007.

\bibitem{clamond2005efficient}
D.~Clamond, D.~Fructus, J.~Grue, and {\O}.~Kristiansen.
\newblock An efficient model for three-dimensional surface wave simulations. {Part II: Generation} and absorption.
\newblock {\em Journal of Computational Physics}, 205(2):686--705, 2005.

\bibitem{clamond2001fast}
D.~Clamond and J.~Grue.
\newblock A fast method for fully nonlinear water-wave computations.
\newblock {\em Journal of Fluid Mechanics}, 447:337--355, 2001.

\bibitem{clauss1999}
G.~F. Clauss and U.~Steinhagen.
\newblock Numerical simulation of nonlinear transient waves and its validation by laboratory data.
\newblock In {\em Proceedings of the 9th International Offshore and Polar Engineering Conference, Brest, France}, pages 368--375, 1999.

\bibitem{craig1993numerical}
W.~Craig and C.~Sulem.
\newblock Numerical simulation of gravity waves.
\newblock {\em Journal of Computational Physics}, 108(1):73--83, 1993.

\bibitem{davies1984surface}
A.~G. Davies and A.~D. Heathershaw.
\newblock Surface-wave propagation over sinusoidally varying topography.
\newblock {\em Journal of Fluid Mechanics}, 144:419--443, 1984.

\bibitem{dean1991water}
R.~G. Dean and R.~A. Dalrymple.
\newblock {\em Water wave mechanics for engineers and scientists}, volume~2.
\newblock world scientific publishing company, 1991.

\bibitem{dold1992efficient}
J.~W. Dold.
\newblock An efficient surface-integral algorithm applied to unsteady gravity waves.
\newblock {\em Journal of Computational Physics}, 103(1):90--115, 1992.

\bibitem{dommermuth1987high}
D.~G. Dommermuth and D.~K.~P. Yue.
\newblock A high-order spectral method for the study of nonlinear gravity waves.
\newblock {\em Journal of Fluid Mechanics}, 184:267--288, 1987.

\bibitem{ducrozet20073}
G.~Ducrozet, F{\'e}. Bonnefoy, D.~Le~Touz{\'e}, and P.~Ferrant.
\newblock {3-D HOS} simulations of extreme waves in open seas.
\newblock {\em Natural Hazards and Earth System Sciences}, 7(1):109--122, 2007.

\bibitem{ducrozet2012modified}
G.~Ducrozet, F{\'e}. Bonnefoy, D.~Le~Touz{\'e}, and P.~Ferrant.
\newblock A modified high-order spectral method for wavemaker modeling in a numerical wave tank.
\newblock {\em European Journal of Mechanics-B/Fluids}, 34:19--34, 2012.

\bibitem{ducrozet2016hos}
G.~Ducrozet, F{\'e}. Bonnefoy, D.~Le~Touz{\'e}, and P.~Ferrant.
\newblock {HOS-ocean}: Open-source solver for nonlinear waves in open ocean based on high-order spectral method.
\newblock {\em Computer Physics Communications}, 203:245--254, 2016.

\bibitem{engsig_kurap2009}
A.~P. Engsig-Karup, H.~B. Bingham, and O.~Lindberg.
\newblock An efficient flexible-order model for {3D} nonlinear water waves.
\newblock {\em Journal of Computational Physics}, 228(6):2100--2118, 2009.

\bibitem{engsig2016stabilised}
A.~P. Engsig-Karup, C.~Eskilsson, and D.~Bigoni.
\newblock A stabilised nodal spectral element method for fully nonlinear water waves.
\newblock {\em Journal of Computational Physics}, 318:1--21, 2016.

\bibitem{engsig_kurap2012}
A.~P. Engsig-Karup, M.~G. Madsen, and S.~L. Glimberg.
\newblock A massively parallel gpu-accelerated model for analysis of fully nonlinear free surface waves.
\newblock {\em International Journal for Numerical Methods in Fluids}, 70(1):20--36, 2012.

\bibitem{Ertekin1986}
R.~C. Ertekin, W.~C. Webster, R.~C. Ertekin, W.~C. Webster, and J.~V. Wehausen.
\newblock {Waves caused by a moving disturbance in a shallow channel of finite width}.
\newblock {\em Journal of Fluid Mechanics}, 169:275--292, 1986.

\bibitem{fenton1988numerical}
J.~D. Fenton.
\newblock The numerical solution of steady water wave problems.
\newblock {\em Computers \& Geosciences}, 14(3):357--368, 1988.

\bibitem{ferrant1997}
P.~Ferrant.
\newblock Simulation of strongly nonlinear wave generation and wave-body interactions using a {3-D MEL} model.
\newblock In {\em Proceedings of the 21st ONR Symposium on Naval Hydrodynamics, Trondheim, Norway}, pages 93--109, 1997.

\bibitem{fochesato2006fast}
C.~Fochesato and F.~Dias.
\newblock A fast method for nonlinear three-dimensional free-surface waves.
\newblock {\em Proceedings of the Royal Society A: Mathematical, Physical and Engineering Sciences}, 462(2073):2715--2735, 2006.

\bibitem{frigo2005design}
M.~Frigo and S.~G. Johnson.
\newblock The design and implementation of fftw3.
\newblock {\em Proceedings of the IEEE}, 93(2):216--231, 2005.

\bibitem{fructus2005}
D.~Fructus, D.~Clamond, J.~Grue, and {\O}.~Kristiansen.
\newblock An efficient model for three-dimensional surface wave simulations: Part i: Free space problems.
\newblock {\em Journal of computational physics}, 205(2):665--685, 2005.

\bibitem{fructus2007}
D.~Fructus and J.~Grue.
\newblock An explicit method for the nonlinear interaction between water waves and variable and moving bottom topography.
\newblock {\em Journal of Computational Physics}, 222(2):720--739, 2007.

\bibitem{gao2021investigation}
J.~Gao, X.~Ma, G.~Dong, H.~Chen, Q.~Liu, and J.~Zang.
\newblock Investigation on the effects of bragg reflection on harbor oscillations.
\newblock {\em Coastal Engineering}, 170:103977, 2021.

\bibitem{goda1976estimation}
Y.~Goda and Y.~Suzuki.
\newblock Estimation of incident and reflected waves in random wave experiments.
\newblock In {\em Coastal Engineering 1976}, pages 828--845. 1976.

\bibitem{goreinov1997matrix}
S.~A. Goreinov, E.~E. Tyrtyshnikov, and A.~Yu. Yeremin.
\newblock Matrix-free iterative solution strategies for large dense linear systems.
\newblock {\em Numerical linear algebra with applications}, 4(4):273--294, 1997.

\bibitem{gouin2016development}
M.~Gouin, G.~Ducrozet, and P.~Ferrant.
\newblock Development and validation of a non-linear spectral model for water waves over variable depth.
\newblock {\em European Journal of Mechanics-B/Fluids}, 57:115--128, 2016.

\bibitem{gouin2017propagation}
M.~Gouin, G.~Ducrozet, and P.~Ferrant.
\newblock Propagation of 3d nonlinear waves over an elliptical mound with a high-order spectral method.
\newblock {\em European Journal of Mechanics-B/Fluids}, 63:9--24, 2017.

\bibitem{Green1974}
A.~E. Green, N.~Laws, and P.~M. Naghdi.
\newblock {On the theory of water waves}.
\newblock {\em Proceedings of the Royal Society A: Maths, Physics, Engineering and Science}, 338:43--55, 1974.

\bibitem{Green1976}
A.~E. Green and P.~M. Naghdi.
\newblock {A derivation of equations for wave propagation in water of variable depth}.
\newblock {\em Journal of Fluid Mechanics}, 78:237--246, 1976.

\bibitem{Green1976a}
A.~E. Green and P.~M. Naghdi.
\newblock {Directed Fluid Sheets}.
\newblock {\em Proceedings of the Royal Society A: Maths, Physics, Engineering and Science}, 347:447--473, 1976.

\bibitem{grilli2001}
S.~T. Grilli, P.~Guyenne, and F.~Dias.
\newblock A fully non-linear model for three-dimensional overturning waves over an arbitrary bottom.
\newblock {\em International Journal for Numerical Methods in Fluids}, 35(7):829--867, 2001.

\bibitem{grilli1997numerical}
S.~T. Grilli and J.~Horrillo.
\newblock Numerical generation and absorption of fully nonlinear periodic waves.
\newblock {\em Journal of engineering mechanics}, 123(10):1060--1069, 1997.

\bibitem{grilli2020fully}
S.~T. Grilli, J.~Horrillo, and S.~Guignard.
\newblock Fully nonlinear potential flow simulations of wave shoaling over slopes: Spilling breaker model and integral wave properties.
\newblock {\em Water Waves}, 2(2):263--297, 2020.

\bibitem{grilli1989efficient}
S.~T. Grilli, J.~Skourup, and I.~A. Svendsen.
\newblock An efficient boundary element method for nonlinear water waves.
\newblock {\em Engineering Analysis with Boundary Elements}, 6(2):97--107, 1989.

\bibitem{guignard2001modeling}
S.~Guignard and S.~T. Grilli.
\newblock Modeling of wave shoaling in a 2d-nwt using a spilling breaker model.
\newblock In {\em The Eleventh International Offshore and Polar Engineering Conference}. OnePetro, 2001.

\bibitem{guyenne2005numerical}
P.~Guyenne and D.~P. Nicholls.
\newblock Numerical simulation of solitary waves on plane slopes.
\newblock {\em Mathematics and Computers in Simulation}, 69(3-4):269--281, 2005.

\bibitem{guyenne2008high}
P.~Guyenne and D.~P. Nicholls.
\newblock A high-order spectral method for nonlinear water waves over moving bottom topography.
\newblock {\em SIAM Journal on Scientific Computing}, 30(1):81--101, 2008.

\bibitem{hackbusch1999sparse}
W.~Hackbusch.
\newblock A sparse matrix arithmetic based on-matrices. part i: Introduction to-matrices.
\newblock {\em Computing}, 62(2):89--108, 1999.

\bibitem{hao2022numerical}
J.~Hao, J.~Li, S.~Liu, and L.~Wang.
\newblock Numerical studies on the generation and propagation of tsunami waves based on the high-order spectral method.
\newblock {\em China Ocean Engineering}, 36(2):268--278, 2022.

\bibitem{harris2022nonlinear}
J.~C. Harris, E.~Dombre, M.~Benoit, S.~T. Grilli, and K.~I. Kuznetsov.
\newblock Nonlinear time-domain wave-structure interaction: A parallel fast integral equation approach.
\newblock {\em International Journal for Numerical Methods in Fluids}, 94(2):188--222, 2022.

\bibitem{henderson1999unsteady}
K.~L. Henderson, D.~H. Peregrine, and J.~W. Dold.
\newblock Unsteady water wave modulations: fully nonlinear solutions and comparison with the nonlinear {Schr\"{o}dinger} equation.
\newblock {\em Wave Motion}, 29(4):341--361, 1999.

\bibitem{higuera2023}
P.~Higuera, J.~Wang, J.~Hu, and Z.~Yang.
\newblock {\em Numerical Modeling of Water Waves in Coastal and Ocean Engineering}.
\newblock World Scientific, 2023.

\bibitem{holthuijsen10}
L.~H. Holthuijsen.
\newblock {\em Waves in oceanic and coastal waters.}
\newblock Cambridge University Press, 2010.

\bibitem{Hsiao2002}
S.-C. Hsiao, P.~L.-F. Liu, and Y.~Chen.
\newblock {Nonlinear water waves propagating over a permeable bed}.
\newblock {\em Proceedings of the Royal Society A: Maths, Physics, Engineering and Science}, 458(2022):1291--1322, 2002.

\bibitem{hsu2007application}
T.-W. Hsu, S.-C. Hsiao, S.-H. Ou, S.-K. Wang, B.-D. Yang, and S.-E. Chou.
\newblock An application of boussinesq equations to bragg reflection of irregular waves.
\newblock {\em Ocean Engineering}, 34(5-6):870--883, 2007.

\bibitem{jacobsen2017waves2foam}
N.~G. Jacobsen.
\newblock waves2foam manual.
\newblock {\em Deltares, The Netherlands}, 2017.

\bibitem{karniadakis2013spectral}
G.~Karniadakis and S.~Sherwin.
\newblock {\em Spectral/hp element methods for computational fluid dynamics}.
\newblock Oxford University Press, 2013.

\bibitem{kennedy2000boussinesq}
A.~B. Kennedy, Q.~Chen, J.~T. Kirby, and R.~A. Dalrymple.
\newblock Boussinesq modeling of wave transformation, breaking, and runup. i: 1d.
\newblock {\em Journal of waterway, port, coastal, and ocean engineering}, 126(1):39--47, 2000.

\bibitem{kennedy1997fully}
A.~B. Kennedy and J.~D. Fenton.
\newblock A fully-nonlinear computational method for wave propagation over topography.
\newblock {\em Coastal engineering}, 32(2-3):137--161, 1997.

\bibitem{kharif2008rogue}
C.~Kharif, E.~Pelinovsky, and A.~Slunyaev.
\newblock {\em Rogue waves in the ocean}.
\newblock Springer Science \& Business Media, 2008.

\bibitem{Kirby1984}
J.~T. Kirby.
\newblock A note on linear surface wave-current interaction.
\newblock {\em Journal of Geophysical Research}, 89:745--747, 1984.

\bibitem{Kirby1986}
J.~T. Kirby.
\newblock A general wave equation for waves over rippled beds.
\newblock {\em Journal of Fluid Mechanics}, 162:171--186, 1986.

\bibitem{kirby1998funwave}
J.~T. Kirby, G.~Wei, Q.~Chen, A.~B. Kennedy, and R.~A. Dalrymple.
\newblock {FUNWAVE 1.0: fully nonlinear Boussinesq wave model - Documentation and user's manual}.
\newblock {\em research report NO. CACR-98-06}, 1998.

\bibitem{komen1996dynamics}
G.~J. Komen, L.~Cavaleri, M.~Donelan, K.~Hasselmann, S.~Hasselmann, and P.~A. E.~M. Janssen.
\newblock {\em Dynamics and modelling of ocean waves}.
\newblock Cambridge University Press, UK, 1996.

\bibitem{korsmeyer1993multipole}
F.~T. Korsmeyer, D.~K.~P. Yue, K.~Nabors, and J.~White.
\newblock Multipole-accelerated preconditioned iterative methods for three-dimensional potential problems.
\newblock {\em WIT Transactions on Modelling and Simulation}, 1, 1993.

\bibitem{kurganov2018}
A.~Kurganov.
\newblock Finite-volume schemes for shallow-water equations.
\newblock {\em Acta Numerica}, 27:289--351, 2018.

\bibitem{Lannes2009}
D.~Lannes and P.~Bonneton.
\newblock {Derivation of asymptotic two-dimensional time-dependent equations for surface water wave propagation}.
\newblock {\em Physics of Fluids}, 21(1):16601, 2009.

\bibitem{li1997three}
B.~Li and C.~A. Fleming.
\newblock A three dimensional multigrid model for fully nonlinear water waves.
\newblock {\em Coastal Engineering}, 30(3-4):235--258, 1997.

\bibitem{liu2019analytical}
H.-W. Liu, X.-F. Li, and P.~Lin.
\newblock Analytical study of bragg resonance by singly periodic sinusoidal ripples based on the modified mild-slope equation.
\newblock {\em Coastal Engineering}, 150:121--134, 2019.

\bibitem{liu1989note}
P.~L.-F. Liu.
\newblock A note on long waves induced by short-wave groups over a shelf.
\newblock {\em Journal of Fluid Mechanics}, 205:163--170, 1989.

\bibitem{Liu1995}
P.~L.-F. Liu, Y.~S. Cho, M.~J. Briggs, U.~Kanoglu, and C.~E. Synolakis.
\newblock {Runup of solitary waves on a circular Island}.
\newblock {\em Journal of Fluid Mechanics}, 302:259--285, 1995.

\bibitem{liu1998generalized}
Y.~Liu and D.~K.~P. Yue.
\newblock On generalized bragg scattering of surface waves by bottom ripples.
\newblock {\em Journal of Fluid Mechanics}, 356:297--326, 1998.

\bibitem{Liu2018}
Z.~B. Liu, K.~Z. Fang, and Y.~Z. Cheng.
\newblock {A new multi-layer irrotational Boussinesq-type model for highly nonlinear and dispersive surface waves over a mildly sloping seabed}.
\newblock {\em Journal of Fluid Mechanics}, 842:323--353, 2018.

\bibitem{longuet1976deformation}
M.~S. Longuet-Higgins and E.~D. Cokelet.
\newblock The deformation of steep surface waves on water - {I. A} numerical method of computation.
\newblock {\em Proceedings of the Royal Society of London. A. Mathematical and Physical Sciences}, 350(1660):1--26, 1976.

\bibitem{longuet1962radiation}
M.~S. Longuet-Higgins and R.~W. Stewart.
\newblock Radiation stress and mass transport in gravity waves, with application to ‘surf beats’.
\newblock {\em Journal of Fluid Mechanics}, 13(4):481--504, 1962.

\bibitem{Lynett2004}
P.~Lynett and P.~L.-F. Liu.
\newblock {A two-layer approach to wave modelling}.
\newblock {\em Proceedings of the Royal Society A}, 460:2637--2669, 2004.

\bibitem{Lynett2004b}
P.~J. Lynett and P.~L.-F. Liu.
\newblock {Linear analysis of the multi-layer model}.
\newblock {\em Coastal Engineering}, 51:439--454, 2004.

\bibitem{ma1997}
G.~X. Ma, Q. W.and~Wu and R.~{Eatock-Taylor}.
\newblock Finite element analysis of non-linear transient waves in a three dimensional long tank.
\newblock In {\em Proceedings of the 12th International Workshop on Water Waves and Floating Bodies, Carry-le-Rouet, France}, 1997.

\bibitem{ma1998}
Q.~W. Ma.
\newblock {\em Numerical simulation of nonlinear interaction between structures and steep waves}.
\newblock PhD thesis, University College London, 1998.

\bibitem{ma2006}
Q.~W. Ma and S.~Yan.
\newblock Quasi {ALE} finite element method for nonlinear water waves.
\newblock {\em Journal of Computational Physics}, 212(1):52--72, 2006.

\bibitem{ma2009}
Q.~W. Ma and S.~Yan.
\newblock {QALE-FEM} for numerical modelling of non-linear interaction between {3D} moored floating bodies and steep waves.
\newblock {\em International Journal for Numerical Methods in Engineering}, 78(6):713--756, 2009.

\bibitem{Massel993}
S.~R. Massel.
\newblock Extended refraction-diffraction equation for surface waves.
\newblock {\em Coastal Engineering}, 19(1-2):97--126, 1993.

\bibitem{mcallister2019laboratory}
M.~L. McAllister, S.~Draycott, T.~A.~A. Adcock, P.~H. Taylor, and T.~S. Van Den~Bremer.
\newblock Laboratory recreation of the draupner wave and the role of breaking in crossing seas.
\newblock {\em Journal of Fluid Mechanics}, 860:767--786, 2019.

\bibitem{mei1984long}
C.~C. Mei and C.~Benmoussa.
\newblock Long waves induced by short-wave groups over an uneven bottom.
\newblock {\em Journal of Fluid Mechanics}, 139:219--235, 1984.

\bibitem{mohanlal2023unified}
S.~Mohanlal, J.~C. Harris, M.~L. Yates, and S.~T. Grilli.
\newblock Unified depth-limited wave breaking detection and dissipation in fully nonlinear potential flow models.
\newblock {\em Coastal Engineering}, page 104316, 2023.

\bibitem{nicholls1998traveling}
D.~P. Nicholls.
\newblock Traveling water waves: spectral continuation methods with parallel implementation.
\newblock {\em Journal of Computational Physics}, 143(1):224--240, 1998.

\bibitem{nwogu1993alternative}
O.~Nwogu.
\newblock Alternative form of boussinesq equations for nearshore wave propagation.
\newblock {\em Journal of waterway, port, coastal, and ocean engineering}, 119(6):618--638, 1993.

\bibitem{onorato2006modulational}
M.~Onorato, A.~R. Osborne, and M.~Serio.
\newblock Modulational instability in crossing sea states: A possible mechanism for the formation of freak waves.
\newblock {\em Physical review letters}, 96(1):014503, 2006.

\bibitem{papoutsellis2019modelling}
C.~E. Papoutsellis, M.~L. Yates, B.~Simon, and M.~Benoit.
\newblock Modelling of depth-induced wave breaking in a fully nonlinear free-surface potential flow model.
\newblock {\em Coastal Engineering}, 154:103579, 2019.

\bibitem{patera1984spectral}
A.~T. Patera.
\newblock A spectral element method for fluid dynamics: laminar flow in a channel expansion.
\newblock {\em Journal of Computational Physics}, 54(3):468--488, 1984.

\bibitem{pelinovsky2008extreme}
E.~Pelinovsky and C.~Kharif.
\newblock {\em Extreme ocean waves}.
\newblock Springer, 2008.

\bibitem{Peregrine1967}
D.~H. Peregrine.
\newblock {Long waves on a beach}.
\newblock {\em Journal of Fluid Mechanics}, 27:815--827, 1967.

\bibitem{peregrine1983water}
D.~H. Peregrine.
\newblock Water waves, nonlinear schr{\"o}dinger equations and their solutions.
\newblock {\em The ANZIAM Journal}, 25(1):16--43, 1983.

\bibitem{potts2003fast}
D.~Potts and G.~Steidl.
\newblock Fast summation at nonequispaced knots by nfft.
\newblock {\em SIAM Journal on Scientific Computing}, 24(6):2013--2037, 2003.

\bibitem{raoult2016validation}
C.~Raoult, M.~Benoit, and M.~L. Yates.
\newblock Validation of a fully nonlinear and dispersive wave model with laboratory non-breaking experiments.
\newblock {\em Coastal Engineering}, 114:194--207, 2016.

\bibitem{schaffer2008comparison}
H.~A. Sch{\"a}ffer.
\newblock Comparison of {Dirichlet--Neumann} operator expansions for nonlinear surface gravity waves.
\newblock {\em Coastal Engineering}, 55(4):288--294, 2008.

\bibitem{schaffer1993boussinesq}
H.~A. Sch{\"a}ffer, P.~A. Madsen, and R.~Deigaard.
\newblock A boussinesq model for waves breaking in shallow water.
\newblock {\em Coastal engineering}, 20(3-4):185--202, 1993.

\bibitem{seiffert2018simulation}
B.~R. Seiffert and G.~Ducrozet.
\newblock Simulation of breaking waves using the high-order spectral method with laboratory experiments: wave-breaking energy dissipation.
\newblock {\em Ocean Dynamics}, 68:65--89, 2018.

\bibitem{shao2014harmonic}
Y.-L. Shao and O.~M. Faltinsen.
\newblock A harmonic polynomial cell (hpc) method for 3d laplace equation with application in marine hydrodynamics.
\newblock {\em Journal of Computational Physics}, 274:312--332, 2014.

\bibitem{Sharmaetal2014}
A.~Sharma, V.~G. Panchang, and J.~M. Kaihatu.
\newblock Modeling nonlinear wave–wave interactions with the elliptic mild slope equation.
\newblock {\em Applied Ocean Research}, 48:114--125, 2014.

\bibitem{Shi2012}
F.~Shi, J.~T. Kirby, J.~C. Harris, J.~D. Geiman, and S.~T. Grilli.
\newblock {A high-order adaptive time-stepping TVD solver for Boussinesq modeling of breaking waves and coastal inundation}.
\newblock {\em Ocean Modelling}, 43-44:36--51, 2012.

\bibitem{Shields1988}
J.~J. Shields and W.~C. Webster.
\newblock {On direct methods in water wave theory}.
\newblock {\em Journal of Fluid Mechanics}, 197:171--199, 1988.

\bibitem{shrira2010makes}
V.~I. Shrira and V.~V. Geogjaev.
\newblock What makes the peregrine soliton so special as a prototype of freak waves?
\newblock {\em Journal of Engineering Mathematics}, 67:11--22, 2010.

\bibitem{shukla2006instability}
P.~K. Shukla, I.~Kourakis, B.~Eliasson, M.~Marklund, and L.~Stenflo.
\newblock Instability and evolution of nonlinearly interacting water waves.
\newblock {\em Physical review letters}, 97(9):094501, 2006.

\bibitem{simon2019comparing}
B.~Simon, C.~E. Papoutsellis, M.~Benoit, and M.~L. Yates.
\newblock Comparing methods of modeling depth-induced breaking of irregular waves with a fully nonlinear potential flow approach.
\newblock {\em Journal of Ocean Engineering and Marine Energy}, 5(4):365--383, 2019.

\bibitem{smith1998operator}
R.~A. Smith.
\newblock An operator expansion formalism for nonlinear surface waves over variable depth.
\newblock {\em Journal of Fluid Mechanics}, 363:333--347, 1998.

\bibitem{sriram2006}
V.~Sriram, S.A. Sannasiraj, and V.~Sundar.
\newblock Simulation of 2-d nonlinear waves using finite element method with cubic spline approximation.
\newblock {\em Journal of Fluids and Structures}, 22(5):663--681, 2006.

\bibitem{svendsen1978wave}
I.~A. Svendsen, P.~A. Madsen, and J.~B. Hansen.
\newblock Wave characteristics in the surf zone.
\newblock In {\em Coastal Engineering 1978}, pages 520--539. 1978.

\bibitem{tanaka1983stability}
M.~Tanaka.
\newblock The stability of steep gravity waves.
\newblock {\em Journal of the Physical Society of Japan}, 52(9):3047--3055, 1983.

\bibitem{tanaka1986stability}
M.~Tanaka.
\newblock The stability of solitary waves.
\newblock {\em The Physics of fluids}, 29(3):650--655, 1986.

\bibitem{toffoli2011extreme}
A.~Toffoli, E.~M. Bitner-Gregersen, A.~R. Osborne, M.~Serio, J.~Monbaliu, and M.~Onorato.
\newblock Extreme waves in random crossing seas: Laboratory experiments and numerical simulations.
\newblock {\em Geophysical Research Letters}, 38(6), 2011.

\bibitem{toffoli2010evolution}
A.~Toffoli, O.~Gramstad, K.~Trulsen, J.~Monbaliu, E.~Bitner-Gregersen, and M.~Onorato.
\newblock Evolution of weakly nonlinear random directional waves: laboratory experiments and numerical simulations.
\newblock {\em Journal of Fluid Mechanics}, 664:313--336, 2010.

\bibitem{Toubouletal2016}
J.~Touboul, J.~Charland, V.~Rey, and K.~Belibassakis.
\newblock Extended mild-slope equation for surface waves interacting with a vertically sheared current.
\newblock {\em Coastal Engineering}, 116:77--88, 2016.

\bibitem{veeramony1998boussinesq}
J.~Veeramony and I.~A. Svendsen.
\newblock A boussinesq model for breaking waves: comparisons with experiments.
\newblock In {\em Coastal Engineering 1998}, pages 258--271. 1998.

\bibitem{wang2022}
J.~Wang and P.~L.-F. Liu.
\newblock Modeling infra-gravity waves using {SCHISM-WWMIII} based on improved formulas and coupling approach.
\newblock In {\em the 37th International Conference on Coastal Engineering, Sydney, Australia}, 2022.

\bibitem{wang2015numerical}
J.~Wang and Q.~W. Ma.
\newblock Numerical investigation on limitation of boussinesq equation for generating focusing waves.
\newblock {\em Procedia Engineering}, 126:597--601, 2015.

\bibitem{wang2015}
J.~Wang and Q.~W. Ma.
\newblock Numerical techniques on improving computational efficiency of spectral boundary integral method.
\newblock {\em International Journal for Numerical Methods in Engineering}, 102(10):1638--1669, 2015.

\bibitem{wang2016}
J.~Wang, Q.~W. Ma, and S.~Yan.
\newblock A hybrid model for simulating rogue waves in random seas on a large temporal and spatial scale.
\newblock {\em Journal of Computational Physics}, 313:279--309, 2016.

\bibitem{wang2018}
J.~Wang, Q.~W. Ma, and S.~Yan.
\newblock A fully nonlinear numerical method for modeling wave--current interactions.
\newblock {\em Journal of Computational Physics}, 369:173--190, 2018.

\bibitem{wang2021extreme}
J.~Wang, Q.~W. Ma, and S.~Yan.
\newblock On extreme waves in directional seas with presence of oblique current.
\newblock {\em Applied Ocean Research}, 112:102586, 2021.

\bibitem{wang2021modeling}
J.~Wang, Q.~W. Ma, S.~Yan, and B.~Liang.
\newblock Modeling crossing random seas by fully non-linear numerical simulations.
\newblock {\em Frontiers in Physics}, 9:593394, 2021.

\bibitem{wang2022two}
J.~Wang, Q.~W. Ma, Z.~Yang, J.~Gao, and G.~Wu.
\newblock Two types of wave-current interactions and their effects on extreme waves in directional seas.
\newblock {\em Ocean Engineering}, 266:112637, 2022.

\bibitem{wang1995efficient}
P.~Wang, Y.~Yao, and M.~P. Tulin.
\newblock An efficient numerical tank for non-linear water waves, based on the multi-subdomain approach with bem.
\newblock {\em International journal for numerical methods in fluids}, 20(12):1315--1336, 1995.

\bibitem{wang2022flexible}
W.~Wang, C.~P{\'a}kozdi, A.~Kamath, S.~Fouques, and H.~Bihs.
\newblock A flexible fully nonlinear potential flow model for wave propagation over the complex topography of the norwegian coast.
\newblock {\em Applied Ocean Research}, 122:103103, 2022.

\bibitem{Wei1995}
G.~Wei, J.~T. Kirby, S.~T. Grilli, and R.~Subramanya.
\newblock {A Fully Nonlinear Boussinesq Model for Surface Waves. Part 1. Highly Nonlinear Unsteady Waves}.
\newblock {\em Journal of Fluid Mechanics}, 294:71--92, 1995.

\bibitem{west1987new}
B.~J. West, K.~A. Brueckner, R.~S. Janda, D.~M. Milder, and R.~L. Milton.
\newblock A new numerical method for surface hydrodynamics.
\newblock {\em Journal of Geophysical Research: Oceans}, 92(C11):11803--11824, 1987.

\bibitem{wu1994}
G.~X. Wu and R.~{Eatock-Taylor}.
\newblock Finite element analysis of two-dimensional non-linear transient water waves.
\newblock {\em Applied Ocean Research}, 16(6):363--372, 1994.

\bibitem{wu1995}
G.~X. Wu and R.~{Eatock-Taylor}.
\newblock Time stepping solutions of the two-dimensional nonlinear wave radiation problem.
\newblock {\em Ocean Engineering}, 22(8):785--798, 1995.

\bibitem{xiao2013rogue}
W.~Xiao, Y.~Liu, G.~Wu, and D.~K.~P. Yue.
\newblock Rogue wave occurrence and dynamics by direct simulations of nonlinear wave-field evolution.
\newblock {\em Journal of Fluid Mechanics}, 720:357--392, 2013.

\bibitem{xing2017}
Y.~Xing.
\newblock Numerical methods for the nonlinear shallow water equations.
\newblock In {\em Handbook of Numerical Analysis}. Elsevier, 2017.

\bibitem{xu2009numerical}
L.~Xu and P.~Guyenne.
\newblock Numerical simulation of three-dimensional nonlinear water waves.
\newblock {\em Journal of Computational Physics}, 228(22):8446--8466, 2009.

\bibitem{xue2001}
M.~Xue, H.~Xü, Y.~Liu, and D.~K.~P. Yue.
\newblock Computations of fully nonlinear three-dimensional wave–wave and wave–body interactions. {Part 1. Dynamics} of steep three-dimensional waves.
\newblock {\em Journal of Fluid Mechanics}, 438:11–39, 2001.

\bibitem{yan2006}
S.~Yan.
\newblock {\em Numerical simulation on nonlinear response of moored floating structures to steep waves}.
\newblock PhD thesis, City University London, 2006.

\bibitem{yan2010}
S.~Yan and Q.~W. Ma.
\newblock {QALE-FEM for modelling 3D overturning waves}.
\newblock {\em International Journal for Numerical Methods in Fluids}, 63(6):743--768, 2010.

\bibitem{yang2020}
Z.~Yang and P.~L.-F. Liu.
\newblock Depth-integrated wave--current models. {Part 1. Two-dimensional} formulation and applications.
\newblock {\em Journal of Fluid Mechanics}, 883:A4, 2020.

\bibitem{yang2022depth}
Z.~Yang and P.~L.-F. Liu.
\newblock Depth-integrated wave--current models. part 2. current with an arbitrary profile.
\newblock {\em Journal of Fluid Mechanics}, 936:A31, 2022.

\bibitem{yates2015accuracy}
M.~L. Yates and M.~Benoit.
\newblock Accuracy and efficiency of two numerical methods of solving the potential flow problem for highly nonlinear and dispersive water waves.
\newblock {\em International Journal for Numerical Methods in Fluids}, 77(10):616--640, 2015.

\bibitem{yokota2011biomolecular}
R.~Yokota, J.~P. Bardhan, M.~G. Knepley, L.~A. Barba, and T.~Hamada.
\newblock Biomolecular electrostatics using a fast multipole bem on up to 512 gpus and a billion unknowns.
\newblock {\em Computer Physics Communications}, 182(6):1272--1283, 2011.

\bibitem{zakharov1968}
V.~E. Zakharov.
\newblock Stability of periodic waves of finite amplitude on the surface of a deep fluid.
\newblock {\em Journal of Applied Mechanics and Technical Physics}, 9(2):190--194, 1968.

\end{thebibliography}
%Use of the above commands will create a bibliography using the .bib file. Shown below is a bibliography built from individual items.

%% End of file `jfm2esam.bib'.

\end{document}